\RequirePackage[l2tabu, orthodox]{nag}
\documentclass[12pt,a4paper]{article}

\usepackage{fixltx2e}

\usepackage[font=small]{caption}
\usepackage[text={450pt,650pt},headheight={14.5pt},centering]{geometry}

\usepackage{lmodern}

\usepackage{graphicx}
\usepackage{booktabs}
\usepackage{subfig}
\usepackage{tikz}
\usetikzlibrary{plotmarks,calc,decorations,decorations.pathmorphing}

\tikzstyle dynkin node=[very thick,shape=circle,draw,inner sep=0pt,minimum size=5mm]
\tikzstyle dynkin line=[very thick]
\tikzstyle inverse line=[gray,line width=1.46pt,line cap=round, dash pattern=on 0pt off 2\pgflinewidth]
\tikzstyle red phase=[red,decoration={snake,amplitude=0.1mm,segment length=1.6mm},decorate]
\tikzstyle blue phase=[blue,decoration={snake,amplitude=0.1mm,segment length=0.9mm},decorate]

\pgfdeclarelayer{background}
\pgfdeclarelayer{foreground}
\pgfsetlayers{background,main,foreground}

\usepackage{dsfont}
\usepackage{collref}

\usepackage{amsmath,amssymb}
\usepackage{mathrsfs}
\usepackage{mathtools}
\usepackage{bbm}
\usepackage{slashed}
\usepackage{braket}

\DeclareMathAlphabet{\mathsfit}{\encodingdefault}{\sfdefault}{m}{sl}

\usepackage{hyperref}

\usepackage{xspace}

\numberwithin{equation}{section}

\makeatletter
 \let\old@startsection=\@startsection
 \let\oldl@section=\l@section
 \renewcommand{\@startsection}[6]{\old@startsection{#1}{#2}{#3}{#4}{#5}{#6\mathversion{bold}}}
 \renewcommand{\l@section}[2]{\oldl@section{\mathversion{bold}#1}{#2}}
\makeatother

\newcommand{\fixedspaceL}[2]{\mathrlap{#2}\phantom{#1}}
\newcommand{\fixedspaceR}[2]{\phantom{#1}\mathllap{#2}}
\renewcommand{\geq}{\geqslant}

\DeclareMathOperator{\diag}{diag}

\DeclareMathOperator{\str}{str}

\def\Xint#1{\mathchoice
  {\XXint\displaystyle\textstyle{#1}}%
  {\XXint\textstyle\scriptstyle{#1}}%
  {\XXint\scriptstyle\scriptscriptstyle{#1}}%
  {\XXint\scriptscriptstyle\scriptscriptstyle{#1}}%
  \!\int}
\def\XXint#1#2#3{{\setbox0=\hbox{$#1{#2#3}{\int}$}
    \vcenter{\hbox{$#2#3$}}\kern-.5\wd0}}
\def\pint{\;\Xint-}

\newcommand{\AdS}{\text{AdS}}
\newcommand{\CFT}{\mathrm{CFT}}
\newcommand{\Sphere}{S}
\newcommand{\Torus}{T}
\newcommand{\CP}{\mathrm{CP}}

\newcommand{\mat}[1]{\mathbb{#1}}
\newcommand{\matId}{\mathds{1}}

\newcommand{\Smat}{\mathcal{S}}

\DeclareMathOperator*{\res}{Res}

\newcommand{\comm}[2]{[#1,#2]}
\newcommand{\acomm}[2]{\{#1,#2\}}

\newcommand{\alg}[1]{\mathfrak{#1}}

\newcommand{\algD}[1]{\alg{d}(2,1;#1)}

\newcommand{\algSL}{\alg{sl}}

\newcommand{\algSU}{\alg{su}}

\newcommand{\algU}{\alg{u}}

\newcommand{\algPSU}{\alg{psu}}

\newcommand{\algGL}{\alg{gl}}

\newcommand{\gen}[1]{\mathfrak{#1}}
\newcommand{\Ygen}[1]{#1}

\newcommand{\genQL}{\gen{Q}^{\smallL}}
\newcommand{\genQR}{\gen{Q}^{\smallR}}
\newcommand{\genSL}{\gen{S}^{\smallL}}
\newcommand{\genSR}{\gen{S}^{\smallR}}
\newcommand{\genHL}{\gen{H}^{\smallL}}
\newcommand{\genHR}{\gen{H}^{\smallR}}

\newcommand{\genP}{\gen{P}}
\newcommand{\genK}{\gen{P}^{\dag}}

\newcommand{\smallL}{\scriptscriptstyle\textit{L}}
\newcommand{\smallR}{\scriptscriptstyle\textit{R}}

\newcommand{\smallLL}{\scriptscriptstyle\textit{LL}}
\newcommand{\smallLR}{\scriptscriptstyle\textit{LR}}

\newcommand{\smallRR}{\scriptscriptstyle\textit{RR}}

\newcommand{\I}{\text{I}}
\newcommand{\II}{\text{II}}

\newcommand{\Integers}{\mathbbm{Z}}

\newcommand{\order}{\mathcal{O}}

\newcommand{\superN}{\mathcal{N}}

\newcommand{\ie}{\textit{i.e.}\xspace}
\newcommand{\eg}{\textit{e.g.}\xspace}

\begin{document}
\thispagestyle{empty}
\begin{flushright}\footnotesize\ttfamily
ITP-UU-13/08\\
SPIN-13/05\\
DMUS-MP-13/08
\end{flushright}
\vspace{4em}

\begin{center}
\textbf{\Large\mathversion{bold} The all-loop integrable spin-chain for strings on $\AdS_3\times \Sphere^3\times \Torus^4$: the massive sector }

\vspace{2em}

\textrm{\large Riccardo Borsato${}^1$, Olof Ohlsson Sax${}^1$, Alessandro Sfondrini${}^1$,\\ Bogdan Stefa\'nski, jr.${}^2$ and Alessandro Torrielli${}^3$} 

\vspace{2em}

\vspace{1em}
\begingroup\itshape
1. Institute for Theoretical Physics and Spinoza Institute, Utrecht University, Leuvenlaan 4, 3584 CE Utrecht, The Netherlands

2. Centre for Mathematical Science, City University London, Northampton Square, EC1V 0HB London, UK

3. Department of Mathematics, University of Surrey, Guildford, GU2 7XH, UK\par\endgroup

\vspace{1em}

\texttt{R.Borsato@uu.nl, O.E.OlssonSax@uu.nl, A.Sfondrini@uu.nl, Bogdan.Stefanski.1@city.ac.uk, a.torrielli@surrey.ac.uk}

%%%%%%%%

\end{center}

\vspace{6em}

\begin{abstract}\noindent
We bootstrap the all-loop dynamic S-matrix for the homogeneous~$\algPSU(1,1|2)^2$ spin-chain believed to correspond to the discretization of the massive modes of string theory on~$\AdS_3\times \Sphere^3\times \Torus^4$. The S-matrix is the tensor product of two copies of the~$\algSU(1|1)^2$ invariant S-matrix constructed recently for the $\algD{\alpha}^2$ chain, and depends on two antisymmetric dressing phases. We write down the crossing equations that these phases have to satisfy. Furthermore, we present the corresponding Bethe Ansatz, which differs from the one previously conjectured, and discuss how our construction matches several recent perturbative calculations. 
\end{abstract}

\newpage

\section{Introduction}
\label{sec:introduction}

The AdS/CFT correspondence~\cite{Maldacena:1997re,Witten:1998qj,Gubser:1998bc} is a remarkable framework for relating quantum gauge and gravity theories. In certain, often highly (super-)symmetric settings, underlying the duality is an integrable structure.\footnote{For a recent review and an extensive list of references see~\cite{Beisert:2010jr}.} The presence of integrability gives a powerful tool for finding the mass spectrum of all perturbative states in the gauge/gravity dual pair in the planar limit. Initially, integrability was best understood in the case of $\mathcal{N}=4$ super-Yang-Mills (SYM) and its type IIB string dual on $\AdS_5\times \Sphere^5$ supported by RR fluxes, as well as its orbifolds, orientifolds and deformations.  Later, the integrability approach was extended to the so-called ABJM super-Chern-Simons theory and its type IIA string dual on $\AdS_4\times \CP^3$.  This provided a second class of examples of a gauge/string dual pair in which the spectral problem is very well understood. The $ \AdS_4/\CFT_3$ dual pair has less supersymmetry than the original $\AdS_5/\CFT_4$,\footnote{The $\AdS_5/\CFT_4$ and $\AdS_4/\CFT_3$  dual pairs have 32 and 24 real supercharges, respectively.} 
and has a number of new features, such as Chern-Simons gauge fields, as well as scalars that transform in the bifundamental representation of the gauge group. It is a remarkable fact that much of the integrability machinery could be extended to this setting too.

Another interesting instance of the correspondence is the case of $\AdS_3/\CFT_2$, which constitutes one of the earliest examples of holography, where the conformal symmetry algebra is infinite dimensional~\cite{Brown:1986nw}. The maximally supersymmetric string backgrounds are $\AdS_3\times \Sphere^3\times \Torus^4$ and  $\AdS_3\times \Sphere^3\times \Sphere^3\times \Sphere^1$, both preserving 16 supercharges. Both backgrounds can be supported by RR or NSNS fluxes. In the particular case of pure NSNS background it was possible to effectively describe the theories using the NSR formalism~\cite{Maldacena:2000hw,Maldacena:2000kv,Maldacena:2001km}. The CFT corresponding to the former background is expected to be a deformation of the $Sym^N(\Torus^4)$ orbifold theory, and have small $\superN = (4,4)$ superconformal symmetry~\cite{Maldacena:1998bw,Seiberg:1999xz,Larsen:1999uk}; the one corresponding to  $\AdS_3\times \Sphere^3\times \Sphere^3\times \Sphere^1$ has large $\superN = (4,4)$ superconformal symmetry but is much less well understood~\cite{Gauntlett:1998kc,Elitzur:1998mm,deBoer:1999rh,Gukov:2004ym}. Their finite dimensional sub-algebras are $\algPSU(1,1|2)^2$ and $\algD\alpha{}^2$ respectively, corresponding to the superisometries of each background.

The success of integrability in the more supersymmetric instances sparked new interest in the $\AdS_3/\CFT_2$ correspondence. Even if integrability in the CFT side remains challenging~\cite{Pakman:2009mi}, it has been shown that the RR string non-linear sigma model is classically integrable~\cite{Babichenko:2009dk,Sundin:2012gc}.\footnote{This was later extended to backgrounds supported by a mixture of NSNS and RR fluxes, see~\cite{Cagnazzo:2012se}.} As a result, one can make progress by discretising the string world-sheet and reducing the integrable system to a set of Bethe Ansatz equations~\cite{Babichenko:2009dk,OhlssonSax:2011ms}. However, this procedure keeps track only of the excitations which remain massive in the BMN limit. Fully incorporating the massless modes remains an open issue.\footnote{For recent progress on this, see~\cite{Sax:2012jv}.}

At weak coupling the Bethe equations yield the spectrum of an integrable spin-chain. The symmetry algebra preserving the vacuum of the spin-chain can be extended by three central charges~\cite{David:2008yk,Babichenko:2009dk} and so one may attempt to construct an S-matrix generalising the  bootstrap method presented in~\cite{Beisert:2005tm} for the S-matrix of fundamental particles of the $\AdS_5/\CFT_4$ correspondence. In~\cite{Borsato:2012ud}, this approach was applied to the alternating spin-chain related to string theory on $\AdS_3\times \Sphere^3\times \Sphere^3\times \Sphere^1$ constructed in~\cite{OhlssonSax:2011ms}, yielding a set of all-loop Bethe equations~\cite{Borsato:2012ss}.

One may be tempted to consider the $\AdS_3\times \Sphere^3\times \Torus^4$ background as a limiting case of the $\AdS_3\times \Sphere^3\times \Sphere^3\times \Sphere^1$ one, where one of the spheres blows up. While this is certainly true for the gauge fixed string Hamiltonian, such a limit (which amounts to sending~$\alpha\to1$) is quite subtle at the level of the symmetry algebra, of its representations, and therefore of the invariant S-matrix. In particular, the notion of fundamental and composite excitations is different in the two cases, so that one cannot simply send $\alpha\to1$ in the S-matrix and Bethe ansatz of~\cite{Borsato:2012ud,Borsato:2012ss}. Instead, one should repeat the bootstrap procedure from scratch for the $\algPSU(1,1|2)^2$ chain of~\cite{OhlssonSax:2011ms}. This is the main aim of this work. 

This paper is organised as follows. First, in section~\ref{sec:psu112-spin-chain} we review the~$\algPSU(1,1|2)^2$ chain, discuss the symmetry algebra of its vacuum, and its central extension. Out of these symmetries we write down the all-loop S-matrix in section~\ref{sec:S-matrix}, where we also show that it solves the Yang-Baxter equation; the S-matrix is fixed up to two antisymmetric scalar factors, which we constrain by imposing crossing symmetry. In section~\ref{sec:diag-BA} we diagonalise the S-matrix, write down the resulting Bethe ansatz, which describes the massive sector of the corresponding string theory, and discuss its weak and strong coupling limits. Finally, in section~\ref{sec:pertubative-S-matrix} we compare our results with perturbative calculations appeared in the literature, and section~\ref{sec:conclusions} is devoted to some concluding remarks. In appendix~\ref{sec:fermionic-duality} we work out the dualities of the Bethe ansatz we found, and in the following two appendices we discuss how integrability for the $\AdS_3/\CFT_2$ correspondence can be framed in the context of Hopf algebras and Yangians.

\section{The \texorpdfstring{$\algPSU(1,1|2)^2$}{psu(1,1|2) x psu(1,1|2)} symmetric spin-chain}
\label{sec:psu112-spin-chain}

In this section we will review the $\algPSU(1,1|2)^2$ homogeneous spin-chain originally constructed in~\cite{OhlssonSax:2011ms}. The two copies of the superalgebra $\algPSU(1,1|2)$ describe the left- and right-moving sectors of string theory in $\AdS_3 \times \Sphere^3 \times \Torus^4$. The weak coupling limit of this spin-chain was analysed in~\cite{OhlssonSax:2011ms}. In this limit, the left- and right-movers decouple. The spectrum is then described by two homogeneous spin-chains with the sites of each transforming in the representation $(-\tfrac{1}{2};\tfrac{1}{2})$ of $\algPSU(1,1|2)$. At higher loops the two sectors couple to each other through local interactions. The full spin-chain will be discussed in more detail below.

\paragraph{The algebra.}
The superalgebra $\algPSU(1,1|2)$ has, in the relevant real form, a bosonic subalgebra $\algSL(2) \otimes \algSU(2)$. The corresponding generators are denoted by $\gen{S}_0$, $\gen{S}_{\pm}$ and $\gen{L}_5$, $\gen{L}_{\pm}$, respectively. Additionally, there are eight supercharges $\gen{Q}_{\pm\pm\pm}$. The commutation relations of $\algPSU(1,1|2)$ algebra read
\begin{equation*}
  \begin{aligned}
    \comm{\gen{S}_0}{\gen{S}_\pm} &= \pm \gen{S}_\pm , &
    \comm{\gen{S}_+}{\gen{S}_-} &= 2 \gen{S}_0 , &
    \comm{\gen{S}_0}{\gen{Q}_{\pm\beta\dot{\beta}}} &= \pm\frac{1}{2} \gen{Q}_{\pm\beta\dot{\beta}} , &
    \comm{\gen{S}_\pm}{\gen{Q}_{\mp\beta\dot{\beta}}} &= \gen{Q}_{\pm\beta\dot{\beta}} , \\
    \comm{\gen{L}_5}{\gen{L}_\pm} &= \pm \gen{L}_\pm , &
    \comm{\gen{L}_+}{\gen{L}_-} &= 2 \gen{L}_5 , &
    \comm{\gen{L}_5}{\gen{Q}_{b\pm\dot{\beta}}} &= \pm\frac{1}{2} \gen{Q}_{b\pm\dot{\beta}} , &
    \comm{\gen{L}_\pm}{\gen{Q}_{b\mp\dot{\beta}}} &= \gen{Q}_{b\pm\dot{\beta}} , \\
  \end{aligned}
\end{equation*}
\begin{equation}
  \begin{aligned}
    \acomm{\gen{Q}_{\pm++}}{\gen{Q}_{\pm--}} &= \pm \gen{S}_{\pm} , \! &
    \acomm{\gen{Q}_{\pm+-}}{\gen{Q}_{\pm-+}} &= \mp \gen{S}_{\pm} , \! &
    \acomm{\gen{Q}_{+\pm\pm}}{\gen{Q}_{-\mp\mp}} &= - \gen{S}_0 \pm \gen{L}_5 , \\
    \acomm{\gen{Q}_{+\pm+}}{\gen{Q}_{-\pm-}} &= \mp \gen{L}_{\pm} , \! &
    \acomm{\gen{Q}_{+\pm-}}{\gen{Q}_{-\pm+}} &= \pm \gen{L}_{\pm} , \! &
    \acomm{\gen{Q}_{+\pm\mp}}{\gen{Q}_{-\mp\pm}} &= + \gen{S}_0 \mp \gen{L}_5 .
  \end{aligned}
\end{equation}
The $\algPSU(1,1|2)$ algebra admits a $\algU(1)$ automorphism generated by $\gen{R}_8$ and acting on the supercharges as\footnote{%
  We denote the automorphism by $\gen{R}_8$ since it is related to one of the $\algSU(2)$ generators of the superalgebra $\algD{\alpha}$. In the limit $\alpha \to 0,1$ the algebra $\algD{\alpha}$ turns into $\algPSU(1,1|2)$. This relation is described in more detail in~\cite{OhlssonSax:2011ms}.%
} %
\begin{equation}
  \comm{\gen{R}_8}{\gen{Q}_{b\beta\pm}} = \pm \frac{1}{2} \gen{Q}_{b\beta\pm} ,
\end{equation}
and commuting with the bosonic charges.

The full symmetry of the spin-chain is $\algPSU(1,1|2) \times \algPSU(1,1|2)$. When we need to distinguish the left- and right-moving copies of the algebra we will add an additional subscript, taking values $L$ and $R$, to the generators.

\paragraph{Serre-Chevalley bases.}

For superalgebras there are in general several inequivalent Dynkin diagrams, corresponding to different choices of simple roots. Each such choice corresponds to a set of Cartan generators $\gen{h}_i$, and corresponding raising and lowering operators $\gen{e}_i$ and $\gen{f}_i$, where the index $i$ takes values from $1$ to the rank of the algebra, which is $3$ for $\algPSU(1,1|2)$. These generators satisfy an algebra of the form
\begin{equation}
  \comm{\gen{h}_i}{\gen{h}_j} = 0 , \qquad
  \comm{\gen{e}_i}{\gen{f}_j} = \delta_{ij} \gen{h}_j , \qquad
  \comm{\gen{h}_i}{\gen{e}_j} = + A_{ij} \gen{e}_j , \qquad
  \comm{\gen{h}_i}{\gen{f}_j} = - A_{ij} \gen{f}_j ,
\end{equation}
where $A_{ij}$ is the Cartan matrix.

In this paper we will mainly consider two gradings of $\algPSU(1,1|2)$. In the $\algSU(2)$ grading the simple roots are given by
\begin{equation}\label{eq:SC-basis-su2}
  \begin{aligned}
    \gen{h}_1 &= -\gen{S}_0 - \gen{L}_5 , \qquad &
    \gen{e}_1 &= +\gen{Q}_{+--} , \qquad &
    \gen{f}_1 &= +\gen{Q}_{-++} , \\
    \gen{h}_2 &= +2\gen{L}_5 , \qquad &
    \gen{e}_2 &= +\gen{L}_+ , \qquad &
    \gen{f}_2 &= +\gen{L}_- , \\
    \gen{h}_3 &= -\gen{S}_0 - \gen{L}_5 , \qquad &
    \gen{e}_3 &= +\gen{Q}_{+-+} , \qquad &
    \gen{f}_3 &= -\gen{Q}_{-+-} .
  \end{aligned}
\end{equation}
This leads to the Cartan matrix
\begin{equation}\label{eq:Cartan-su2}
  \begin{pmatrix}
     0 & -1 &  0 \\
    -1 & +2 & -1 \\
     0 & -1 &  0
  \end{pmatrix}.
\end{equation}
The corresponding Dynkin diagram is shown in figure~\ref{fig:dynkin-su22}~\subref{fig:dynkin-su22-su}.
\begin{figure}
  \centering

  \subfloat[\label{fig:dynkin-su22-su}]{
    \begin{tikzpicture}
      [
      thick,
      node/.style={shape=circle,draw,thick,inner sep=0pt,minimum size=5mm}
      ]

      \useasboundingbox (-1.5cm,-1cm) rectangle (1.5cm,1cm);

      \node (v1) at (-1.1cm, 0cm) [node] {};
      \node (v2) at (  0.0cm, 0cm) [node] {};
      \node (v3) at (  1.1cm, 0cm) [node] {};

      \draw (v1.south west) -- (v1.north east);
      \draw (v1.north west) -- (v1.south east);

      \draw (v3.south west) -- (v3.north east);
      \draw (v3.north west) -- (v3.south east);

      \draw (v1) -- (v2);
      \draw (v2) -- (v3);

      \node at (v2.south) [anchor=north] {$+1$};
    \end{tikzpicture}
  }
  \hspace{1cm}
  \subfloat[\label{fig:dynkin-su22-fff}]{
    \begin{tikzpicture}
      [
      thick,
      node/.style={shape=circle,draw,thick,inner sep=0pt,minimum size=5mm}
      ]

      \useasboundingbox (-1.5cm,-1cm) rectangle (1.5cm,1cm);

      \node (v1) at (-1.1cm, 0cm) [node] {};
      \node (v2) at (  0.0cm, 0cm) [node] {};
      \node (v3) at (  1.1cm, 0cm) [node] {};

      \draw (v1.south west) -- (v1.north east);
      \draw (v1.north west) -- (v1.south east);

      \draw (v2.south west) -- (v2.north east);
      \draw (v2.north west) -- (v2.south east);

      \draw (v3.south west) -- (v3.north east);
      \draw (v3.north west) -- (v3.south east);

      \draw (v1) -- (v2);
      \draw (v2) -- (v3);

      \node at (v2.south) [anchor=north] {$\pm 1$};
    \end{tikzpicture}
  }
  \hspace{1cm}
  \subfloat[\label{fig:dynkin-su22-sl}]{
    \begin{tikzpicture}
      [
      thick,
      node/.style={shape=circle,draw,thick,inner sep=0pt,minimum size=5mm}
      ]

      \useasboundingbox (-1.5cm,-1cm) rectangle (1.5cm,1cm);

      \node (v1) at (-1.1cm, 0cm) [node] {};
      \node (v2) at (  0.0cm, 0cm) [node] {};
      \node (v3) at (  1.1cm, 0cm) [node] {};

      \draw (v1.south west) -- (v1.north east);
      \draw (v1.north west) -- (v1.south east);

      \draw (v3.south west) -- (v3.north east);
      \draw (v3.north west) -- (v3.south east);

      \draw (v1) -- (v2);
      \draw (v2) -- (v3);

      \node at (v2.south) [anchor=north] {$-1$};
    \end{tikzpicture}
  }
  
  \caption{Three Dynkin diagrams for $\algPSU(1,1|2)$.}
  \label{fig:dynkin-su22}
\end{figure}

In the $\algSL(2)$ grading we have
\begin{equation}\label{eq:SC-basis-sl2}
  \begin{aligned}
    \hat{\gen{h}}_1 &= +\gen{S}_0 + \gen{L}_5 , \qquad &
    \hat{\gen{e}}_1 &= -\gen{Q}_{-++} , \qquad &
    \hat{\gen{f}}_1 &= +\gen{Q}_{+--} , \\
    \hat{\gen{h}}_2 &= -2\gen{S}_0 , \qquad &
    \hat{\gen{e}}_2 &= +\gen{S}_+ , \qquad &
    \hat{\gen{f}}_2 &= -\gen{S}_- , \\
    \hat{\gen{h}}_3 &= +\gen{S}_0 + \gen{L}_5 , \qquad &
    \hat{\gen{e}}_3 &= -\gen{Q}_{-+-} , \qquad &
    \hat{\gen{f}}_3 &= -\gen{Q}_{+-+} ,
  \end{aligned}
\end{equation}
with the Cartan matrix
\begin{equation}\label{eq:Cartan-sl2}
  \begin{pmatrix}
     0 & +1 &  0 \\
    +1 & -2 & +1 \\
     0 & +1 &  0
  \end{pmatrix}.
\end{equation}
The Dynkin diagram for the $\algSL(2)$ grading is shown in figure~\ref{fig:dynkin-su22}~\subref{fig:dynkin-su22-sl}.

There are also fermionic gradings of~$\algPSU(1,1|2)$, in which all three raising operators $\gen{e}_i$ are odd. In particular we can choose them to be either
\begin{equation*}
  \gen{Q}_{+-+}, \quad \gen{Q}_{++-}, \quad \gen{Q}_{-++} \,, \qquad 
  \text{or} \qquad
  \gen{Q}_{-+-}, \quad \gen{Q}_{--+}, \quad \gen{Q}_{+--} \,.\,
\end{equation*}
This leads to the Cartan matrices
\begin{equation}\label{eq:Cartan-ferm}
    \begin{pmatrix}
     0 & +1 &  0 \\
    +1 & 0 & -1 \\
     0 & -1 &  0
  \end{pmatrix} , \qquad
  \text{and} \qquad
    \begin{pmatrix}
     0 & -1 &  0 \\
    -1 & 0 & +1 \\
     0 & +1 &  0
  \end{pmatrix},
\end{equation}
respectively, corresponding to the Dynkin diagram in figure~\ref{fig:dynkin-su22}~\subref{fig:dynkin-su22-fff}.

\paragraph{The spin-chain representation.}

The sites of the $\algPSU(1,1|2)$ spin-chain transform in the infinite dimensional representation $(-\tfrac{1}{2};\tfrac{1}{2})$, consisting of the bosonic $\algSU(2)$ doublet $\phi^{(n)}_{\pm}$ and the two fermionic $\algSU(2)$ singlets $\psi^{(n)}_{\pm}$, where the index $n$ indicates the $\algSL(2)$ quantum number. The action of the generators on these states is given by
\begin{equation}\label{eq:su112-representation}
  \begin{gathered}
      \gen{L}_5 \ket{\phi_{\pm}^{(n)}} = \pm \frac{1}{2} \ket{\phi_{\pm}^{(n)}} , \qquad
      \gen{L}_+ \ket{\phi_{-}^{(n)}} = \ket{\phi_{+}^{(n)}} , \qquad
      \gen{L}_- \ket{\phi_{+}^{(n)}} = \ket{\phi_-^{(n)}} , \\
    \begin{aligned}
      \gen{S}_0 \ket{\phi_{\beta}^{(n)}} &= - \left( \tfrac{1}{2} + n \right) \ket{\phi_{\beta}^{(n)}} , &
      \gen{S}_0 \ket{\psi_{\dot\beta}^{(n)}} &= - \left( 1 + n \right) \ket{\psi_{\dot\beta}^{(n)}} , \\
      \gen{S}_+ \ket{\phi_{\beta}^{(n)}} &= +n \ket{\phi_{\beta}^{(n-1)}} , &
      \gen{S}_+ \ket{\psi_{\dot\beta}^{(n)}} &= +\sqrt{(n + 1)n} \ket{\psi_{\dot\beta}^{(n-1)}} , \\
      \gen{S}_- \ket{\phi_{\beta}^{(n)}} &= -(n+1)\ket{\phi_{\beta}^{(n+1)}} , &
      \gen{S}_- \ket{\psi_{\dot\beta}^{(n)}} &= -\sqrt{(n + 2) (n + 1)} \ket{\psi_{\dot\beta}^{(n+1)}} ,
    \end{aligned} \\
    \begin{aligned}
      \gen{Q}_{-\pm\dot\beta} \ket{\phi_{\mp}^{(n)}} &= \pm \sqrt{n+1} \ket{\psi_{\dot\beta}^{(n)}} , &
      \gen{Q}_{+\pm\dot\beta} \ket{\phi_{\mp}^{(n)}} &= \pm \sqrt{n} \ket{\psi_{\dot\beta}^{(n-1)}} , \\
      \gen{Q}_{-\beta\pm} \ket{\psi_{\mp}^{(n)}} &= \mp \sqrt{n+1} \ket{\phi_{\beta}^{(n+1)}} , &
      \gen{Q}_{+\beta\pm} \ket{\psi_{\mp}^{(n)}} &= \mp \sqrt{n+1} \ket{\phi_{\beta}^{(n)}} .
    \end{aligned}
  \end{gathered}
\end{equation}
The highest weight state $\ket{\phi^{(0)}_+}$ is annihilated by the $\algSU(2)$ grading raising operators $\gen{Q}_{+\pm\pm}$, as well as by the two generators $\gen{Q}_{-+\pm}$. Hence, the representation $(-\tfrac{1}{2};\tfrac{1}{2})$ is a short representation, satisfying the shortening conditions
\begin{equation}
  \acomm{\gen{Q}_{+-\mp}}{\gen{Q}_{-+\pm}} \ket{\phi^{(0)}_+} = \mp (\gen{S}_0 + \gen{L}_5 ) \ket{\phi^{(0)}_+} = 0.
\end{equation}
%The sites of the left- and right-moving spin-chains transform in identical modules. To distinguish the states we denote the states in the right-moving sector by additional bar, so that the above basis is written as $\bar{\phi}^{(n)}_{\pm}$ and $\bar{\psi}^{(n)}_{\pm}$.

\paragraph{The ground state.}

The states of the left- and right-moving spin-chains of length\footnote{%
  The symbol $L$ is used to denote the length and to denote left-moving generators and excitations. Hopefully the meaning will be clear from the context.%
} %
$L$ transform in the $L$-fold tensor product of the above representation. The ground state of the full spin-chain is given by
\begin{equation}
  \ket{0}_L = \Ket{(\phi^{(0)}_+)^L} \otimes \Ket{(\phi^{(0)}_+)^L}.
\end{equation}
This is the highest weight state of the short 1/2-BPS representation $(-\tfrac{L}{2};\tfrac{L}{2}) \otimes (-\tfrac{L}{2};\tfrac{L}{2})$ of $\algPSU(1,1|2) \times \algPSU(1,1|2)$. In each sector this state is constructed from the spin-chain module discussed above. Hence, the ground state is preserved by eight supercharges $\gen{Q}_i^{\scriptscriptstyle I}$ and $\gen{S}_i^{\scriptscriptstyle I}$, with $i=1,2$ and $I=L,R$, as well as two central charges $\gen{H}^{\scriptscriptstyle I}$, which in terms of the $\algPSU(1,1|2)$ generators are given by
\begin{equation}
  \begin{gathered}
    \gen{Q}_1^{\scriptscriptstyle I} = +\gen{Q}_{-++}^{\scriptscriptstyle I} , \quad
    \gen{Q}_2^{\scriptscriptstyle I} = -\gen{Q}_{-+-}^{\scriptscriptstyle I} , \quad
    \gen{S}_1^{\scriptscriptstyle I} = +\gen{Q}_{+--}^{\scriptscriptstyle I} , \quad
    \gen{S}_2^{\scriptscriptstyle I} = +\gen{Q}_{+-+}^{\scriptscriptstyle I} , \\
    \gen{H}^{\scriptscriptstyle I} = -\gen{S}_0^{\scriptscriptstyle I}- \gen{L}_5^{\scriptscriptstyle I},
  \end{gathered}
\end{equation}
This forms two $\algPSU(1|1)^2 \ltimes \algU(1)$ algebras
\begin{equation}\label{eq:su11-su11-algebra}
  \acomm{\gen{Q}_i^{\scriptscriptstyle I}}{\gen{S}_j^{\scriptscriptstyle J}} = \delta_{ij} \delta^{\scriptscriptstyle IJ} \gen{H}^{\scriptscriptstyle I}.
\end{equation}
The charges $\genHL$ and $\genHR$ are the left- and right-moving spin-chain Hamiltonians. It is useful to introduce the combinations
\begin{equation}
  \gen{H} = \genHL + \genHR , \qquad
  \gen{M} = \genHL - \genHR .
\end{equation}
The Hamiltonian $\gen{H}$ gives the energy of a spin-chain state, and depends on the momenta of the spin-chain excitations. The central charge $\gen{M}$ measures the angular momentum in $\AdS_3$ and should be independent of the spin-chain momentum.

We can introduce two additionally generators $\gen{B}_1$ and $\gen{B}_2$ acting as outer automorphisms on the above algebra. These can be constructed from the $\algPSU(1,1|2)$ generators $\gen{L}_5^{\scriptscriptstyle I}$ and the automorphisms $\gen{R}_8^{\scriptscriptstyle I}$,
\begin{equation}\label{eq:su11-automorphisms-def}
  \gen{B}_1 = - (\gen{R}_8^{\smallL} - \gen{R}_8^{\smallR}) - (\gen{L}_5^{\smallL} - \gen{L}_5^{\smallR}), \qquad
  \gen{B}_2 = + (\gen{R}_8^{\smallL} - \gen{R}_8^{\smallR}) - (\gen{L}_5^{\smallL} - \gen{L}_5^{\smallR}) .
\end{equation}
The above linear combinations are chosen so that $\gen{B}_1$ commutes with $\gen{Q}_2^{\scriptscriptstyle I}$ and $\gen{S}_2^{\scriptscriptstyle I}$, while $\gen{B}_2$ commutes with $\gen{Q}_1^{\scriptscriptstyle I}$ and $\gen{S}_1^{\scriptscriptstyle I}$. The commutation relations involving the supercharges then read
\begin{equation}
  \begin{aligned}
    \comm{\gen{B}_i}{\genQL_j} &= - \delta_{ij} \genQL_i , \qquad &
    \comm{\gen{B}_i}{\genSL_j} &= + \delta_{ij} \genSL_i , \\
    \comm{\gen{B}_i}{\genQR_j} &= + \delta_{ij} \genQR_i , \qquad &
    \comm{\gen{B}_i}{\genSR_j} &= - \delta_{ij} \genSR_i .
  \end{aligned}
\end{equation}
Taking the generators $\gen{B}_i$ into account it is useful to regroup the symmetry algebra into two copies of $\algU(1) \ltimes \algSU(1|1)^2$, with the generators given by
\begin{equation}
  \bigl\{\genQL_1,\genSL_1,\genQR_1,\genSR_1,\genHL,\genHR,\gen{B}_1\bigr\} , 
  \qquad \text{and} \qquad
  \bigl\{\genQL_2,\genSL_2,\genQR_2,\genSR_2,\genHL,\genHR,\gen{B}_2\bigr\} ,
\end{equation}
respectively. Since the central charges $\genHL$ and $\genHR$ are shared between the two $\algSU(1|1)^2$ algebras, the full symmetry preserving the ground state can be written as
\begin{equation}
  \left[ \algU(1) \ltimes \algPSU(1|1)^2 \right]^2 \ltimes \algU(1)^2 .
\end{equation}

\paragraph{Excitations.}

\begin{figure}
  \centering
  
  \begin{tikzpicture}[%
    box/.style={outer sep=1pt},
    Q node/.style={inner sep=1pt,outer sep=0pt},
    arrow/.style={-latex}
    ]%

    \node [box] (PhiM) at ( 0  , 1.5cm) {\small $\ket{\phi_-^{(0)}}$};
    \node [box] (PsiP) at (-1.5cm, 0cm) {\small $\ket{\psi_+^{(0)}}$};
    \node [box] (PsiM) at (+1.5cm, 0cm) {\small $\mathllap{-}\ket{\psi_-^{(0)}}$};
    \node [box] (PhiP) at ( 0  ,-1.5cm) {\small $\ket{\phi_+^{(1)}}$};

    \newcommand{\horshift}{0.09cm,0cm}
    \newcommand{\vershift}{0cm,0.10cm}
 
    \draw [arrow] ($(PhiM.west) +(\vershift)$) -- ($(PsiP.north)-(\horshift)$) node [pos=0.5,anchor=south east,Q node] {\scriptsize $+\gen{Q}_1$};
    \draw [arrow] ($(PsiP.north)+(\horshift)$) -- ($(PhiM.west) -(\vershift)$) node [pos=0.5,anchor=north west,Q node] {\scriptsize $+\gen{S}_1$};

    \draw [arrow] ($(PsiM.south)-(\horshift)$) -- ($(PhiP.east) +(\vershift)$) node [pos=0.5,anchor=south east,Q node] {\scriptsize $+\gen{Q}_1$};
    \draw [arrow] ($(PhiP.east) -(\vershift)$) -- ($(PsiM.south)+(\horshift)$) node [pos=0.5,anchor=north west,Q node] {\scriptsize $+\gen{S}_1$};

    \draw [arrow] ($(PhiM.east) -(\vershift)$) -- ($(PsiM.north)-(\horshift)$) node [pos=0.5,anchor=north east,Q node] {\scriptsize $+\gen{Q}_2$};
    \draw [arrow] ($(PsiM.north)+(\horshift)$) -- ($(PhiM.east) +(\vershift)$) node [pos=0.5,anchor=south west,Q node] {\scriptsize $+\gen{S}_2$};

    \draw [arrow] ($(PsiP.south)-(\horshift)$) -- ($(PhiP.west) -(\vershift)$) node [pos=0.5,anchor=north east,Q node] {\scriptsize $-\gen{Q}_2$};
    \draw [arrow] ($(PhiP.west) +(\vershift)$) -- ($(PsiP.south)+(\horshift)$) node [pos=0.5,anchor=south west,Q node] {\scriptsize $-\gen{S}_2$};
  \end{tikzpicture}

  \caption{The action of the supercharges $\gen{Q}_i$ and $\gen{S}_i$ on the bi-fundamental representation~\eqref{eq:bi-fund-fields}.}
  \label{fig:representation}
\end{figure}
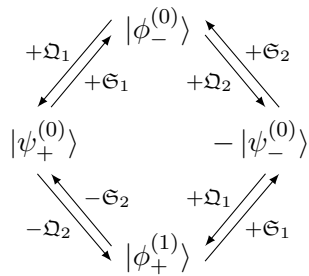

To construct excited spin-chain states we replace one or more of the ground state sites by any other state in the same module. We can classify these excitations by their eigenvalues under the left- and right-moving spin-chain Hamiltonians $\gen{H}^{\smallL}$ and $\gen{H}_{\smallR}$ at zero coupling. Let us consider excitations in the left-moving sector. Replacing one of the highest weight states $\phi^{(0)}_+$ by the scalar $\phi^{(n)}_-$ or $\phi^{(n)}_+$ increases the eigenvalue of $\gen{H}^{\smallL}$ by $n$ or $n+1$, respectively. Similarly, insertion of a fermion $\psi^{(n)}_{\pm}$ also adds $n$ to the energy. The lightest excitations are therefore
\begin{equation*}
  \phi_-^{(0)} , \qquad \psi_+^{(0)} , \qquad \psi_-^{(0)} , \qquad \text{and} \qquad \phi_+^{(1)}.
\end{equation*}
These states form a four-dimensional bi-fundamental representation of the left-moving $\algPSU(1|1)^2$ algebra~\eqref{eq:su11-su11-algebra}, as illustrated in figure~\ref{fig:representation}. To emphasize this we introduce the notation
\begin{equation}\label{eq:bi-fund-fields}
  \Phi^{+\dot{+}} = +\phi_-^{(0)} , \qquad
  \Phi^{-\dot{-}} = +\phi_+^{(1)} , \qquad
  \Phi^{-\dot{+}} = +\psi_+^{(0)} , \qquad
  \Phi^{+\dot{-}} = -\psi_-^{(0)} .
\end{equation}

We also introduce the symbols $\epsilon_{\gamma}$ and $\epsilon_{\dot{\gamma}}$ to keep track of the grading
\begin{equation}
  (-1)^{\epsilon_{\pm}} = \pm 1 , \qquad
  (-1)^{\epsilon_{\dot{\pm}}} = \pm 1 .
\end{equation}
The excitation $\Phi^{\gamma\dot{\gamma}}$ then has statistics $(-1)^{\epsilon_{\gamma}+\epsilon_{\dot{\gamma}}}$. 

To make the bi-fundamental nature of the representation above more explicit, we introduce an auxiliary fundamental $\algSU(1|1)$ representation with basis $(\phi|\psi)$, and the generators $\gen{Q}$, $\gen{S}$ and $\gen{H}$ acting as
\begin{equation}\label{eq:su11-repr}
  \gen{Q} \ket{\phi} = a \ket{\psi} , \qquad
  \gen{S} \ket{\psi} = b \ket{\phi} , \qquad
  \gen{H} \ket{\phi} = ab \ket{\phi} , \qquad
  \gen{H} \ket{\psi} = ab \ket{\psi} .
\end{equation}
We can then identify
\begin{equation}\label{eq:Phi-tensor}
  \begin{aligned}
    \Phi^{+\dot{+}} = \phi \otimes \phi , \qquad
    \Phi^{+\dot{-}} = \phi \otimes \psi , \qquad
    \Phi^{-\dot{+}} = \psi \otimes \phi , \qquad
    \Phi^{-\dot{-}} = \psi \otimes \psi .
  \end{aligned}
\end{equation}
The supercharges $\genQL_1$ and $\genSL_1$ only act on the first index of $\Phi^{\gamma\dot{\gamma}}$, while $\genQL_2$ and $\genSL_2$ act on the second index. Hence the action on the tensor products~\eqref{eq:Phi-tensor} is given by
\begin{equation}
  \begin{aligned}
    \genQL_1 = \gen{Q} \otimes 1 , \qquad
    \genSL_1 = \gen{S} \otimes 1 , \qquad
    \genQL_2 = 1 \otimes \gen{Q} , \qquad
    \genSL_2 = 1 \otimes \gen{S} .
  \end{aligned}
\end{equation}
Note that we get a minus sign from commuting a supercharge through a fermion when we act with the charges of the second type on a state with a fermion in the first part of the tensor product. Hence, the left-moving generators act as
\begin{equation}\label{eq:representation-LL}
  \begin{aligned}
    \genQL_1 \ket{\Phi^{+\dot{+}}} &= +a \ket{\Phi^{-\dot{+}}} , \qquad &
    \genQL_1 \ket{\Phi^{+\dot{-}}} &= +a \ket{\Phi^{-\dot{-}}} , \\
    \genSL_1 \ket{\Phi^{-\dot{+}}} &= +b \ket{\Phi^{+\dot{+}}} , &
    \genSL_1 \ket{\Phi^{-\dot{-}}} &= +b \ket{\Phi^{+\dot{-}}} , \\
    \genQL_2 \ket{\Phi^{+\dot{+}}} &= +a \ket{\Phi^{+\dot{-}}} , &
    \genQL_2 \ket{\Phi^{-\dot{+}}} &= -a \ket{\Phi^{-\dot{-}}} , \\
    \genSL_2 \ket{\Phi^{+\dot{-}}} &= +b \ket{\Phi^{+\dot{+}}} , &
    \genSL_2 \ket{\Phi^{-\dot{-}}} &= -b \ket{\Phi^{-\dot{+}}} ,
  \end{aligned}
\end{equation}
with the right-moving charges acting trivially. Comparing the above representation with~\eqref{eq:su112-representation} we find that the central charge $\genHL$ has eigenvalue $ab=1$.

Similarly, we introduce the right-moving excitations $\bar{\Phi}^{\pm\dot{\pm}}$, transforming in a bi-fundamental representation of the right-moving $\algPSU(1|1)^2$ algebra.

\paragraph{The centrally extended algebra.}

When we take quantum corrections into account, the spin-chain Hamiltonian $\gen{H}$ should depend on the coupling constant and on the momentum of the excitations. This requires the bi-fundamental representations discussed above to be deformed. However, this should be done in such a way that the angular momentum $\gen{M}$ remains undeformed. This means that we need to consider a generalized symmetry algebra in which the right-moving generators act nontrivially on the left-moving excitations, and \textit{vice versa}. Hence, we introduce two additional central charges $\gen{P}$ and $\gen{P}^\dag$ appearing in the anti-commutator between a left- and a right-moving supercharge
\begin{equation}
  \acomm{\genQL_i}{\genQR_j} = \delta_{ij} \gen{P} , \qquad
  \acomm{\genSL_i}{\genSR_j} = \delta_{ij} \gen{P}^{\dag} .
\end{equation}
Exactly like the central charges $\genHL$ and $\genHR$, these charges are shared between the two copies of $\algU(1) \ltimes \algSU(1|1)^2$. Hence, the extended algebra can be written as\footnote{%
  The role of the central extensions for $\algPSU(1|1)^2$ was originally discussed in~\cite{Beisert:2007sk}, in the context of $AdS_5/CFT_4$ duality, and in the case of $AdS_3/CFT_2$ in~\cite{David:2008yk,Babichenko:2009dk}.
  The same symmetry algebra was found in the analysis of the Pohlmeyer reduced sigma model of the $\AdS_3 \times \Sphere^3$ in~\cite{Hoare:2011fj}, and more recently in the world-sheet analysis of string theory in $\AdS_3 \times \Sphere^3 \times \Torus^4$ in~\cite{Hoare:2013pma}.%
} %
\begin{equation}\label{eq:su11-4-centrally-extended-algebra}
  \left[ \algU(1) \ltimes \algPSU(1|1)^2 \right]^2 \ltimes \algU(1)^4 .
\end{equation}
This is the maximal central extension of $\algPSU(1|1)^4$.

Since the charges in general are momentum dependent we will consider spin-chain states in which the excitations carry specific momenta. A one-excitation state can then be written as a plane wave
\begin{equation}
  \ket{\mathcal{X}_p} = \sum_{n=1}^{L} e^{ipn} \ket{ Z^{n-1} \mathcal{X} Z^{L-n} } ,
\end{equation}
where $\mathcal{X}$ is any left- or right-moving excitation, and we have introduced the shorthand notation $Z$ for a ground-state site. It is now straightforward to generalize this form to the case of multiple excitations. Note that we always consider \emph{asymptotic} states, where the spin-chain is considered to be very long and the excitations are well separated. The interactions are then described by the S-matrix permuting the order of excitations along the chain. 

In order to construct nontrivial representations of the extended algebra we need to allow the supercharges to have a \emph{length-changing} action on the spin-chain excitations. We therefore introduce two additional symbols $Z^\pm$ indicating the insertion or removal of a vacuum site next to an excitation. Writing out the plane waves we can commute these symbols through an excitation of momentum $p$ by picking up an extra phase factor
\begin{equation}
  \ket{Z^\pm \Phi^{\beta\dot{\beta}}_p} = e^{\mp ip} \ket{\Phi^{\beta\dot{\beta}}_p Z^\pm}
\end{equation}
Using these relations we can always shift any insertions of $Z^\pm$ through all excitations and collect them at the right end of the state.

A centrally extended $\algSU(1|1)^2$ algebra with dynamic spin-chain representations was considered in~\cite{Borsato:2012ud}. The supercharges act on the left-moving excitations by
\begin{equation}\label{eq:chiral-rep}
  \begin{aligned}
    \gen{Q}^{\smallL} \ket{\phi_p} &= a_p \ket{\psi_p} , \qquad &
    \gen{Q}^{\smallL} \ket{\psi_p} &= 0 , \\
    \gen{S}^{\smallL} \ket{\phi_p} &= 0 , \qquad &
    \gen{S}^{\smallL} \ket{\psi_p} &= b_p \ket{\phi_p} , \\
    \gen{Q}^{\smallR} \ket{\phi_p} &= 0 , \qquad &
    \gen{Q}^{\smallR} \ket{\psi_p} &= c_p \ket{\phi_p\,Z^+} , \\
    \gen{S}^{\smallR} \ket{\phi_p} &= d_p \ket{\psi_p\,Z^-} , \qquad &
    \gen{S}^{\smallR} \ket{\psi_p} &= 0 .
  \end{aligned}
\end{equation}
The left-moving charges in the above expressions act in the same way as in~\eqref{eq:su11-repr}. Hence, the bi-fundamental representations in~\eqref{eq:representation-LL} can be deformed to a representation of the centrally extended algebra ~\eqref{eq:su11-4-centrally-extended-algebra} by considering a tensor product of the representation \eqref{eq:chiral-rep}. We find that the 
left-moving generators act on the left-movers $\Phi^{\pm\dot{\pm}}$ in the same way as in~\eqref{eq:representation-LL}, but with the coefficients $a$ and $b$ depending on the momentum of the excitation. The action of the right-moving supercharges is given by
\begin{equation}
  \begin{aligned}
    \genQR_1 \ket{\Phi_p^{-\dot{-}}} &= +c_p \ket{\Phi_p^{+\dot{-}} Z^+} , \qquad &
    \genQR_1 \ket{\Phi_p^{-\dot{+}}} &= +c_p \ket{\Phi_p^{+\dot{+}} Z^+} , \\
    \genSR_1 \ket{\Phi_p^{+\dot{+}}} &= +d_p \ket{\Phi_p^{-\dot{+}} Z^-} , \qquad &
    \genSR_1 \ket{\Phi_p^{+\dot{-}}} &= +d_p \ket{\Phi_p^{-\dot{-}} Z^-} , \\
    \genQR_2 \ket{\Phi_p^{-\dot{-}}} &= -c_p \ket{\Phi_p^{-\dot{+}} Z^+} , \qquad &
    \genQR_2 \ket{\Phi_p^{+\dot{-}}} &= +c_p \ket{\Phi_p^{+\dot{+}} Z^+} , \\
    \genSR_2 \ket{\Phi_p^{+\dot{+}}} &= +d_p \ket{\Phi_p^{+\dot{-}} Z^-} , \qquad &
    \genSR_2 \ket{\Phi_p^{-\dot{+}}} &= -d_p \ket{\Phi_p^{-\dot{-}} Z^-} .
  \end{aligned}
\end{equation}
Closure of the algebra requires the central charges to act on the left-movers as
\begin{equation}
  \begin{aligned}
    \genHL \ket{\Phi_p^{\pm\dot{\pm}}} &= a_p b_p \ket{\Phi_p^{\pm\dot{\pm}}} , \qquad &
    \genP \ket{\Phi_p^{\pm\dot{\pm}}} &= a_p c_p \ket{\Phi_p^{\pm\dot{\pm}} Z^+} , \\
    \genHR \ket{\Phi_p^{\pm\dot{\pm}}} &= c_p d_p \ket{\Phi_p^{\pm\dot{\pm}}} , \qquad &
    \genK \ket{\Phi_p^{\pm\dot{\pm}}} &= b_p d_p \ket{\Phi_p^{\pm\dot{\pm}} Z^-} .
  \end{aligned}
\end{equation}

The central charges $\genP$ and $\genK$ are not part of the symmetries of the $\algPSU(1,1|2)^2$ spin-chain. They therefore need to vanish on a physical state. Acting with $\genP$ on a spin-chain state with two left-moving excitations, we obtain
\begin{equation}
  \genP \ket{\Phi^{\beta\dot{\beta}}_p \Phi^{\gamma\dot{\gamma}}_q} = (e^{-iq} a_p c_p + a_q c_q) \ket{\Phi^{\beta\dot{\beta}}_p \Phi^{\gamma\dot{\gamma}}_q}.
\end{equation}
Setting
\begin{equation}
  a_p c_p = h (e^{-ip} - 1) ,
\end{equation}
we find
\begin{equation}
  e^{-iq} a_p c_p + a_q c_q = h ( e^{-i(p+q)} - 1 ) ,
\end{equation}
which vanishes provided the excitations satisfy the momentum constraint $e^{i(p+q)} = 1$.
Similarly, requiring the charge $\genK$ to vanish on a physical state leads us to
\begin{equation}
  b_p d_p = h (e^{+ip} - 1) ,
\end{equation}
We furthermore require that the generator $\gen{M}$ remains undeformed when the central excitations are turned on, which gives
\begin{equation}
  a_p b_p - c_p d_p = 1 .
\end{equation}
The above conditions on the coefficients of the representation together uniquely determines the dispersion relation
\begin{equation}
  E(p) = a_p b_p + c_p d_p = \sqrt{1 + 16h^2\sin^2\frac{p}{2}}.
\end{equation}

To write down the explicit form of the coefficients $a_p,\dotsc,d_p$ it is useful to introduce the \emph{spectral parameters} $x^\pm$ satisfying~\cite{Beisert:2005tm}
\begin{equation}\label{eq:shortening}
  \frac{x_p^+}{x_p^-} = e^{ip} , \qquad
  \left(x_p^+ + \frac{1}{x_p^+}\right) - \left(x_p^- + \frac{1}{x_p^-}\right) = \frac{i}{h}.
\end{equation}
The coefficients can then be written as
\begin{align}
  a_p &= +\sqrt{h} \, \eta_p , \qquad &
  b_p &= +\sqrt{h} \, \eta_p , \qquad &
  c_p &= -\sqrt{h} \, \frac{i \eta_p}{x_p^+} , \qquad &
  d_p &= +\sqrt{h} \, \frac{i \eta_p}{x_p^-} ,
\end{align}
where 
\begin{equation}\label{eq:eta-p-def}
  \eta_p = \sqrt{i(x_p^- - x_p^+)}.
\end{equation}
The above expressions are essentially the same as those given in~\cite{Borsato:2012ud}, apart from a rescaling of the coupling constant $h$ by a factor $2$ and setting the parameter $s$ labeling the representations there to $s=1$ as appropriate in $\AdS_3 \times \Sphere^3 \times \Torus^4$.

\bigskip

\noindent The right-movers again transform in a similar representation in which the roles of the left- and right-moving generators have been interchanged.

\section{S-matrix}
\label{sec:S-matrix}

In order to derive the S-matrix $\Smat$ for the excitations discussed above we will follow closely the procedure of~\cite{Borsato:2012ud}. We focus on the two-particle S-matrix,
\begin{equation}
\ket{\mathcal{Y}_q^{\text{(out)}}\,\mathcal{X}_p^{\text{(out)}}}=\Smat\,\ket{\mathcal{X}_p^{\text{(in)}}\,\mathcal{Y}_q^{\text{(in)}}} .
\end{equation}
where $\Smat$ acts on the spin-chain state by permuting the excitations. First of all, we require that $\Smat$ commutes with the whole centrally extended symmetry algebra, \ie, for any generator $\gen{J}$
\begin{equation}
\label{eq:fundamentalcommrel-S}
  \comm{\gen{J}_1+\gen{J}_2}{\Smat_{12}} = 0 .
\end{equation}
Furthermore, the S-matrix should satisfy the unitarity condition
\begin{equation}
\Smat_{12}\,\Smat_{12}=\mathbf{1} ,
\end{equation}
and physical unitarity, which means that the S-matrix should be unitary as a matrix, so that if $\mathbb{S}$ is the matrix form of $\Smat$ on a basis of asymptotic two-particle  states, we have
\begin{equation}
\mathbb{S}_{pq}\cdot\left(\mathbb{S}_{pq}\right)^\dag=\left(\mathbb{S}_{pq}\right)^\dag\cdot\mathbb{S}_{pq}=\matId\otimes\matId .
\end{equation}
 On top of this we will also require that there is a $\Integers_2$ left-right symmetry, so that matrix elements that differ by interchanging left and right chiralities should be equal. 

In this way we find two solutions, corresponding to \emph{pure transmission} and \emph{pure reflection} of the ``left'' and ``right'' flavors. Consistence with the string theory results~\cite{Rughoonauth:2012qd,Sundin:2013ypa} forces us to choose the pure transmission S-matrix, like in~\cite{Borsato:2012ud}. The similarity with those results is not accidental, and in fact much deeper, since the excitations of the $\algPSU(1,1|2)^2$ chain transform under two copies of the centrally extended $\algSU(1|1)^2$ algebra discussed in~\cite{Borsato:2012ud}. In fact this makes the whole S-matrix factorize into two copies of the $\algSU(1|1)^2$ invariant one,
\begin{equation}
\label{eq:S-mat-tensor-prod}
  \Smat = \Smat_{\algSU(1|1)^2}\, \hat{\otimes}\, \Smat_{\algSU(1|1)^2}\, ,
\end{equation}
where the hat denotes a graded tensor product, so that in components we have
\begin{equation}
\mathbb{S}_{mn}^{kl}\equiv\mathbb{S}_{M\dot{M},N\dot{N}}^{K\dot{K},L\dot{L}}=(-1)^{\epsilon_{\dot{M}}\epsilon_N+\epsilon_{\dot{K}}\epsilon_L}\left(\mathbb{S}_{\algSU(1|1)^2}\right)_{MN}^{KL}\left(\mathbb{S}_{\algSU(1|1)^2}\right)_{\dot{M}\dot{N}}^{\dot{M}\dot{L}},
\end{equation}
where the $\epsilon$ symbol is 0 for bosons and 1 for fermions. This tensor product structure is very similar to the one coming from the centrally extended $\algPSU(2|2) \times \algPSU(2|2)$ symmetry of~\cite{Beisert:2005tm}.

It is convenient to rewrite here the explicit form of $\Smat_{\algSU(1|1)^2}$, in a slightly different normalization with respect to~\cite{Borsato:2012ud}. We have
\begin{equation}
  \begin{aligned}
    \Smat_{\algSU(1|1)^2} \ket{\fixedspaceL{\psi_p \psi_q}{\phi_p \phi_q}} 
    &= \fixedspaceR{D^{\smallLL}_{pq}}{A^{\smallLL}_{pq}} \ket{\fixedspaceL{\psi_p \psi_q}{\phi_q \phi_p}} , \qquad &
    \Smat_{\algSU(1|1)^2} \ket{\fixedspaceL{\psi_p \psi_q}{\phi_p \psi_q}} 
    &= \fixedspaceR{D^{\smallLL}_{pq}}{B^{\smallLL}_{pq}} \ket{\fixedspaceL{\psi_p \psi_q}{\psi_q \phi_p}} + C^{\smallLL}_{pq} \ket{\fixedspaceL{\psi_p \psi_q}{\phi_q \psi_p}}, \\
    \Smat_{\algSU(1|1)^2} \ket{\fixedspaceL{\psi_p \psi_q}{\psi_p \psi_q}} &= \fixedspaceR{D^{\smallLL}_{pq}}{F^{\smallLL}_{pq}} \ket{\fixedspaceL{\psi_p \psi_q}{\psi_q \psi_p}} , \qquad &
    \Smat_{\algSU(1|1)^2} \ket{\fixedspaceL{\psi_p \psi_q}{\psi_p \phi_q}} 
    &= \fixedspaceR{D^{\smallLL}_{pq}}{D^{\smallLL}_{pq}} \ket{\fixedspaceL{\psi_p \psi_q}{\phi_q \psi_p}} + E^{\smallLL}_{pq} \ket{\fixedspaceL{\psi_p \psi_q}{\psi_q \phi_p}}, \\
  \end{aligned}
\end{equation}
where the coefficients take the form
\begin{equation}
  \begin{aligned}
    A^{\smallLL}_{pq} &= S_{pq} , \qquad &
    B^{\smallLL}_{pq} &= S_{pq}  \frac{x_q^+ - x_p^+}{x_q^+ - x_p^-} , \qquad &
    C^{\smallLL}_{pq} &= S_{pq} \frac{x_q^+ - x_q^-}{x_q^+ - x_p^-} \frac{\eta_p}{\eta_q} , \\
    F^{\smallLL}_{pq} &= - S_{pq} \frac{x_q^- - x_p^+}{x_q^+ - x_p^-} , \qquad &
    D^{\smallLL}_{pq} &= S_{pq} \frac{x_q^- - x_p^-}{x_q^+ - x_p^-} , \qquad &
    E^{\smallLL}_{pq} &= S_{pq} \frac{x_p^+ - x_p^-}{x_q^+ - x_p^-} \frac{\eta_q}{\eta_p} ,
  \end{aligned}
\end{equation}
and $S_{pq}$ is an antisymmetric phase that cannot be fixed by symmetries and unitarity alone. Then we have, \eg,
\begin{equation}
  \begin{aligned}
    \Smat \ket{\Phi^{+\dot{+}}_p \Phi^{+\dot{+}}_q} &= A^{\smallLL}_{pq} A^{\smallLL}_{pq} \ket{\Phi^{+\dot{+}}_q \Phi^{+\dot{+}}_p} , \\
    \Smat \ket{\Phi^{+\dot{+}}_p \Phi^{-\dot{-}}_q} &= 
    B^{\smallLL}_{pq} B^{\smallLL}_{pq} \ket{\Phi^{-\dot{-}}_q \Phi^{+\dot{+}}_p} + C^{\smallLL}_{pq} C^{\smallLL}_{pq} \ket{\Phi^{+\dot{+}}_q \Phi^{-\dot{-}}_p}\\
    &\ + B^{\smallLL}_{pq} C^{\smallLL}_{pq}\left(\ket{\Phi^{-\dot{+}}_q \Phi^{+\dot{-}}_p} - \ket{\Phi^{+\dot{-}}_q \Phi^{-\dot{+}}_p}\right), \\
%    &\qquad B^{\smallLL}_{pq} C^{\smallLL}_{pq} \ket{\Phi^{-\dot{+}}_q \Phi^{+\dot{-}}_p} - C^{\smallLL}_{pq} B^{\smallLL}_{pq} \ket{\Phi^{+\dot{-}}_q \Phi^{-\dot{+}}_p} .
  \end{aligned}
\end{equation}
and so on.

In the LR-sector the scattering is reflectionless and we have
\begin{equation}
  \begin{aligned}
    \Smat_{\algSU(1|1)^2} \ket{\fixedspaceL{\psi_p\bar{\psi}_q}{\phi_p \bar{\phi}_q}} 
    &= \fixedspaceR{A^{\smallLR}_{pq}}{A^{\smallLR}_{pq}} \ket{\fixedspaceL{\bar{\psi}_q\psi_p}{\bar{\phi}_q\phi_p}} + \fixedspaceR{A^{\smallLR}_{pq}}{B^{\smallLR}_{pq}} \ket{\bar{\psi}_q \psi_p Z^-}, \quad &
    \Smat_{\algSU(1|1)^2} \ket{\fixedspaceL{\psi_p\bar{\psi}_q}{\phi_p\bar{\psi}_q}} 
    &= \fixedspaceR{A^{\smallLR}_{pq}}{C^{\smallLR}_{pq}} \ket{\fixedspaceL{\bar{\psi}_q\psi_p}{\bar{\psi}_q\phi_p}}, \\
    \Smat_{\algSU(1|1)^2} \ket{\fixedspaceL{\psi_p\bar{\psi}_q}{\psi_p\bar{\psi}_q}} 
    &= \fixedspaceR{A^{\smallLR}_{pq}}{E^{\smallLR}_{pq}} \ket{\fixedspaceL{\bar{\psi}_q\psi_p}{\bar{\psi}_q\psi_p}} + \fixedspaceR{A^{\smallLR}_{pq}}{F^{\smallLR}_{pq}} \ket{\bar{\phi}_q\phi_p Z^+}, \quad &
    \Smat_{\algSU(1|1)^2} \ket{\fixedspaceL{\psi_p\bar{\psi}_q}{\psi_p\bar{\phi}_q}} 
    &= \fixedspaceR{A^{\smallLR}_{pq}}{D^{\smallLR}_{pq}} \ket{\fixedspaceL{\bar{\psi}_q\psi_p}{\bar{\phi}_q\psi_p}}.
  \end{aligned}
\end{equation}
We normalize the S-matrix so that the elements are
\begin{equation}\label{eq:T-solution-LR}
  \begin{aligned}
    A^{\smallLR}_{pq} &= +\tau^{\smallLR}_{pq} \frac{1-\frac{1}{x_p^+ x_q^-}}{1-\frac{1}{x_p^- x_q^-}}, \quad &
    B^{\smallLR}_{pq} &= -\tau^{\smallLR}_{pq} \frac{\eta_p \eta_q}{x_p^- x_q^-} \frac{1}{1-\frac{1}{x_p^- x_q^-}}, \quad &
    C^{\smallLR}_{pq} &= \tau^{\smallLR}_{pq} , \\
    E^{\smallLR}_{pq} &= -\tau^{\smallLR}_{pq} \frac{1-\frac{1}{x_p^- x_q^+}}{1-\frac{1}{x_p^- x_q^-}}, \quad &
    F^{\smallLR}_{pq} &= -\tau^{\smallLR}_{pq} \frac{\eta_p \eta_q}{x_p^+ x_q^+} \frac{1}{1-\frac{1}{x_p^- x_q^-}}, \quad &
    D^{\smallLR}_{pq} &= \tau^{\smallLR}_{pq} \frac{1-\frac{1}{x_p^+ x_q^+}}{1-\frac{1}{x_p^- x_q^-}}.
  \end{aligned}
\end{equation}
where
\begin{equation}\label{eq:tau-LR-RL}
  \tau^{\smallLR}_{pq} = \zeta_{pq}\,\widetilde{S}_{pq} ,\qquad\zeta_{pq}=\sqrt{\frac{1-\frac{1}{x_p^- x_q^-}}{1-\frac{1}{x_p^+ x_q^+}}},
\end{equation}
and $\widetilde{S}_{pq}$ is an undetermined antisymmetric phase. Using \eqref{eq:S-mat-tensor-prod} it is easy to work out the scattering in the LR-sector of the $\algPSU(1,1|2)^2$ chain, finding, \eg,
\begin{equation}
\label{eq:LRscattering}
  \begin{aligned}
    \Smat \ket{\Phi^{+\dot{+}}_p \bar{\Phi}^{+\dot{+}}_q} &= 
    A^{\smallLR}_{pq} A^{\smallLR}_{pq} \ket{\bar{\Phi}^{+\dot{+}}_q \Phi^{+\dot{+}}_p} - B^{\smallLR}_{pq} B^{\smallLR}_{pq} \ket{\bar{\Phi}^{-\dot{-}}_q \Phi^{-\dot{-}}_p Z^- Z^-}  \\
    &\ + A^{\smallLR}_{pq} B^{\smallLR}_{pq}\left( \ket{\bar{\Phi}^{+\dot{-}}_q \Phi^{+\dot{-}}_p Z^-} +  \ket{\bar{\Phi}^{-\dot{+}}_q \Phi^{-\dot{+}}_p Z^-}\right) , \\
%    &\qquad A^{\smallLR}_{pq} B^{\smallLR}_{pq} \ket{\bar{\Phi}^{+-}_q \Phi^{+-}_p Z^-} + B^{\smallLR}_{pq} A^{\smallLR}_{pq} \ket{\bar{\Phi}^{-+}_q \Phi^{-+}_p Z^-} , \\
    \Smat \ket{\Phi^{+\dot{+}}_p \bar{\Phi}^{-\dot{-}}_q} &= C^{\smallLR}_{pq} C^{\smallLR}_{pq} \ket{\bar{\Phi}^{-\dot{-}}_q \Phi^{+\dot{+}}_p} ,
  \end{aligned}
\end{equation}
where we explicitly wrote down the length-changing effects. 

Due to the discrete LR-symmetry, the S-matrix in the RL and RR sectors can be easily found from the previous expressions, again just like in~\cite{Borsato:2012ud}.  Due to this symmetry, the only unknown scalar factors in the S-matrix are $S_{pq}$ and $\widetilde{S}_{pq}$.

An early attempt of deriving the S-matrix for the $\AdS_3\times\Sphere^3\times\Torus^4$ massive modes was made in~\cite{David:2010yg}. However, there the interaction between left- and right-moving sectors was not analyzed in full detail, and as a result the S-matrix contained only one dressing phase. 

More recently another proposal was made for the $\AdS_3\times\Sphere^3\times\Torus^4$ S-matrix~\cite{Ahn:2012hw}. This has not used the central extensions we discussed in section~\ref{sec:psu112-spin-chain} above. The resulting S-matrix and Bethe ansatz therefore differ from the ones presented here, in particular in the LR-sector.

\subsection{The Yang-Baxter equation}
\label{sec:Yang-Baxter}

For an integrable theory, an $N$-body scattering process can be broken down into a sequence of two-body scattering events. The condition that ensures that this can be done in a consistent way is the Yang-Baxter equation, which amounts to requiring that the two ways in which a 3-body scattering can be decomposed are equivalent. In terms of the operator $\Smat$ this reads
\begin{equation}
\label{eq:YBEspinchain}
\Smat_{23}\,\Smat_{12}\,\Smat_{23} =
\Smat_{12}\,\Smat_{23}\,\Smat_{12} ,
\end{equation}
where we recall that $\Smat$ permutes the excitations.

Our S-matrix satisfies the Yang-Baxter equation~\eqref{eq:YBEspinchain}. However, to check this we must not forget that in our spin-chain picture $\Smat$ may add or remove vacuum sites after a two particle excitiation, as in~\eqref{eq:LRscattering}. If we want to rewrite the Yang-Baxter equation on a basis of asymptotic excitations, we must take into account these vacuum sites by shifting them to the far right of the spin chain, as explained in section~\ref{sec:psu112-spin-chain}. This results in additional factors of $e^{\pm i p}$ where $p$ is the momentum of the rightmost excitation. Therefore the Yang-Baxter equation reads, in matrix form,
 \begin{equation}\label{eq:YB-mat-twist}
  \matId\otimes\mat{S}_{pq} \, \cdot \,
  \left(\mat{F}_q^{\phantom{1}}\mat{S}_{pr}\mat{F}_q^{-1}\right) \otimes \matId \, \cdot \,
  \matId\otimes\mat{S}_{qr}
  =
  \left(\mat{F}_p^{\phantom{1}}\mat{S}_{qr}\mat{F}_p^{{-1}}\right) \otimes \matId \, \cdot \,
  \matId\otimes\mat{S}_{pr} \, \cdot \,
  \left(\mat{F}_r^{\phantom{1}} \mat{S}_{pq}\mat{F}_r^{{-1}}\right) \otimes \matId ,
\end{equation}
where the transformation $\mat{F}$ implements a twist depending on the momentum of the third excitation. It is convenient to introduce a new S-matrix $\widehat{\mathbb{S}}$ that, unlike $\mathbb{S}$, obeys the untwisted Yang-Baxter equation~\cite{Arutyunov:2006yd,Borsato:2012ud}
 \begin{equation}\label{eq:YB-stringframe}
  \matId\otimes\widehat{\mat{S}}_{pq} \, \cdot \,
  \widehat{\mat{S}}_{pr} \otimes \matId \, \cdot \,
  \matId\otimes\widehat{\mat{S}}_{qr}
  =
  \widehat{\mat{S}}_{qr} \otimes \matId \, \cdot \,
  \matId\otimes\widehat{\mat{S}}_{pr} \, \cdot \,
   \widehat{\mat{S}}_{pq} \otimes \matId\, .
\end{equation}
 This can be done by means of a nonlocal transformation on the two-particle basis in terms of a twist operator $\mat{U}(p)$ defined as
 \begin{equation}\label{eq:twistZF}
  \widehat{\mat{S}}_{pq} = \mat{U}^\dag(p) \otimes \matId \, \cdot \, \mat{S}_{pq} \, \cdot \, \mat{U}(q) \otimes \matId .
\end{equation}
The factor $\mat{F}_p$ in~\eqref{eq:YB-mat-twist} is then related to $\mat{U}(p)$ by
\begin{equation}
    \mat{F}_p = \mat{U}(p) \otimes \mat{U}(p)\, .
\end{equation}
More specifically, if we let our one-particle basis be  given by
\begin{equation}\label{eq:1partbasis}
\left(\Phi^{+\dot{+}},\,\Phi^{+\dot{-}},\,\Phi^{-\dot{+}},\,\Phi^{-\dot{-}},\,\bar{\Phi}^{+\dot{+}},\,\bar{\Phi}^{+\dot{-}},\,\bar{\Phi}^{-\dot{+}},\,\bar{\Phi}^{-\dot{-}}\right) ,
\end{equation}
the operator $\mat{U}(p)$ is given by\footnote{%
  We have fixed the twist matrix by requiring that also in the string frame it is true that $\widehat{\mat{S}} = \widehat{\mat{S}}_{\algSU(1|1)^2}\, \hat{\otimes}\, \widehat{\mat{S}}_{\algSU(1|1)^2}$, where $ \widehat{\mat{S}}_{\algSU(1|1)^2}$ is the string frame S-matrix of~\cite{Borsato:2012ud}.%
} %
\begin{equation}
 \mat{U}(p)=\diag \left(e^{-i\,p},e^{-i\,p/2},e^{-i\,p/2},\,1,\ e^{-i\,p},e^{-i\,p/2},e^{-i\,p/2},\,1\right) .
\end{equation}
It is now easy to check that our spin chain S-matrix $\mat{S}$ satisfies~\eqref{eq:YB-mat-twist} or equivalently that $\widehat{\mat{S}}$, which is sometimes called the ``string frame'' S-matrix, satisfies~\eqref{eq:YB-stringframe}. This points to the integrability of the underlying theory.

\subsection{Crossing symmetry}
\label{sec:crossing}

The integrable S-matrix that we have found depends on two undetermined  antisymmetric phases $S_{pq}$ and $\widetilde{S}_{pq}$, which  we now want to constrain. One way to do so is, as in~\cite{Borsato:2012ud}, to exploit the fact that there exists two-particle configurations (``singlets'') that are annihilated by the whole symmetry algebra,
\begin{equation}
\label{eq:singletcondition}
\gen{J}\,\ket{\mathsf{1}_{pq}}=0 .
\end{equation}
 This feature was first noticed  for $\AdS_5$ strings \cite{Beisert:2005tm} and also there it can be employed to find constraints on the S-matrix~\cite{Arutyunov:2009ga,Vieira:2010kb}. In fact, since $\ket{\mathsf{1}_{pq}}$ is completely neutral, its scattering with any excitation should be trivial (see equation~\eqref{eq:triv-scatt-singl} below). This yields a constraint on the product of pairs of S-matrix elements. Furthermore, since $\ket{\mathsf{1}_{pq}}$ should have zero energy, it follows that either $p$ or $q$ cannot be physical. In fact it turns out that they must be related by \emph{crossing}, that is in term of the Zhukovski variables
\begin{equation}
x^\pm(p)=x^\pm(\bar{q})=\frac{1}{x^{\pm}(q)} .
\end{equation}
As discussed in~\cite{Arutyunov:2009ga} for $\AdS_5$ strings, it is indeed possible to relate the triviality of scattering by a singlet with the requirement of crossing symmetry. In what follows we will obtain the crossing equations for $S_{pq}$ and $\widetilde{S}_{pq}$ first by considering the scattering of a singlet with an arbitrary excitation, and later by requiring crossing invariance for the (string frame) S-matrix.

Solving \eqref{eq:singletcondition} yields two singlets, related to each other by LR-symmetry. These can be constructed by tensoring two copies of singlet discussed in~\cite{Borsato:2012ud}. They are\footnote{%
  In what follows we write them taking the momenta $p,q$ to be in the physical region. Crossing is indicated by $\bar{p},\bar{q}$, which corresponds to a shift on the rapidity torus by half of its imaginary period~\cite{Janik:2006dc}. Antisymmetry requires shift in the first and second variable to be performed in opposite directions, see appendix~\ref{sec:hopf-algebra}, which leaves us with two seemingly equivalent prescriptions. The preferred choice can be fixed by comparing perturbative results with the dressing phases that solve the crossing equations~\cite{upcoming}.
  \label{footnote:crossing}
} %
\begin{equation*}
  \begin{aligned}
    \ket{\mathsf{1}^{LR}_{\bar{p}p}} &= 
    \ket{\Phi^{+\dot{+}}_{\bar{p}} \bar{\Phi}^{+\dot{+}}_p \, Z^+Z^+} + \Xi_{\bar{p}p} \ket{\Phi^{-\dot{+}}_{\bar{p}} \bar{\Phi}^{-\dot{+}}_p \,  Z^+} + 
    \Xi_{\bar{p}p} \ket{\Phi^{+\dot{-}}_{\bar{p}} \bar{\Phi}^{+\dot{-}}_p \,  Z^+} - (\Xi_{\bar{p}p})^2 \ket{\Phi^{-\dot{-}}_{\bar{p}} \bar{\Phi}^{-\dot{-}}_p } , \\
    \ket{\mathsf{1}^{RL}_{\bar{p}p}} &= 
    \ket{\bar{\Phi}^{+\dot{+}}_{\bar{p}} \Phi^{+\dot{+}}_p  \, Z^+Z^+} + \Xi_{\bar{p}p} \ket{\bar{\Phi}^{-\dot{+}}_{\bar{p}} \Phi^{-\dot{+}}_p \,  Z^+} + 
    \Xi_{\bar{p}p} \ket{\bar{\Phi}^{+\dot{-}}_{\bar{p}} \Phi^{+\dot{-}}_{\bar{p}} \,  Z^+} - (\Xi_{\bar{p}p})^2 \ket{\bar{\Phi}^{-\dot{-}}_{\bar{p}} \Phi^{-\dot{-}}_p } ,
  \end{aligned}
\end{equation*}
where
\begin{equation}
  \Xi_{\bar{p}p} = i \, x^+_p \frac{\eta_{\bar{p}}}{\eta_p},\qquad x_{\bar{p}}^\pm = \frac{1}{x_p^\pm} .
\end{equation}
Requiring that the scattering with any excitation $\ket{\mathcal{X}_p}$ is trivial 
\begin{equation}\label{eq:triv-scatt-singl}
  \Smat_{23}\, \Smat_{12} \ket{\mathcal{X}_p^{} \textsf{1}^{LR}_{\bar{q}q}} = \ket{\textsf{1}^{LR}_{\bar{q}q} \mathcal{X}_p^{} },\qquad
  \Smat_{23}\, \Smat_{12} \ket{\mathcal{X}_p^{} \textsf{1}^{RL}_{\bar{q}q}} = \ket{\textsf{1}^{RL}_{\bar{q}q} \mathcal{X}_p^{} },
\end{equation}
gives the crossing equations. 

In order for the S-matrices to have the standard $\algSU(2)$ and $\algSL(2)$ form we rewrite the scalar factors as
\begin{equation}
S_{pq}^{-2}=\frac{x_p^+ - x_q^-}{x_p^- - x_q^+} \frac{1- \frac{1}{x_p^+ x_q^-}}{1- \frac{1}{x_p^- x_q^+}}\, \sigma^2_{pq} ,\qquad
\widetilde{S}_{pq}^{-2}= \frac{1- \frac{1}{x_p^+ x_q^-}}{1- \frac{1}{x_p^- x_q^+}}\, \widetilde{\sigma}^2_{pq} ,
\end{equation}
where $\sigma^2_{pq}$ and $\widetilde{\sigma}^2_{pq}$ are two new phases, which we will refer to as the ``dressing phases''.\footnote{%
  The appearance of two independent dressing phase factors was previously observed in~\cite{Hoare:2011fj} for the case of Pohlmeyer reduced strings on $\AdS_3 \times \Sphere^3$.%
} % 
They satisfy the crossing equations
\begin{equation}\label{eq:crossingeq-chain}
  \begin{aligned}
    \sigma_{pq}^2\,\widetilde{\sigma}_{p\bar{q}}^2 &= \left(\frac{x_p^+}{x_p^-}\right)^2\frac{(x^-_p-x^+_q)^2}{(x^-_p-x^-_q)(x^+_p-x^+_q)}\frac{1-\frac{1}{x^-_px^+_q}}{1-\frac{1}{x^+_px^-_q}},\\
    \sigma_{p\bar{q}}^2\,\widetilde{\sigma}_{pq}^2 &= \left(\frac{x_p^+}{x_p^-}\right)^2\frac{\left(1-\frac{1}{x^-_px^-_q}\right)\left(1-\frac{1}{x^+_px^+_q}\right)}{\left(1-\frac{1}{x^+_px^-_q}\right)^2}\frac{x^-_p-x^+_q}{x^+_p-x^-_q}.
  \end{aligned}
\end{equation}
It is easy to check that, in both the finite-gap and near-BMN limits, these equations are satisfied if we take both phases to be equal to the AFS one~\cite{Arutyunov:2004vx} at leading order,
\begin{equation}
\label{eq:dressing-AFS}
\sigma(p,q)=\sigma_{AFS}(p,q)+O\left(\frac{1}{h^2}\right),\qquad
\widetilde{\sigma}(p,q)=\sigma_{AFS}(p,q)+O\left(\frac{1}{h^2}\right),
\end{equation} 
where
\begin{equation}
\sigma_{AFS}(x_p,x_q)=\left( \frac{1- \frac{1}{x_p^- x_q^+}}{1- \frac{1}{x_p^+ x_q^-}} \right) \left( \frac{1- \frac{1}{x_p^+ x_q^-}}{1- \frac{1}{x_p^+ x_q^+}}  \, \frac{1- \frac{1}{x_p^- x_q^+}}{1- \frac{1}{x_p^- x_q^-}} \right)^{i h \left(x_p+1/x_p-x_q-1/x_q\right)}.
\end{equation}
We plan to discuss an all-loop solution to these crossing equations in an upcoming publication~\cite{upcoming}.

The same set of equations can be found also by the usual field-theoretic considerations~\cite{Arutyunov:2009ga}. It is convenient to work in the string frame, and for this purpose let us transform the charges of the symmetry algebra as
\begin{equation}
\widehat{\mat{J}}_{pq}=\mat{U}_q^\dag\otimes\matId\cdot\mat{J}_{pq}\cdot\mat{U}_q\otimes\matId ,
\end{equation} 
where $\mat{J}_{pq}$ is the matrix representation of the generator $\gen{J}$ on a two-particle state and $\widehat{\mat{J}}_{pq}$ is its string frame counterpart. Then the action on a two-particle states can be understood in terms of a nontrivial coproduct
\begin{equation}
\label{eq:stringframe-coproduct}
  \hat{\mat{J}}_{pq}=\mat{J}_p\otimes e^{\pm i\,q/2}+\Sigma\otimes\mat{J}_q ,
\end{equation}
where one should pick the positive sign in the exponent for the supercharges $\gen{S}^{\scriptscriptstyle I}_i$ and the negative one for $\gen{Q}^{\scriptscriptstyle I}_i$, and $\Sigma$ takes into account the fermion signs. In particular, in the basis~\eqref{eq:1partbasis}, we have
\begin{equation}
\Sigma=\diag\left(\,1,-1,-1,\,1,\ 1,-1,-1,\,1\right) .
\end{equation}
By taking the supertranspose of the charges $\mat{J}$ we find another representation related to the original one by charge conjugation as
\begin{equation}\label{eq:chargeconj}
C^{-1}\cdot\mat{J}(z+\omega)^{\text{st}}\cdot C=-e^{\mp i\, p/2}\,\mat{J}(z) ,
\end{equation}
where $C$ is the charge conjugation matrix which in our basis reads\footnote{There are more general solutions to~\eqref{eq:chargeconj}. We fixed our choice by picking~$C\cdot C^\dag= C^\dag \cdot C=\matId$ and~$C\cdot C=\matId$. The crossing equations do not depend on this choice.}
\begin{equation}
C=\left(\begin{array}{cc}
0&-i\\
i&0
\end{array}\right)\otimes \left(
\begin{array}{cccc}
1&0&0&0\\
0&-1&0&0\\
0&0&-1&0\\
0&0&0&-1
\end{array}\right) .
\end{equation}
To simplify our notation it is useful to rewrite the S-matrix in terms of an operator that does not permute the excitations. To this end we introduce
\begin{equation}
\mat{R}_{pq}=\Pi\cdot\mat{S}_{pq} ,
\end{equation}
where $\Pi$ is the permutation matrix. In terms of $\mat{R}_{pq}$ the fundamental invariance property~\eqref{eq:fundamentalcommrel-S} becomes
\begin{equation}
\label{eq:fundamentalcommrel-R}
\mat{R}_{pq}\cdot\left(\mat{J}_p\otimes e^{\pm iq/2}+\Sigma\otimes \mat{J}_q\right)=\left(\mat{J}_p\otimes \Sigma+e^{\pm ip/2}\otimes \mat{J}_q\right)\cdot\mat{R}_{pq}\,.
\end{equation}
We can now follow a standard route~\cite{Arutyunov:2009ga} to derive the crossing equations for $\mat{R}_{pq}$, taking the transpose of~\eqref{eq:fundamentalcommrel-R} with respect to either factor of the tensor product, and exploiting the charge conjugation~\eqref{eq:chargeconj}.\footnote{To this end it is useful to rewrite \eqref{eq:chargeconj} as $\mat{J}(z)^{\text{t}}=-e^{\pm ip/2} C\cdot\mat{J}(z-\omega)\cdot C^{-1}\cdot\Sigma$, where $\omega$ is half of the imaginary period of the rapidity torus, and use that $[\Sigma\otimes\Sigma,\mat{R}_{pq}]=0$.} The crossing equations then read
\begin{equation}
  \label{eq:crossingeq-string}
   C^{-1}\otimes \matId\cdot\mat{R}^{\text{t}_1}_{\bar{q}p}\cdot  C\otimes\matId\cdot\mat{R}_{qp} = \matId\otimes \matId , \qquad 
    \matId\otimes C^{-1}\cdot \mat{R}^{\text{t}_2}_{p\bar{q}}\cdot\matId\otimes C\cdot\mat{R}_{pq} = \matId\otimes \matId ,
\end{equation}
where the superscript $\text{t}_{1,2}$ denotes transposition in the first or second factor, or 
\begin{equation}
   \Sigma\, C^{-1}\otimes \matId\cdot\mat{R}^{\text{t}_1}_{\bar{q}p}\cdot   C\,\Sigma\otimes\matId\cdot\mat{R}_{qp} = \matId\otimes \matId , \qquad
   \matId\otimes \Sigma\, C^{-1}\cdot \mat{R}^{\text{t}_2}_{p\bar{q}}\cdot\matId\otimes  C\,\Sigma\cdot\mat{R}_{pq} = \matId\otimes \matId ,
\end{equation}
depending on whether we shift down or up the first variable of the S-matrix under crossing, see footnote~\ref{footnote:crossing}. 
We can take \eg the second equation in~\eqref{eq:crossingeq-string} and evaluate it in terms of the S-matrix elements in the string frame finding that it is satisfied if, \eg,
\begin{equation}
\label{eq:stringframeS-crossing}
\mathcal{A}_{pq}\,\widetilde{\mathcal{A}}_{p\bar{q}}=1 ,\qquad \mathcal{B}_{p\bar{q}}\,\widetilde{\mathcal{C}}_{pq}=1 ,
\end{equation}
where
\begin{equation}
\label{eq:stringframeS-elems}
\begin{aligned}
\mathcal{A}_{pq}=\bra{\Phi^{+\dot{+}}_{q}\Phi^{+\dot{+}}_{p}}\widehat{\mat{S}}_{pq}\ket{\Phi^{+\dot{+}}_{p}\Phi^{+\dot{+}}_{q}},&\qquad&
\widetilde{\mathcal{A}}_{pq}=\bra{\bar{\Phi}^{+\dot{+}}_{q}\Phi^{+\dot{+}}_{p}}\widehat{\mat{S}}_{pq}\ket{\Phi^{+\dot{+}}_{p}\bar{\Phi}^{+\dot{+}}_{q}},\\
\mathcal{B}_{pq}=\bra{\Phi^{-\dot{-}}_{q}\Phi^{+\dot{+}}_{p}}\widehat{\mat{S}}_{pq}\ket{\Phi^{+\dot{+}}_{p}\Phi^{-\dot{-}}_{q}},&\qquad&
\widetilde{\mathcal{C}}_{pq}=\bra{\bar{\Phi}^{-\dot{-}}_{q}\Phi^{+\dot{+}}_{p}}\widehat{\mat{S}}_{pq}\ket{\Phi^{+\dot{+}}_{p}\bar{\Phi}^{-\dot{-}}_{q}}.
\end{aligned}
\end{equation}
Using the explicit form of the S-matrix elements yields~\eqref{eq:crossingeq-chain}. The first equation in~\eqref{eq:crossingeq-string} gives similar equations where the crossed momentum is in the first argument of the S-matrix; these can also be found from considering scattering with a singlet, but with a particle incoming from the right.

Let us remark that, up to a different choice of normalization, the crossing equations are the square of the ones found in~\cite{Borsato:2012ud} for~$\Smat_{\algSU(1|1)^2}$, see also appendix~\ref{sec:hopf-algebra}.
In section~\ref{sec:constr-sc-fact} we will discuss the solutions at strong coupling of these crossing equations and further constraints on the scalar factors.

It is also interesting to notice that the action of the charges on the two-particle states~\eqref{eq:stringframe-coproduct}  takes the form of a nontrivial coproduct, similar to the one appearing in other instances of AdS/CFT integrability~\cite{Gomez:2006va,Plefka:2006ze}. This suggests an alternative route to obtain the all-loop S-matrix: rather than bootstrapping it out of its symmetries, and obtaining an integrable S-matrix as a result, we could have postulated the existence of an underlying Hopf algebra structure. This is yet another route to obtain the crossing equations~\eqref{eq:crossingeq-chain}, which we discuss in appendix~\ref{sec:hopf-algebra}. Furthermore, and in contrast with what happens in the case of $\AdS_5$, here we could have in principle obtained the whole S-matrix from a universal R-matrix (the one of $\algGL(1|1)$) by imposing Yangian symmetry. We refer the reader to appendix~\ref{sec:Yangian} for details on this construction.

%%%%%%%%%%%%%%%
\section{S-matrix diagonalisation and Bethe ansatz}
\label{sec:diag-BA}
Since the all-loop S-matrix~$\Smat_{pq}$  obtained in the previous section satisfies the Yang-Baxter equation, we can use it to obtain the asymptotic Bethe ansatz for the spin chain.

\subsection{Diagonalising the S-matrix}
\label{sec:diag}

In this section we show how to construct asymptotic eigenstates of the spin-chain Hamiltonian $\gen{H}$. We will follow closely the procedure used in~\cite{Borsato:2012ss} to diagonalise the S-matrix of the $\algD{\alpha}^2$ spin-chain. This procedure is  standard~\cite{Beisert:2005wm,Beisert:2005tm,Beisert:2005fw,deLeeuw:2007uf} and we will only present the key steps.

We will construct asymptotic eigenstates (\ie eigenstates for a chain of infinite length) out of the two-particle S-matrix. One such eigenstate containing $K$ excitations will be of the form
\begin{equation}
\ket{\Psi}=\sum_{\pi\in S_K}\Smat_{\pi}\ket{\Psi}^{\I},\qquad \Smat_{\pi} = \prod_{(k,l)\in\pi}\Smat_{kl},
\end{equation}
where $\pi\in S_K$ is a permutation, $\ket{\Psi}^{\I}$ is a wavefunction and we used the fact that the scattering factorizes to write the $K$-body S-matrix as a product of two-body ones, which act by 
\begin{equation}
\Smat_{\pi} \ket{\Psi} = S_{\pi}  \ket{\Psi}_{\pi}.
\end{equation}
Since not all of the quantum numbers scatter by pure transmission  we need to employ the so-called nesting procedure to perform the diagonalisation.

The idea is to introduce a level-I vacuum $\ket{0}^{\I}$, which is just given by $\ket{Z^L}$. Then, rather than considering all of its possible excitations at once, we restrict to the maximal set of excitations that scatter diagonally. This will give the  level-II vacuum $\ket{0}^{\II}$. The remaining fields are then considered as level-II excitations on top of such a vacuum. This will be enough to diagonalise the whole S-matrix in our case.

\paragraph{Level-I vacuum.}

The level-I vacuum is just $\ket{0}^{\I} \equiv \ket{Z^L}$. The S-matrix $\Smat$ given in the previous section can be thought of as the level-I S-matrix, and we will call it $\Smat^{\I}$ in this section.

%%%%%%%%

\paragraph{Level-II vacuum.}

We need to choose a maximal set of excitations that scatter with pure transmission among each other. An $N$-particle state made out of only this kind of excitations will be automatically an eigenstate. From the structure of $\mathcal{S}^{\I}$ leads to four possible choices
\begin{equation}\label{eq:level-II-vacuum}
  \begin{aligned}
    V^{\II}_A &= \{ \Phi^{+\dot{+}},\bar{\Phi}^{-\dot{-}}\}, \qquad 
    V^{\II}_B &= \{ \Phi^{-\dot{-}},\bar{\Phi}^{+\dot{+}}\}, \\
    V^{\II}_C &= \{ \Phi^{+\dot{-}},\bar{\Phi}^{-\dot{+}}\}, \qquad 
    V^{\II}_D &= \{ \Phi^{-\dot{+}},\bar{\Phi}^{+\dot{-}}\}. 
\end{aligned}
\end{equation}
Each candidate level-II vacuum is composed of one left and one right excitation, that are either both bosonic or both fermionic.

In the following we will choose the set $V^{\II}_A$ to construct the level-II vacuum. In appendix \ref{sec:fermionic-duality} we will show how the other possible choices are related by fermionic dualities, allowing us to write all-loop Bethe equations in four different gradings.

\paragraph{Propagation.}

To consider also the other types of fields, we can view them as level-II excitations on the level-II vacuum.
We have four supercharges at our disposal to create other types of fields, starting from the ones of $V^{\II}_A$ (see table \ref{tab:lev-II-exc} for their explicit action).
\begin{table}
  \centering
  \begin{tabular}{r@{\hspace{1.5em}}cccc}
    \toprule
    & $\genQL_1$ & $\genQL_2$ & $\genSR_1$ & $\genSR_2$  \\
    \midrule
    $\Phi^{+\dot{+}}$    &$\Phi^{-\dot{+}}$    &$\Phi^{+\dot{-}}$    &$\Phi^{-\dot{+}} Z^-$    &$\Phi^{+\dot{-}}Z^-$     \\
    $\bar{\Phi}^{-\dot{-}}$ & $\bar{\Phi}^{+\dot{-}}Z^-$    &$\bar{\Phi}^{-\dot{+}}Z^-$    &$\bar{\Phi}^{+\dot{-}}$    &$\bar{\Phi}^{-\dot{+}}$    \\
    \bottomrule
  \end{tabular}
  \caption{%
    Action of the lowering operators on the states of the level-II vacuum $V^{\II}_A$.
  }
  \label{tab:lev-II-exc}
\end{table}

Note that the fields $\Phi^{-\dot{-}}, \bar{\Phi}^{+\dot{+}}$ do not explicitly appear in the Bethe ansatz, since from that point of view they are considered as composite excitations (\ie, one can respectively create them by consecutively applying $\genQL_1$ and $\genQL_2$ on $\Phi^{+\dot{+}}$ and $\genSR_1$ and $\genSR_2$ on $\bar{\Phi}^{-\dot{-}}$).

As in~\cite{Borsato:2012ss}, we can derive the level-II S-matrix by requiring compatibility of the level-I S-matrix with the states in which we allow for one level-II excitation.
In the following we consider two-particle states and we write the wave function that solves the compatibility condition with $\Smat^{\I}$.
The factorization of scattering allows us to extend these results to N-particle excitations, when one level-II excitation is allowed.

Starting from a level-II vacuum defined as $\ket{0}^{\II}_{22}=\ket{\Phi^{+\dot{+}}_p \Phi^{+\dot{+}}_q}$ we can consider level-II excitations created by the action of $\genQL_1$ as\footnote{We use the index 2 when we consider level-II excitations on the field $\Phi^{+\dot{+}}$, while the index $\bar{2}$ for level-II excitations on the field $\bar{\Phi}^{-\dot{-}}$.}
\begin{equation}
  \begin{aligned}
    \ket{\mathcal{Y}_y}^{\II}_{(22)} &= f_2(y,p) \ket{\Phi^{-\dot{+}}_p \Phi^{+\dot{+}}_q} + f_{2}(y,q) S_{{22}}^{\II,\I}(y,p) \ket{\Phi^{+\dot{+}}_p \Phi^{-\dot{+}}_q}, \\
    \ket{\mathcal{Y}_y}^{\II}_{(22), \pi} &=  f_{2}(y,q) \ket{\Phi^{-\dot{+}}_q \Phi^{+\dot{+}}_p} + f_{2}(y,p) S_{{22}}^{\II,\I}(y,q) \ket{\Phi^{+\dot{+}}_q \Phi^{-\dot{+}}_p}.
  \end{aligned}
\end{equation}
The compatibility equation
\begin{equation}
\Smat^{\I}_\pi \ket{\mathcal{Y}_y}^{\II}_{(22)} = A^{\smallLL}_{p q}A^{\smallLL}_{p q} \ket{\mathcal{Y}_y}^{\II}_{(22), \pi}
\end{equation}
is solved by
\begin{equation}
  f_{2}(y,p) = g_{2}(y) \frac{\eta_p}{h_{2}(y) - x_p^+} , \qquad
  S_{{22}}^{\II,\I}(y,p) = \frac{h_{2}(y) - x_p^-}{h_{2}(y) - x_p^+},
\end{equation}
where $h_{1}(y), g_{1}(y)$ are arbitrary functions of $y$.
Starting instead from a level-II vacuum defined as $\ket{0}^{\II}_{\bar{2}\bar{2}}=\ket{\bar{\Phi}^{-\dot{-}}_p \bar{\Phi}^{-\dot{-}}_q}$ we can consider level-II excitations created by the action of $\genQL_1$ as
\begin{equation}
  \begin{aligned}
    \ket{\mathcal{Y}_y}^{\II}_{(\bar{2}\bar{2})} &= f_{\bar{2}}(y,p) \ket{\Phi^{+\dot{-}}_p Z^+ \Phi^{-\dot{-}}_q} + f_{\bar{2}}(y,q) S_{{\bar{2}\bar{2}}}^{\II,\I}(y,p) \ket{\Phi^{-\dot{-}}_p \Phi^{+\dot{-}}_q}, \\
    \ket{\mathcal{Y}_y}^{\II}_{(\bar{2}\bar{2}), \pi} &=  f_{\bar{2}}(y,q) \ket{\Phi^{+\dot{-}}_q Z^+ \Phi^{-\dot{-}}_p} + f_{\bar{2}}(y,p) S_{{\bar{2}\bar{2}}}^{\II,\I}(y,q) \ket{\Phi^{-\dot{-}}_q \Phi^{+\dot{-}}_p}.
  \end{aligned}
\end{equation}
The compatibility equation
\begin{equation}
\Smat^{\I}_\pi \ket{\mathcal{Y}_y}^{\II}_{(\bar{2}\bar{2})} = F^{\smallRR}_{p q}F^{\smallRR}_{p q} \ket{\mathcal{Y}_y}^{\II}_{(\bar{2}\bar{2}), \pi}
\end{equation}
is solved by
\begin{equation}
  f_{\bar{2}}(y,p) = \frac{-i g_{\bar{2}}(y)}{x_p^+} \frac{\eta_p}{1 - \frac{1}{h_{\bar{2}}(y) \ x_p^-}} , \qquad
  S_{{\bar{2}\bar{2}}}^{\II,\I}(y,p) = \frac{1 - \frac{1}{h_{\bar{2}}(y) \ x_p^+}}{1 - \frac{1}{h_{\bar{2}}(y) \  x_p^-}}.
\end{equation}
As before $h_{\bar{2}}(y), g_{\bar{2}}(y)$ are generic functions of $y$.
The last step is to start from the level-II vacuum $\ket{0}^{\II}_{2\bar{2}}=\ket{\Phi^{+\dot{+}}_p \bar{\Phi}^{-\dot{-}}_q}$.
For the level-II excitation we can write
\begin{equation}
  \begin{aligned}
    \ket{\mathcal{Y}_y}^{\II}_{(2 \bar{2})} &= f_{2}(y,p) \ket{\Phi^{-\dot{+}}_{p} \bar{\Phi}^{-\dot{-}}_{q}} + f_{\bar{2}}(y,q) S_{{\bar{2} 2}}^{\II,\I}(y,p) \ket{\Phi^{+\dot{+}}_{p} \bar{\Phi}^{+\dot{-}}_{q} Z^+}, \\
    \ket{\mathcal{Y}_y}^{\II}_{(2 \bar{2}), \pi} &= f_{\bar{2}}(y,q) \ket{\bar{\Phi}^{+\dot{-}}_{q} Z^+ \Phi^{+\dot{+}}_{p}} + f_{2}(y,p) S_{{2 \bar{2}}}^{\II,\I}(y,q) \ket{\bar{\Phi}^{-\dot{-}}_{q} \Phi^{-\dot{+}}_{p}}.
  \end{aligned}
\end{equation}
and we can solve the equation
\begin{equation}
\Smat^{\I}_\pi \ket{\mathcal{Y}_y}^{\II}_{(2 \bar{2})} = C^{{\smallLR}}_{p q}C^{{\smallLR}}_{p q} \ket{\mathcal{Y}_y}^{\II}_{(2 \bar{2}), \pi}
\end{equation}
by
\begin{equation}\label{eq:sol-nest-mix}
  \begin{aligned}
    h_{\bar{2}} (y) &=h_{2}(y) \equiv y, \qquad &
    g_{\bar{2}}(y) &=-\frac{g_{2}(y)}{h_{2}(y)}, \\
    S_{{\bar{2} 2}}^{\II,\I}(y,p) &=S_{{22}}^{\II,\I}(y,p), \qquad &
    S_{{2 \bar{2}}}^{\II,\I}(y,p) &=S_{{\bar{2}\bar{2}}}^{\II,\I}(y,p).
  \end{aligned}
\end{equation}
These calculations are exactly equivalent to the ones already performed in~\cite{Borsato:2012ss}. This is not surprising, since in the diagonalisation procedure we have to consider doublets of $\algSU(1|1)$ (e.g. $(\Phi^{+\dot{+}} | \Phi^{-\dot{+}})$ in the present case and $(\phi^1|\psi^1)$ in ~\cite{Borsato:2012ss}).
This makes clear that the diagonalisation procedure works in a similar way for the other level-II excitations.

\paragraph{Scattering.}

All the level-II excitations scatter trivially amongst each other. We show the explicit example in which we start from the level-II vacuum $\ket{0}^{\II}_{22}=\ket{\Phi^{+\dot{+}}_p \Phi^{+\dot{+}}_q}$ and we create two level-II excitations by acting with the charge $\genQL_1$.
The two-particle states are
\begin{equation}
  \begin{aligned}
    \ket{\mathcal{Y}_{y_1} \mathcal{Y}_{y_2}}^{\II}_{(22)} &= f_2(y_1,p) f_2(y_2,q) S^{\II,\I}_{22}(y_2,p) \ket{\Phi^{-\dot{+}}_p \Phi^{-\dot{+}}_q} \\
    &\qquad + f_2(y_2,p) f_2(y_1,q) S^{\II,\I}_{22}(y_1,p) S^{\II,\II}_{22}(y_1,y_2) \ket{\Phi^{-\dot{+}}_p \Phi^{-\dot{+}}_q} , \\
    \ket{\mathcal{Y}_{y_1} \mathcal{Y}_{y_2}}^{\II}_{(22),\pi} &= f_2(y_1,q) f_2(y_2,p) S^{\II,\I}_{22}(y_2,q) \ket{\Phi^{-\dot{+}}_q \Phi^{-\dot{+}}_p} \\
    &\qquad + f_2(y_2,q) f_2(y_1,p) S^{\II,\I}_{22}(y_1,q) S^{\II,\II}_{22}(y_1,y_2) \ket{\Phi^{-\dot{+}}_q \Phi^{-\dot{+}}_p} .
  \end{aligned}
\end{equation}
Requiring the equation
\begin{equation}
\Smat^{\I}_\pi \ket{\mathcal{Y}_{y_1} \mathcal{Y}_{y_2}}^{\II}_{(22)} = A^{\smallLL}_{p q}A^{\smallLL}_{p q} \ket{\mathcal{Y}_{y_1} \mathcal{Y}_{y_2}}^{\II}_{(22),\pi}
\end{equation}
and using the previous results, we find that
\begin{equation}
S^{\II,\II}_{22}(y_1,y_2)=-1,
\end{equation}
which confirms the trivial scattering.
One can repeat the calculation by starting with a different level-II vacuum and by acting with different supercharges to create the level-II excitations and still find trivial scattering.

\subsection{Bethe equations}
\label{sec:BAE}

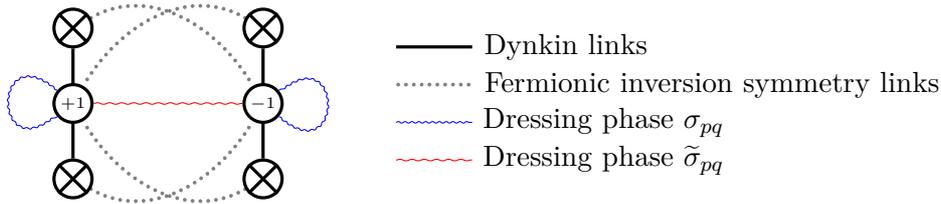
\begin{figure}
  \centering
  
\begin{tikzpicture}
  \begin{scope}
    \coordinate (m) at (0cm,0cm);

    \node (v1L) at (-1.25cm, 1cm) [dynkin node] {};
    \node (v2L) at (-1.25cm, 0cm) [dynkin node] {$\scriptscriptstyle +1$};
    \node (v3L) at (-1.25cm,-1cm) [dynkin node] {};

    \draw [dynkin line] (v1L) -- (v2L);
    \draw [dynkin line] (v2L) -- (v3L);
    
    \node (v1R) at (+1.25cm, 1cm) [dynkin node] {};
    \node (v2R) at (+1.25cm, 0cm) [dynkin node] {$\scriptscriptstyle -1$};
    \node (v3R) at (+1.25cm,-1cm) [dynkin node] {};

    \draw [dynkin line] (v1L.south west) -- (v1L.north east);
    \draw [dynkin line] (v1L.north west) -- (v1L.south east);

    \draw [dynkin line] (v3L.south west) -- (v3L.north east);
    \draw [dynkin line] (v3L.north west) -- (v3L.south east);
    
    \draw [dynkin line] (v1R.south west) -- (v1R.north east);
    \draw [dynkin line] (v1R.north west) -- (v1R.south east);

    \draw [dynkin line] (v3R.south west) -- (v3R.north east);
    \draw [dynkin line] (v3R.north west) -- (v3R.south east);

    \draw [dynkin line] (v1R) -- (v2R);
    \draw [dynkin line] (v2R) -- (v3R);

    \begin{pgfonlayer}{background}
      \draw [inverse line] [out=  0+30,in= 120] (v1L) to (v2R);
      \draw [inverse line] [out=  0-30,in=240] (v3L) to (v2R);
      \draw [inverse line] [out=180-30,in= 60] (v1R) to (v2L);
      \draw [inverse line] [out=180+30,in=300] (v3R) to (v2L);
    \end{pgfonlayer}

    \draw [red phase] [out=0,in=180] (v2L) to (v2R);
%    \draw [red phase] [out=270-110,in= 90-70] (v2R) to (v2L);

    \draw [blue phase] [out=180-40,in= 180+40,loop] (v2L) to (v2L);
    \draw [blue phase] [out=-40,in=40,loop] (v2R) to (v2R);
  \end{scope}

  \begin{scope}[xshift=+3cm,yshift=-0.75cm]
    \draw [dynkin line]  (0cm,1.5cm) -- (1cm,1.5cm) node [anchor=west,black] {\small Dynkin links};
    \draw [inverse line] (0cm+0.75pt,1.0cm) -- (1cm,1.0cm) node [anchor=west,black] {\small Fermionic inversion symmetry links};
    \draw [blue phase]   (0cm,0.5cm) -- (1cm,0.5cm) node [anchor=west,black] {\small Dressing phase $\sigma_{pq}$};
    \draw [red phase]    (0cm,0.0cm) -- (1cm,0.0cm) node [anchor=west,black] {\small Dressing phase $\widetilde{\sigma}_{pq}$};
  \end{scope}
\end{tikzpicture}

  \caption{The Dynkin diagram for $\algPSU(1,1|2)^2$ with the various interaction terms appearing in the Bethe ansatz indicated. The label $\pm 1$ inside the middle Dynkin nodes indicate the $\algSU(2)$ and $\algSL(2)$ gradings of the left- and right-moving sectors.}
  \label{fig:bethe-equations}
\end{figure}

To obtain the Bethe ansatz equations we impose periodic boundary conditions on a spin-chain of finite length $L$ and use the S-matrix in its diagonal form. The central nodes of the two Dynkin diagrams~\ref{fig:dynkin-su22}~\subref{fig:dynkin-su22-su} and~\ref{fig:dynkin-su22}~\subref{fig:dynkin-su22-sl} correspond to the positive roots of $\algSU(2) \subset \algPSU(1,1|2)_L$ and $\algSL(2) \subset \algPSU(1,1|2)_R$, respectively, and give the momentum carrying nodes. We denote the corresponding variables by $x^\pm$ and $\bar{x}^\pm$, respectively.
The number of the corresponding excitations is denoted by $K_2$, $K_{\bar{2}}$. We have two auxiliary ``left'' roots denoted by $y_1,y_3$, corresponding respectively to the action of the supercharges $\genQL_1$, $\genQL_2$. The two auxiliary ``right'' roots are denoted by $y_{\bar{1}},y_{\bar{3}}$ and they correspond respectively to the action of the supercharges $\genSR_1$, $\genSR_2$. The number of the corresponding excitations is denoted by $K_1$, $K_3$, $K_{\bar{1}}$, and $K_{\bar{3}}$.
The Bethe equations then read
\begin{align}
    \label{eq:BA-1}
    1 &= 
    \prod_{j=1}^{K_2} \frac{y_{1,k} - x_j^+}{y_{1,k} - x_j^-}
    \prod_{j=1}^{K_{\bar{2}}} \frac{1 - \frac{1}{y_{1,k} \bar{x}_j^-}}{1- \frac{1}{y_{1,k} \bar{x}_j^+}} , \\
    \begin{split}
    \label{eq:BA-2}
      \left(\frac{x_k^+}{x_k^-}\right)^L &=
      \prod_{\substack{j = 1\\j \neq k}}^{K_2} \frac{x_k^+ - x_j^-}{x_k^- - x_j^+} \frac{1- \frac{1}{x_k^+ x_j^-}}{1- \frac{1}{x_k^- x_j^+}} \sigma^2(x_k,x_j)
      \prod_{j=1}^{K_1} \frac{x_k^- - y_{1,j}}{x_k^+ - y_{1,j}}
      \prod_{j=1}^{K_3} \frac{x_k^- - y_{3,j}}{x_k^+ - y_{3,j}}
      \\ &\phantom{\ = \ }\times
      \prod_{j=1}^{K_{\bar{2}}} \frac{1- \frac{1}{x_k^+ \bar{x}_j^+}}{1- \frac{1}{x_k^- \bar{x}_j^-}} \frac{1- \frac{1}{x_k^+ \bar{x}_j^-}}{1- \frac{1}{x_k^- \bar{x}_j^+}} \tilde{\sigma}^2(x_k,\bar{x}_j)
      \prod_{j=1}^{K_{\bar{1}}} \frac{1 - \frac{1}{x_k^- y_{\bar{1},j}}}{1- \frac{1}{x_k^+ y_{\bar{1},j}}}
      \prod_{j=1}^{K_{\bar{3}}} \frac{1 - \frac{1}{x_k^- y_{\bar{3},j}}}{1- \frac{1}{x_k^+ y_{\bar{3},j}}} ,
    \end{split} \\
    \label{eq:BA-3}
    1 &= 
    \prod_{j=1}^{K_2} \frac{y_{3,k} - x_j^+}{y_{3,k} - x_j^-}
    \prod_{j=1}^{K_{\bar{2}}} \frac{1 - \frac{1}{y_{3,k} \bar{x}_j^-}}{1- \frac{1}{y_{3,k} \bar{x}_j^+}} , \\
    \label{eq:BA-1b}
    1 &= 
    \prod_{j=1}^{K_{\bar{2}}} \frac{y_{\bar{1},k} - \bar{x}_j^-}{y_{\bar{1},k} - \bar{x}_j^+}
    \prod_{j=1}^{K_2} \frac{1 - \frac{1}{y_{\bar{1},k} x_j^+}}{1- \frac{1}{y_{\bar{1},k} x_j^-}} , \\
    \begin{split}
    \label{eq:BA-2b}
      \left(\frac{\bar{x}_k^+}{\bar{x}_k^-}\right)^L &=
      \prod_{\substack{j = 1\\j \neq k}}^{K_{\bar{2}}} \frac{\bar{x}_k^- - \bar{x}_j^+}{\bar{x}_k^+ - \bar{x}_j^-} \frac{1- \frac{1}{\bar{x}_k^+ \bar{x}_j^-}}{1- \frac{1}{\bar{x}_k^- \bar{x}_j^+}} \sigma^2(\bar{x}_k,\bar{x}_j)
      \prod_{j=1}^{K_{\bar{1}}} \frac{\bar{x}_k^+ - y_{\bar{1},j}}{\bar{x}_k^- - y_{\bar{1},j}}
      \prod_{j=1}^{K_{\bar{3}}} \frac{\bar{x}_k^+ - y_{\bar{3},j}}{\bar{x}_k^- - y_{\bar{3},j}}
      \\ &\phantom{\ = \ }\times
      \prod_{j=1}^{K_2} \frac{1- \frac{1}{\bar{x}_k^- x_j^-}}{1- \frac{1}{\bar{x}_k^+ x_j^+}} \frac{1- \frac{1}{\bar{x}_k^+ x_j^-}}{1- \frac{1}{\bar{x}_k^- x_j^+}} \tilde{\sigma}^2(\bar{x}_k,x_j)
      \prod_{j=1}^{K_{1}} \frac{1 - \frac{1}{\bar{x}_k^+ y_{1,j}}}{1- \frac{1}{\bar{x}_k^- y_{1,j}}}
      \prod_{j=1}^{K_{3}} \frac{1 - \frac{1}{\bar{x}_k^+ y_{3,j}}}{1- \frac{1}{\bar{x}_k^- y_{3,j}}} ,
    \end{split} \\
    \label{eq:BA-3b}
    1 &= 
    \prod_{j=1}^{K_{\bar{2}}} \frac{y_{\bar{3},k} - \bar{x}_j^-}{y_{\bar{3},k} - \bar{x}_j^+}
    \prod_{j=1}^{K_2} \frac{1 - \frac{1}{y_{\bar{3},k} x_j^+}}{1- \frac{1}{y_{\bar{3},k} x_j^-}} .
\end{align}
The couplings appearing in the Bethe equations are graphically summarized in figure~\ref{fig:bethe-equations}.%

The level matching condition is equivalent to the requirement that the total momentum of the system must vanish
\begin{equation}
\prod_j^{K_2} \frac{x^+_j}{x^-_j} \, \prod_j^{K_{\bar{2}}} \frac{\bar{x}^+_j}{\bar{x}^-_j}=1.
\end{equation}
The total energy of a multi-excitation state that satisfies the Bethe equations and the level matching condition is given by
\begin{equation}
E=E_2+E_{\bar{2}}, \qquad E_j=\sum_{k=1}^{K_j} \sqrt{1+16 h^2 \sin^2{\frac{p_k}{2}}} .
\end{equation}

We note that the Bethe equations above differ from the ones conjectured in~\cite{Babichenko:2009dk} in two respects. Firstly the Bethe ansatz above contains symmetric phases that couple interactions of excitations with opposite chirality. In particular, equation~\eqref{eq:BA-2} contains an interaction between roots of type $2$ and $\bar{2}$ of the form
\begin{equation}
  \frac{1 - \frac{1}{x^+_k \bar{x}_j^+}}{1 - \frac{1}{x^-_k \bar{x}_j^-}}
\end{equation}
which is symmetric under the exchange of the momenta $p_k$ and $\bar{p}_j$ corresponding to the roots in the two sectors. Furthermore, equation~\eqref{eq:BA-2b} contains a factor that is the inverse of the above. These phases cannot be easily conjectured from the finite gap limit, but are necessary to ensure unitarity of the underlying S-matrix. 

Secondly, we have seen that within our construction it is not possible to choose the grading used in~\cite{Babichenko:2009dk} at the level of the all-loop Bethe equations. On the other hand the grading there was chosen arbitrarily from the finite gap equation, where there is complete freedom to do so. Indeed we will see in section~\ref{sec:finite-gap-limit} that the finite gap limit of our equation coincides with the construction of~\cite{Babichenko:2009dk} for an appropriate Cartan matrix, as we will see later.

Here we have constructed the Bethe equations in a particular grading, corresponding to choosing the first level-II vacuum in~\eqref{eq:level-II-vacuum}. Bethe equations corresponding to the three other choices can be obtained through a set of duality transformations. This is further discussed in appendix~\ref{sec:fermionic-duality}.

\subsection{Small \texorpdfstring{$h$}{h} limit and Cartan matrix}
\label{sec:small-h-limit}

In the weak coupling limit we expect the BA equation for the $l$-th node  to take the form
\begin{equation}
  \label{eq:BE-one-loop}
  \left( \frac{u_{l,i} + \frac{i}{2} w_l}{u_{l,i} - \frac{i}{2} w_l} \right)^L
  = \prod_{\substack{k = 1\\k \neq i}}^{K_l} \frac{u_{l,i} - u_{l,k} + \frac{i}{2} A_{ll}}{u_{l,i} - u_{l,k} - \frac{i}{2} A_{ll}}
  \prod_{l' \neq l} \prod_{k=1}^{K_{l'}} 
  \frac{u_{l,i} - u_{l',k} + \frac{i}{2} A_{ll'}}{u_{l,i} - u_{l,k} - \frac{i}{2} A_{ll'}},
\end{equation}
where $w_l$ are weights and $A_{ll'}$ is an element of the Cartan matrix of $\algPSU(1,1|2)^2$.
When $h\ll1$, let us expand
\begin{equation}
x^\pm\approx \frac{u_x\pm i/2}{h},\qquad
y\approx\frac{u_y}{h},
\end{equation}
in the left sector, where $u_i$ are finite as $h\to0$, and similarly  in the right sector.

If we assume that the dressing phases $\sigma$ and $\tilde{\sigma}$ expand trivially in this limit, we indeed find that the Bethe ansatz takes the form~\eqref{eq:BE-one-loop} and we can read off the resulting Cartan matrix
\begin{equation}\label{eq:cartan}
  A = \begin{pmatrix}
    0  & -1 &0 &0 & 0 &0 \\
    -1  & +2 & -1 & 0 & 0 & 0 \\
    0 & -1 & 0 &0 & 0 &0 \\
    0 & 0 &0 & 0 & +1  & 0 \\
    0 & 0 & 0 & +1  & -2 & +1 \\
    0 & 0 &0 & 0 & +1 & 0
  \end{pmatrix}.
\end{equation}
Comparing this with the Cartan matrices in~\eqref{eq:Cartan-su2} and~\eqref{eq:Cartan-sl2}, we see that this indeed corresponds to $\algPSU(1,1|2)^2$, with different gradings for the two factors of the algebra. 

After dualization of the nodes 1 and $\bar{1}$ the Bethe equations are written in a different grading, where all the nodes of the Dynkin diagrams are fermionic. From the weak coupling expansion we get the Cartan matrix
\begin{equation}\label{eq:cartan-dual}
  \tilde{A} =
  \begin{pmatrix}
    0  & 1   &0 &0 & 0 &0 \\
    1  & 0 & -1 & 0 & 0 & 0 \\
    0 & -1 & 0 &0 & 0 &0 \\
    0 & 0 &0 & 0 & -1  & 0 \\
    0 & 0 & 0 & -1  & 0 & 1 \\
    0 & 0 &0 & 0 & 1 & 0
  \end{pmatrix},
\end{equation}
corresponding to the fermionic gradings in~\eqref{eq:Cartan-ferm}. If we had dualized the nodes 3 and $\bar{3}$ instead we would have found the Cartan matrix $-\tilde{A}$. The consecutive dualization of 1, $\bar{1}$ and 3, $\bar{3}$ gives the Cartan matrix $-A$.

\subsection{Global charges}
\label{sec:constr-sc-fact}

By expanding the Bethe equations around large values of the spectral parameter we should obtain the global charges of the symmetry algebra~\cite{Beisert:2005fw,OhlssonSax:2011ms}. In doing so we will assume that the phases $\sigma_{pq}$ and $\widetilde{\sigma}_{pq}$ do not contribute to the charges. This is consistent with the phases being given at the leading order by the AFS phase in~\eqref{eq:dressing-AFS}. 

As we have seen above, the left- and right-moving sectors of the Bethe equations are naturally written using different gradings of the $\algPSU(1,1|2)$ algebra. The Dynkin labels $r_1$, $r_2$ and $r_3$ for the left-movers therefore give the eigenvalues of the Cartan generators $\gen{h}_i$ given in~\eqref{eq:SC-basis-su2}, while the labels $r_{\bar{1}}$, $r_{\bar{2}}$ and $r_{\bar{3}}$ for the right-movers correspond to the generators $\hat{\gen{h}}_i$ in~\eqref{eq:SC-basis-sl2}.
Expanding the Bethe equations we find
\begin{equation}
  \begin{aligned}
    \label{eq:dynkinlab}
    r_1 &= r_3 = + K_2 + \tfrac{1}{2} \delta D , \qquad & 
    r_2 &= L+K_1-2K_2+K_3 , \\
    r_{\bar{1}} &= r_{\bar{3}} =- K_{\bar{2}}  -\tfrac{1}{2} \delta D , \qquad &
    r_{\bar{2}} &= L - K_{\bar{1}} + 2K_{\bar{2}} - K_{\bar{3}} + \delta D,
  \end{aligned} 
\end{equation}
where the anomalous dimension $\delta D$ is given by
\begin{equation}\label{eq:anom-dim-def}
  \delta D = E_2 + E_{\bar{2}} - K_2 - K_{\bar{2}} = 
  2ih \sum_{k=1}^{K_2} \left(\frac{1}{x_k^-} - \frac{1}{x_k^+}\right) + 2ih \sum_{k=1}^{K_{\bar{2}}} \left(\frac{1}{\bar{x}_k^-} - \frac{1}{\bar{x}_k^+}\right) .
\end{equation}

A representation of $\algPSU(1,1|2)^2$ can be labeled by the eigenvalues of the highest weight state under the four generators $\gen{S}_0^L$, $\gen{S}_0^R$, $\gen{L}_5^L$ and $\gen{L}_5^R$. It is useful to combine them into the charges
\begin{equation}\label{eq:cartan-generators}
  \begin{aligned}
    \gen{D} &= -(\gen{S}_0^L + \gen{S}_0^R) , \qquad &
    \gen{J} &= \gen{L}_5^L + \gen{L}_5^R , \\
    \gen{S} &= -(\gen{S}_0^L - \gen{S}_0^R) , \qquad &
    \gen{K} &= \gen{L}_5^L - \gen{L}_5^R .
  \end{aligned}
\end{equation}
The spin-chain Hamiltonian can then be written as
\begin{equation}
  \gen{H} = \gen{D} - \gen{J} .
\end{equation}
We can now express the eigenvalues of the generators~\eqref{eq:cartan-generators} in terms of the excitation numbers $K_i$ as
\begin{equation}
  \begin{aligned}
    D &= + K_{\bar{2}} + \tfrac{1}{2}( K_1 + K_3 - K_{\bar{1}} - K_{\bar{3}}) + L + \delta D , \\
    S &= - K_{\bar{2}} + \tfrac{1}{2}( K_1 + K_3 + K_{\bar{1}} + K_{\bar{3}}) , \\
    J &= - K_{2} + \tfrac{1}{2}( K_1 + K_3 - K_{\bar{1}} - K_{\bar{3}}) + L , \\
    K &= - K_{2} + \tfrac{1}{2}( K_1 + K_3 + K_{\bar{1}} + K_{\bar{3}}) .
  \end{aligned}
\end{equation}
Note that the anomalous dimension $\delta D$ only contributes to the eigenvalue $D$ of the dilatation operator. The eigenvalue of the Hamiltonian now takes the form
\begin{equation}
  E = K_2 + K_{\bar{2}} + \delta D,
\end{equation}
as expected from equation~\eqref{eq:anom-dim-def}.

\subsection{Finite gap limit}
\label{sec:finite-gap-limit}

We are now interested in the finite gap limit of the Bethe equations. 
We consider the case of a long spin chain and large number of excitations $K_i$ with $L\approx K_i\gg1$. The semiclassical limit is achieved by requiring also large values of $h$. In this limit the Bethe roots condense on the cuts that appear in the finite gap equations~\cite{Kazakov:2004qf}.
In terms of the Bethe roots, we define the densities by
\begin{equation}
\rho_i(x)=\sum_{k=1}^{K_i}\frac{x^2}{x^2-1}\delta(x-x_{i,k}),\quad\quad i=1,2,3,\bar{1},\bar{2},\bar{3}.
\end{equation}
where the excitation numbers are large $K_i\gg1$ and we make use of the expansion
\begin{equation}
\label{eq:FGexpansion}
x_i^\pm \approx x_i\pm \frac{i}{2 h}\frac{x^2}{x^2-1}.
\end{equation}
With this prescription, we find the following finite gap equations
\begin{equation}
\begin{aligned}
2\pi n_1 &= - \int\frac{\rho_2(y)}{x-y}dy- \int\frac{\rho_{\bar{2}}(y)}{x-1/y}\frac{dy}{y^2} \\
2\pi n_2&=-\frac{x}{x^2-1}2\pi\mathcal{E}-\int\frac{\rho_1(y)}{x-y}dy + 2 \pint\frac{\rho_2(y)}{x-y}dy -\int\frac{\rho_3(y)}{x-y}dy\\
&\phantom{={}} +\int\frac{\rho_{\bar{1}}(y)}{x-1/y}\frac{dy}{y^2} + \int\frac{\rho_{\bar{3}}(y)}{x-1/y}\frac{dy}{y^2}+\frac{1}{x^2-1}\mathcal{M}, \\
2\pi n_3&= - \int\frac{\rho_2(y)}{x-y}dy- \int\frac{\rho_{\bar{2}}(y)}{x-1/y}\frac{dy}{y^2} \\
%\end{aligned}
%\end{equation}
%\begin{equation}
%\begin{aligned}
%%
\label{eq:FGlimit}
2\pi n_{\bar{1}}&= \int\frac{\rho_{2}(y)}{x-1/y}\frac{dy}{y^2}+\int\frac{\rho_{\bar{2}}(y)}{x-y}dy  \\
2\pi n_{\bar{2}}&=-\frac{x}{x^2-1}2\pi\mathcal{E}-\int\frac{\rho_1(y)}{x-1/y}\frac{dy}{y^2}  -\int\frac{\rho_3(y)}{x-1/y}\frac{dy}{y^2}\\
&\phantom{={}} +\int\frac{\rho_{\bar{1}}(y)}{x-y}dy -2 \pint\frac{\rho_{\bar{2}}(y)}{x-y}dy + \int\frac{\rho_{\bar{3}}(y)}{x-y}dy+\frac{1}{x^2-1}\mathcal{M}, \\
2\pi n_{\bar{3}}&= \int\frac{\rho_{2}(y)}{x-1/y}\frac{dy}{y^2}+\int\frac{\rho_{\bar{2}}(y)}{x-y}dy . \\ 
\end{aligned}
\end{equation}
Here $\mathcal{E}$ corresponds to the residue of the quasi-momentum and it is given by
\begin{equation}
\mathcal{E} = \frac{1}{2 \pi} (L -\epsilon_1 +2 \epsilon_2 -\epsilon_3 +\epsilon_{\bar{1}} +\epsilon_{\bar{3}}),
\end{equation}
where
\begin{equation}
\epsilon_i = \int \frac{\rho_i(x)}{x^2} dx.
\end{equation}
The quantity $\mathcal{M}$ has the meaning of winding of the corresponding solutions and it is given by
\begin{equation}\label{eq:winding-finite-gap}
\mathcal{M}=\mathcal{P}_1+\mathcal{P}_3-\mathcal{P}_{\bar{1}}+2 \mathcal{P}_{\bar{2}}- \mathcal{P}_{\bar{3}}=\mathcal{P}_1 - \mathcal{P}_{2} +\mathcal{P}_3-\mathcal{P}_{\bar{1}}+ \mathcal{P}_{\bar{2}}- \mathcal{P}_{\bar{3}},
\end{equation}
where
\begin{equation}
\mathcal{P}_i = \int \frac{\rho_i (x)}{x} dx.
\end{equation}
The last equality in~\eqref{eq:winding-finite-gap} is possible thanks to the level matching condition that reads
\begin{equation}
\mathcal{P}_2+\mathcal{P}_{\bar{2}}=0.
\end{equation}
The finite gap equations that we derived are apparently different but equivalent to the ones in~\cite{Babichenko:2009dk,Zarembo:2010yz}, and indeed the same construction performed there with a different choice of the grading, such as~\eqref{eq:cartan}, would have given precisely~\eqref{eq:FGlimit}.

The incompatibility between the coset construction of the finite gap equations for $\AdS_3 \times \Sphere^3 \times \Sphere^3 \times \Sphere^1$ proposed in~\cite{Babichenko:2009dk} and the near-BMN expansion performed in~\cite{Rughoonauth:2012qd} was highlighted in~\cite{Borsato:2012ss}.
In the discussion there it was clear that the problem is related to the presence of modes of mass $\alpha$ and $1-\alpha$ at the same time. In the $\alpha \to 1$ limit all the massive excitations have the same mass and we find no mismatch between the finite gap and the near-BMN descriptions for $\AdS_3 \times \Sphere^3 \times \Torus^4$.

\section{Comparison with perturbative results}
\label{sec:pertubative-S-matrix}

Recently several perturbative computations for the $\AdS_3\times \Sphere^3\times \Torus^4$ superstring have been performed \cite{David:2010yg,Abbott:2012dd,Rughoonauth:2012qd,Beccaria:2012kb,Sundin:2013ypa,Hoare:2013pma}. In particular, in~\cite{Beccaria:2012kb} by Beccaria, Levkovich-Maslyuk, Macorini and Tseytlin (BLMMT) a prediction for the scalar factors was derived from finite gap calculations, up to~$O(1/h^2)$. In~\cite{Sundin:2013ypa} Sundin and Wulff (SW) worked out several S-matrix elements at tree level in the near-BMN limit, and some at one-loop~$O(1/h^2)$ in the near-flat-space (or Maldacena-Swanson) limit~\cite{Maldacena:2006rv}. In~\cite{Hoare:2013pma} Hoare and Tseytlin (HT) wrote down, among other things, the whole near-BMN tree-level S-matrix.

Here we will discuss how our results compare with these.

\subsection{The HT tree-level near-BMN S-matrix}
It is immediate to see that the scattering processes allowed in the HT S-matrix are those that are non-zero in the one we derived in section~
\ref{sec:S-matrix}. In particular, both S-matrices are reflectionless, and both come from the tensor product of two $\algSU(1|1)^2$ S-matrices.

In fact we can compare both with the full S-matrix~(4.1) and with each factor~(4.8). To do so, we identify the fundamental fields of~\cite{Hoare:2013pma} with the ones of our $\phi,\psi,\bar{\phi},\bar{\psi}$ as follows
\begin{equation}
  \ket{\phi}=\ket{\phi_+} , \qquad
  \ket{\psi}=e^{i\pi/4}\ket{\psi_+} , \qquad
  \ket{\bar{\phi}}=\ket{\phi_-} , \qquad
  \ket{\bar{\psi}}=e^{i\pi/4}\ket{\psi_-} ,
\end{equation}
and (consistently with the tensor product structure) we identify the composite fields as
\begin{equation}
  \begin{aligned}
    \ket{\Phi^{+\dot{+}}} &= \ket{y_+}, \; &
    \ket{\Phi^{+\dot{-}}} &= e^{i\pi/4}\ket{\zeta_+}, \; &
    \ket{\Phi^{-\dot{+}}} &= e^{i\pi/4}\ket{\chi_+}, \; &
    \ket{\Phi^{-\dot{-}}} &= e^{i\pi/2}\ket{z_+},\\
    \ket{\bar{\Phi}^{+\dot{+}}} &= \ket{y_-}, \; &
    \ket{\bar{\Phi}^{+\dot{-}}} &= e^{i\pi/4}\ket{\zeta_-}, \; &
    \ket{\bar{\Phi}^{-\dot{+}}} &= e^{i\pi/4}\ket{\chi_-}, \; &
    \ket{\bar{\Phi}^{-\dot{-}}} &= e^{i\pi/2}\ket{z_-}.
  \end{aligned}
\end{equation}

Furthermore, we should take into account fermion signs arising from permuting the final states.

We can now expand our string frame S-matrix in the near-BMN limit where the momentum of the excitation scales as $p \sim \mathsfit{p}/h$ and the Zhukovsky variables expand as
\begin{equation}
  x^{\pm}_{\mathsfit{p}} = \left(1 \pm \frac{i \mathsfit{p}}{4 h}\right) \frac{(1+\omega_{\mathsfit{p}})}{\mathsfit{p}} + \order(1/h^2),\qquad \omega_{\mathsfit{p}}=\sqrt{1+{\mathsfit{p}}^2} ,
\end{equation}
consistently with the conditions in~\eqref{eq:shortening}.

Then, up to a rescaling of the expansion parameter  $h\to h/2$ and fixing the gauge parameter at $a=0$,\footnote{We can match our results with the general general $a$-gauge if allow for an additional (crossing invariant) factor   
$e^{i\,(p\, E_q-q\,E_p)\,\frac{a}{2}}$
in our S-matrix.} we reproduce perfectly the elements in~(4.8) of~\cite{Hoare:2013pma}, and consequently the ones of~(4.1) there.

% We also reproduce the elements of~(4.1) there up to some mismatches in the boson-fermion scattering, namely in the fourth line and in the signs in front of~$l_4(p,q)$ int he second column. We believe these to be misprints in the preprint.

\subsection{The SW tree level and one-loop results}
At tree-level, the comparison with~\cite{Sundin:2013ypa} follows the one of the previous section. Since in SW some computations are performed for the more general $\AdS_3\times \Sphere^3\times \Sphere^3\times \Sphere^1$ string theory, we should take $\alpha=1$ everywhere to recover the $\Torus^4$ background. Here the identifications necessary are,
\begin{equation}
  \begin{aligned}
    \ket{\Phi^{+\dot{+}}} &= \ket{y_2}, \; &
    \ket{\Phi^{+\dot{-}}} &= e^{-i\pi/4}\ket{\chi_2}, \; &
    \ket{\Phi^{-\dot{+}}} &= e^{i\pi/4}\ket{\chi_1}, \; &
    \ket{\Phi^{-\dot{-}}} &= \ket{y_1}, \\
    \ket{\bar{\Phi}^{+\dot{+}}} &= \ket{\bar{y}_2}, \; &
    \ket{\bar{\Phi}^{+\dot{-}}} &= e^{-i\pi/4}\ket{\bar{\chi}_2}, \; &
    \ket{\bar{\Phi}^{-\dot{+}}} &= e^{i\pi/4}\ket{\bar{\chi}_1}, \; &
    \ket{\bar{\Phi}^{-\dot{-}}} &= \ket{\bar{y}_1}.
  \end{aligned}
\end{equation}
Then, up to the redefinition $h\to-h/2$, we match the results presented there in the gauge~$a=0$.

Sundin and Wulff~\cite{Sundin:2013ypa} also computed certain one-loop elements in the near-flat-space limit. They correspond to the elements $\mathcal{A},\widetilde{\mathcal{A}},\mathcal{B},\widetilde{\mathcal{C}}$ of~\eqref{eq:stringframeS-elems}, and read\footnote{The tree level expression for $\mathcal{B}$ and $\widetilde{\mathcal{C}}$ was not given in~\cite{Sundin:2013ypa}. We thank the authors for communicating it to us privately.}
\begin{align}
\label{eq:1-loop-S-elements-1st}
\mathcal{A}_{pq} &= 1-\frac{i}{4h}\frac{p_-q_-(p_-+q_-)}{p_--q_-}\\
\nonumber
& \phantom{{}={}} \!{} 
+\frac{1}{32h^2}\frac{p_-^2q_-^2}{q_-^2-p_-^2}
\left(\frac{i}{\pi}(p_-+q_-)^2-\frac{2i}{\pi}\frac{q_-p_-(q_-+p_-)}{q_--p_-}\log\frac{q_-}{p_-}-\frac{(q_-+p_-)^3}{q_--p_-}\right)\\
\widetilde{\mathcal{A}}_{pq} &= 1-\frac{i}{4h}\frac{p_-q_-(p_--q_-)}{p_-+q_-}\\
\nonumber
& \phantom{{}={}} \!{} 
-\frac{1}{32h^2}\frac{p_-^2q_-^2}{q_-^2-p_-^2}
\left(\frac{i}{\pi}(p_--q_-)^2+\frac{2i}{\pi}\frac{q_-p_-(q_--p_-)}{q_-+p_-}\log\frac{q_-}{p_-}+\frac{(q_-^2+p_-^2)(q_--p_-)}{q_-+p_-}\right)\\
\mathcal{B}_{pq} &= 1-\frac{i}{4h}p_-q_-\\
\nonumber
& \phantom{{}={}} \!{} 
+\frac{1}{32h^2}\frac{p_-^2q_-^2}{q_-^2-p_-^2}
\left(\frac{i}{\pi}(p_-+q_-)^2-\frac{2i}{\pi}\frac{q_-p_-(q_-+p_-)}{q_--p_-}\log\frac{q_-}{p_-}-\frac{(q_-^2+p_-^2)(q_-+p_-)}{q_--p_-}\right)\\
\label{eq:1-loop-S-elements-last}
\widetilde{\mathcal{C}}_{pq} &= 1-\frac{i}{4h}p_-q_-\\
\nonumber
& \phantom{{}={}} \!{}
-\frac{1}{32h^2}\frac{p_-^2q_-^2}{q_-^2-p_-^2}
\left(\frac{i}{\pi}(p_--q_-)^2+\frac{2i}{\pi}\frac{q_-p_-(q_--p_-)}{q_-+p_-}\log\frac{q_-}{p_-}+(q_-^2-p_-^2)\right),
\end{align}
where we explicitly used the coupling constant~$h$ as a loop-counting parameter.  A first nontrivial requirement of our construction is that these elements satisfy the crossing equations~\eqref{eq:stringframeS-crossing}. It is easy to check that this is actually the case, recalling that in lightcone coordinate crossing $p\to\bar{p}$ amounts to $p_-\to \bar{p}_-=-p_-$, and taking everywhere the upper branch of the logarithm.

As a further check, we can explicitly expand our S-matrix elements at one-loop in the near-flat-space limit to match the results of~\cite{Sundin:2013ypa}. To do this we need to specify what the dressing factors $\sigma_{pq}$ and $\widetilde{\sigma}_{pq}$ are in that limit.\footnote{The dressing factors contribute to the imaginary part of the one loop terms in~\eqref{eq:1-loop-S-elements-1st}--\eqref{eq:1-loop-S-elements-last}, as discussed in~\cite{Sundin:2013ypa}.} Following SW, we use the near-flat-space expansion of the BLMMT factors~\cite{Beccaria:2012kb}, and in this way we find perfect agreement with all of~\eqref{eq:1-loop-S-elements-1st}--\eqref{eq:1-loop-S-elements-last}.

\subsection{The BLMMT dressing factors}
In~\cite{Beccaria:2012kb} a proposal for the one-loop dressing factors $\sigma_{pq}$ and $\widetilde{\sigma}_{pq}$ was put forward, which we have checked to be compatible with our crossing equations in the near-flat-space limit in the previous section.
However, the expressions~(6.8) and~(6.9) in~\cite{Beccaria:2012kb} are written in the finite-gap limit, which contains more information than the near-flat-space one. We can check whether these phases satisfy our crossing equations~\eqref{eq:crossingeq-chain} in the finite gap limit using the expansion~\eqref{eq:FGexpansion} and the crossing transformation  $x\to\bar{x}=1/x$.

When we plug in $\vartheta(x,y)$ and $\widetilde{\vartheta}(x,y)$ from~\cite{Beccaria:2012kb} in our crossing relations we find a mismatch, as the two phases match the imaginary part of the crossing equation, but not the real one. This discrepancy is not entirely surprising, since in~\cite{Beccaria:2012kb} the phases were computed by first working out the semiclassical energy shifts and then comparing with the Bethe ansatz of~\cite{Babichenko:2009dk} which differs from ours.\footnote{%
  Recall that there the interaction terms between particles of type $2$ and $\bar{2}$ did not include the symmetric phase present in~\eqref{eq:BA-2}.%
} 

It is interesting to notice that one can trace the mismatch to the fact that  the rational parts of $\vartheta(x,y)$ and $\widetilde{\vartheta}(x,y)$ do not satisfy
\begin{equation}
\vartheta_{\text{rational}}(x,y)+\widetilde{\vartheta}_{\text{rational}}(x,1/y)=0,
\end{equation}
which is also a natural generalization of what happens, \eg, in the $\AdS_5$ case, where the rational part of the Hernandez-Lopez phase is crossing-symmetric~\cite{Hernandez:2006tk}.\footnote{We thank Arkady Tseytlin for discussions on the one-loop dressing phases.}

\section{Conclusions}
\label{sec:conclusions}
We have constructed the all-loop S-matrix for a $\algPSU(1,1|2)^2$ spin-chain dual to strings on $\AdS_3\times \Sphere^3\times \Torus^4$ out of bootstrap. Due to its centrally extended $(\algPSU(1|1)^2\times\algU(1))^2$ symmetry, the resulting S-matrix is the (graded) tensor product of two copies of the $\algSU(1|1)^2$ invariant S-matrix discussed in~\cite{Borsato:2012ud}. It is completely determined up to antisymmetric ``dressing phases'' and we determined the crossing relations that these phases have to satisfy. Furthermore, the S-matrix satisfies the Yang-Baxter equation. This points to integrability of the underlying theory and allowed us to write down a set of Bethe ansatz equations for the asymptotic energy spectrum. These modify the ones originally conjectured in~\cite{Babichenko:2009dk} based on a discretisation of the finite-gap equations.

We have shown that at leading order in the strongly coupled regime the crossing equations are solved by setting both phases to be the same as the AFS phase~\cite{Arutyunov:2004vx}.  This allowed us to successfully compare our proposal with several independent perturbative calculations. First, we reproduce the finite-gap equations of~\cite{Babichenko:2009dk}, up to a different choice of grading. Furthermore, we reproduce the near-BMN tree-level results of~\cite{Rughoonauth:2012qd,Sundin:2013ypa,Hoare:2013pma} and the one-loop near-flat-space results of~\cite{Sundin:2013ypa}. Beyond the near-flat-space limit, our crossing equations are not compatible with the one-loop phases constructed in~\cite{Beccaria:2012kb}. However, such phases were found from matching with the Bethe ansatz of~\cite{Babichenko:2009dk} with a semiclassical calculation of the energy shifts for certain solutions, and the mismatch may just reflect the fact that the BA we constructed out of bootstrap does not coincide with that one. It would be very interesting to repeat that calculation for our BA and see whether the mismatch is resolved.

There are still several questions left to investigate, on which we hope to return soon. The most obvious one is finding an all-loop expression for the dressing phases that satisfies crossing, both in this theory\footnote{We plan to present an all-loop solution for the crossing equations presented here in an upcoming publication~\cite{upcoming}.} and in the case of the $\algD{\alpha}^2$ chain corresponding to $\AdS_3\times \Sphere^3\times \Sphere^3\times \Sphere^1$ strings~\cite{Borsato:2012ud,Borsato:2012ss}. Recall that such backgrounds are related to the ones considered here by blowing up one of the spheres (which amounts to $\alpha\to1$). Even if this limit is singular at the level of the symmetry algebra and the S-matrix, it should not be so on the physical observables, \eg, the spectrum. Indeed this should be made evident when comparing the BA of~\cite{Borsato:2012ss} with the one proposed here, once the dressing factors are established.

Additionally, let us stress again that both here and in~\cite{Babichenko:2009dk,OhlssonSax:2011ms,Borsato:2012ud,Borsato:2012ss} the integrability machinery has been applied only to the massive excitations of the spectrum. The two massless modes of $\AdS_3\times \Sphere^3\times \Sphere^3\times \Sphere^1$ and the four of $\AdS_3\times \Sphere^3\times \Torus^4$ are still missing from the picture. Following the results of~\cite{Sax:2012jv}, investigating the decompactification limit ($\alpha\to1$) may provide important insights. Once the massless modes are included in the fundamental particle spectrum, the S-matrix should be extended to allow for scattering of all possible bound states, see \eg~\cite{Ahn:2010ka}, thus completing the bootstrap program in the spirit of~\cite{Dorey:1996gd}.

In~\cite{Cagnazzo:2012se}, it has been shown that classical integrability holds also when both mentioned backgrounds are supported by a mixture of RR and NSNS fluxes. Recently, for such a mixed $\AdS_3\times \Sphere^3\times \Torus^4$ background the near-BMN S-matrix was put forward~\cite{Hoare:2013pma}, and it displayed a surprisingly simple form.  It would be very interesting to understand the spin-chain and S-matrix of the backgrounds with mixed NS-NS and R-R fluxes away from the near-BMN limit. If possible, this would provide an intriguing bridge between integrable models and the CFT current algebra techniques used to solve the pure NS-NS theory

\section*{Acknowledgments}

We thank Gleb Arutyunov, Davide Fioravanti, Ben Hoare, Per Sundin, Andrea Prinsloo, Arkady Tseytlin, Dima Volin, and Kostya Zarembo for interesting discussions,
and Gleb Arutyunov, Arkady Tseytlin and Kostya Zarembo for their comments on the manuscript.
R.B., O.O.S.\@ and A.S.\@ acknowledge support by the Netherlands Organization for Scientific Research (NWO) under the VICI grant 680-47-602; their work is also part of the ERC Advanced grant research programme No. 246974, ``Supersymmetry: a window to non-perturbative physics''. 
B.S.\@ acknowledges funding support from an EPSRC Advanced Fellowship and an STFC Consolidated Grant "Theoretical Physics at City University" ST/J00037X/1. He would also like to thank the CERN Theory division for hospitality during the final stages of this project.
A.T.\@ thanks EPSRC for funding under the First Grant project EP/K014412/1 "Exotic quantum groups, Lie superalgebras and integrable systems".

\appendix

\section{Fermionic duality}
\label{sec:fermionic-duality}

As argued in section \ref{sec:diag}, there are four different possible gradings in which we can write the all-loop Bethe equations.
A way to relate them is to perform fermionic dualities on the nodes corresponding to auxiliary roots.
In order to do that, let us define the following polynomial of degree $n=K_2+K_{\bar{2}}-1$
\begin{equation}
P(\xi)= \prod_{j=1}^{K_2} (\xi - x_j^+) \prod_{j=1}^{K_{\bar{2}}} (\xi - \frac{1}{\bar{x}_j^-}) -\prod_{j=1}^{K_2}(\xi - x_j^-) \prod_{j=1}^{K_{\bar{2}}}(\xi- \frac{1}{ \bar{x}_j^+}).
\end{equation}
The Bethe equations for auxiliary roots $y_1,y_3,y_{\bar{1}},y_{\bar{3}}$ can be written respectively as
\begin{equation}
P(y_1)=0, \qquad P(y_3)=0, \qquad P(1/y_{\bar{1}})=0, \qquad P(1/y_{\bar{3}})=0.
\end{equation}
We can choose to dualize either the auxiliary roots $y_1, y_{\bar{1}}$ or $y_3, y_{\bar{3}}$. In the first case we consider a set of dual $\tilde{K}_1,\tilde{K}_{\bar{1}}$ roots such that $K_1+ \tilde{K}_1=K_2 -1$ and $K_{\bar{1}}+\tilde{K}_{\bar{1}}=K_{\bar{2}}-1$. The polynomial can thus be rewritten as
\begin{equation}
  P(\xi) =  \xi  \prod_{j=1}^{K_1} (\xi - y_{1,j}) \prod_{j=1}^{\tilde{K}_1} (\xi - \tilde{y}_{1,j})  \prod_{j=1}^{K_{\bar{1}}} \left(\xi - \frac{1}{y_{\bar{1},j}}\right)  \prod_{j=1}^{\tilde{K}_{\bar{1}}} \left(\xi - \frac{1}{\tilde{y}_{\bar{1},j}}\right) 
\end{equation}
Evaluating the quantity $\frac{P(x^+_k)}{P(x^-_k)}$ we get the identity
\begin{multline}
  \left(\frac{x^+_k}{x^-_k}\right)^{K_{\bar{2}}-K_{\bar{1}}-\tilde{K}_{\bar{1}}-1} \prod_{\substack{j = 1\\j \neq k}}^{K_2}\frac{x^+_k - x_j^-}{x^-_k - x_j^+} \prod_{j=1}^{K_{\bar{2}}} \frac{1- \frac{1}{x^+_k \bar{x}_j^+}}{1 - \frac{1}{x^-_k \bar{x}_j^-}} \prod_{j=1}^{K_1} \frac{x^-_k - y_{1,j}}{x^+_k - y_{1,j}}   \prod_{j=1}^{K_{\bar{1}}} \frac{1 - \frac{1}{x^-_k y_{\bar{1},j}}}{1 - \frac{1}{x^+_k y_{\bar{1},j}}} = \\
  \prod_{j=1}^{\tilde{K}_1} \frac{x^+_k - \tilde{y}_{1,j}}{x^-_k - \tilde{y}_{1,j}}  \prod_{j=1}^{\tilde{K}_{\bar{1}}} \frac{1 - \frac{1}{x^+_k \tilde{y}_{\bar{1},j}}}{1 - \frac{1}{x^-_k \tilde{y}_{\bar{1},j}}} 
\end{multline}
Similarly, considering $\frac{P(1/\bar{x}^-_k)}{P(1/\bar{x}^+_k)}$ we get
\begin{multline}
  \left(\frac{\bar{x}^+_k}{\bar{x}^-_k}\right)^{K_{\bar{2}}-K_{\bar{1}}-\tilde{K}_{\bar{1}}-1} \prod_{j=1}^{K_2} \frac{x_j^-}{x_j^+}  \prod_{j=1}^{K_{\bar{2}}} \frac{\bar{x}^-_j}{\bar{x}^+_j} \prod_{j=1}^{K_2} \frac{1-\frac{1}{\bar{x}^-_k x_j^- }}{1-\frac{1}{\bar{x}^+_k x_j^+ }} \prod_{\substack{j = 1\\j \neq k}}^{K_{\bar{2}}}  \frac{\bar{x}^-_k - \bar{x}_j^+}{\bar{x}^+_k - \bar{x}_j^-}  \\
  \times \prod_{j=1}^{K_1} \frac{1 - \frac{1}{\bar{x}^+_k y_{1,j}}}{1 -\frac{1}{\bar{x}^-_k y_{1,j}}}  \prod_{j=1}^{K_{\bar{1}}} \frac{\bar{x}^+_k - y_{\bar{1},j}}{\bar{x}^-_k - y_{\bar{1},j}} =
  \prod_{j=1}^{\tilde{K}_1} \frac{1 - \frac{1}{\bar{x}^-_k \tilde{y}_{1,j}}}{1 -\frac{1}{\bar{x}^+_k \tilde{y}_{1,j}}} \prod_{j=1}^{\tilde{K}_{\bar{1}}} \frac{\bar{x}^-_k - \tilde{y}_{\bar{1},j}}{\bar{x}^+_k - \tilde{y}_{\bar{1},j}}
\end{multline}
With the help of these identities we can write the dualized Bethe equations
\begin{align}\label{eq:BA-ferm-first}
    1 &= 
    \prod_{j=1}^{K_2} \frac{\tilde{y}_{1,k} - x_j^-}{\tilde{y}_{1,k} - x_j^+}
    \prod_{j=1}^{K_{\bar{2}}} \frac{1 - \frac{1}{\tilde{y}_{1,k} \bar{x}_j^+}}{1- \frac{1}{\tilde{y}_{1,k} \bar{x}_j^-}} , \\
    \begin{split}
      \left(\frac{x_k^+}{x_k^-}\right)^L &=
      \prod_{\substack{j = 1\\j \neq k}}^{K_2} \frac{1- \frac{1}{x_k^+ x_j^-}}{1- \frac{1}{x_k^- x_j^+}} \sigma^2(x_k,x_j)
      \prod_{j=1}^{\tilde{K}_1} \frac{x^+_k - \tilde{y}_{1,j}}{x^-_k - \tilde{y}_{1,j}}  
	\prod_{j=1}^{K_3} \frac{x_k^- - y_{3,j}}{x_k^+ - y_{3,j}}
      \\ &\phantom{\ = \ }\times
      \prod_{j=1}^{K_{\bar{2}}}  \frac{1- \frac{1}{x_k^+ \bar{x}_j^-}}{1- \frac{1}{x_k^- \bar{x}_j^+}} \tilde{\sigma}^2(x_k,\bar{x}_j)
      \prod_{j=1}^{\tilde{K}_{\bar{1}}} \frac{1 - \frac{1}{x^+_k \tilde{y}_{\bar{1},j}}}{1 - \frac{1}{x^-_k \tilde{y}_{\bar{1},j}}} 
	\prod_{j=1}^{K_{\bar{3}}} \frac{1 - \frac{1}{x_k^- y_{\bar{3},j}}}{1- \frac{1}{x_k^+ y_{\bar{3},j}}},
    \end{split} \\
     1 &= 
    \prod_{j=1}^{K_2} \frac{y_{3,k} - x_j^+}{y_{3,k} - x_j^-}
    \prod_{j=1}^{K_{\bar{2}}} \frac{1 - \frac{1}{y_{3,k} \bar{x}_j^-}}{1- \frac{1}{y_{3,k} \bar{x}_j^+}} , \\
    1 &= 
    \prod_{j=1}^{K_{\bar{2}}} \frac{\tilde{y}_{\bar{1},k} - \bar{x}_j^+}{\tilde{y}_{\bar{1},k} - \bar{x}_j^-}
    \prod_{j=1}^{K_2} \frac{1 - \frac{1}{\tilde{y}_{\bar{1},k} x_j^-}}{1- \frac{1}{\tilde{y}_{\bar{1},k} x_j^+}} , \\
    \begin{split}
      \left(\frac{\bar{x}_k^+}{\bar{x}_k^-}\right)^L &=
      \prod_{\substack{j = 1\\j \neq k}}^{K_{\bar{2}}} \frac{1- \frac{1}{\bar{x}_k^+ \bar{x}_j^-}}{1- \frac{1}{\bar{x}_k^- \bar{x}_j^+}} \sigma^2(\bar{x}_k,\bar{x}_j)
      \prod_{j=1}^{\tilde{K}_{\bar{1}}} \frac{\bar{x}^-_k - \tilde{y}_{\bar{1},j}}{\bar{x}^+_k - \tilde{y}_{\bar{1},j}} 
	 \prod_{j=1}^{K_{\bar{3}}} \frac{\bar{x}_k^+ - y_{\bar{3},j}}{\bar{x}_k^- - y_{\bar{3},j}}
      \\ &\phantom{\ = \ }\times
      \prod_{j=1}^{K_2} \frac{1- \frac{1}{\bar{x}_k^+ x_j^-}}{1- \frac{1}{\bar{x}_k^- x_j^+}} \tilde{\sigma}^2(\bar{x}_k,x_j)
      \prod_{j=1}^{\tilde{K}_1} \frac{1 - \frac{1}{\bar{x}^-_k \tilde{y}_{1,j}}}{1 -\frac{1}{\bar{x}^+_k \tilde{y}_{1,j}}} 
	\prod_{j=1}^{K_{3}} \frac{1 - \frac{1}{\bar{x}_k^+ y_{3,j}}}{1- \frac{1}{\bar{x}_k^- y_{3,j}}} ,
    \end{split} \\
	\label{eq:BA-ferm-last}
     1 &= 
    \prod_{j=1}^{K_{\bar{2}}} \frac{y_{\bar{3},k} - \bar{x}_j^-}{y_{\bar{3},k} - \bar{x}_j^+}
    \prod_{j=1}^{K_2} \frac{1 - \frac{1}{y_{\bar{3},k} x_j^+}}{1- \frac{1}{y_{\bar{3},k} x_j^-}} .
\end{align}
The above equations are the ones that can be obtained by choosing $\Phi^{-\dot{+}},\bar{\Phi}^{+\dot{-}}$ to be the fields that compose the level-II vacuum.
Similarly, one could have started by dualizing the auxiliary roots $y_3, y_{\bar{3}}$ and obtain Bethe equations  corresponding to the choice of $\Phi^{+\dot{-}},\bar{\Phi}^{-\dot{+}}$ in the level-II vacuum. We do not write them, since they are equal to the ones written above after exchanging 1 and 3.
Two consecutive dualizations of first $y_1, y_{\bar{1}}$ and then $y_3, y_{\bar{3}}$ (or the opposite order) give Bethe equations  corresponding to the choice of $\Phi^{-\dot{-}},\bar{\Phi}^{+\dot{+}}$ in the level-II vacuum. They are equal to the Bethe equations derived in section \ref{sec:BAE} after exchanging left and right.

\section{Hopf algebra}
\label{sec:hopf-algebra}

In this section, we construct a Hopf algebra for the $\AdS_3$ scattering problem, following \cite{Plefka:2006ze}. Hopf algebras are a very convenient framework where to express several properties of integrable systems and their scattering problems. For reviews of the $\AdS_5$ treatment containing references to the relevant quantum group literature we refer to \cite{Torrielli:2010kq,Torrielli:2011gg}. The main object in question is the so-called \emph{coproduct} map on the Hopf algebra $A$
\begin{equation}
\Delta: \, A \rightarrow \, A \otimes A,
\end{equation}
which can be thought of as the symmetry action on two-particle \emph{in} states. The opposite coproduct $\Delta^{op} \equiv \Pi^g \, \Delta$, with $\Pi^g$ the graded permutation, acts then on \emph{out} states, with the S-matrix (R-matrix in the Mathematics literature) there to provide a canonical transformation between the asymptotic scattering bases. 

As it is the case for the $\AdS_5$ superstring, the coproduct can assume different forms according to the \emph{frame} (choice of basis) one uses, and different frames are related by (possibly non-local) field redefinitions. Few of the forms which one encounters correspond to the following pictures:
\begin{itemize}

\item \textit{String frame.} This gives rise to the form of the coproduct reported in the main text, see section~\ref{sec:crossing}. 

\item \textit{Spin-chain frame.} This is closer in nature to the expression of the symmetry generators as operators on a spin-chain \cite{Borsato:2012ud}. A nontrivial braiding of the coproduct originates from the length-changing nature of the symmetry. 
By the mechanism described in \cite{Gomez:2006va,Plefka:2006ze,Torrielli:2010kq,Torrielli:2011gg} applied to the left-moving representation in section 4 of \cite{Borsato:2012ud}, one begins by deducing
\begin{equation}
  \label{eq:coprod}
  \begin{aligned}
    \Delta(\gen{P}) &= \gen{P} \otimes e^{-ip} + \matId \otimes \gen{P}, \qquad & 
    \Delta(\gen{P}^\dag) &=\gen{P}^\dag \otimes e^{ip} + \matId \otimes \gen{P}^\dag,  \\
    \Delta(\gen{H}_{\smallL}) &=\gen{H}_{\smallL} \otimes \matId + \matId \otimes \gen{H}_{\smallL}, \qquad &
    \Delta(\gen{H}_{\smallR}) &=\gen{H}_{\smallR} \otimes \matId + \matId \otimes \gen{H}_{\smallR}, \\
    \Delta(\gen{Q}_{\smallL}) &=\gen{Q}_{\smallL} \otimes \matId + \matId \otimes \gen{Q}_{\smallL}, \qquad &
    \Delta(\gen{S}_{\smallL}) &=\gen{S}_{\smallL} \otimes \matId + \matId \otimes \gen{S}_{\smallL}, \\
    \Delta(\gen{Q}_{\smallR}) &=\gen{Q}_{\smallR} \otimes e^{-ip} + \matId \otimes \gen{Q}_{\smallR}, \qquad & 
    \Delta(\gen{S}_{\smallR}) &=\gen{S}_{\smallR} \otimes e^{ip} + \matId \otimes \gen{S}_{\smallR}.
  \end{aligned}
\end{equation}
from which by standard means - see equation \eqref{eq:rule} below - one obtains
\begin{equation}
\begin{aligned}
\label{eq:antio}
  \mathscr{S} (\gen{P}) &=- e^{ip} \gen{P}, \qquad &
  \mathscr{S} (\gen{P}^\dag) &=- e^{-ip} \, \gen{P}^\dag, \\
  \mathscr{S} (\gen{H}_{\smallL}) &=- \gen{H}_{\smallL}, \qquad &
  \mathscr{S} (\gen{H}_{\smallR}) &=- \gen{H}_{\smallR}, \\
  \mathscr{S} (\gen{Q}_{\smallL}) &=- \gen{Q}_{\smallL}, \qquad &
  \mathscr{S} (\gen{S}_{\smallL}) &=- \gen{S}_{\smallL}, \\
  \mathscr{S} (\gen{Q}_{\smallR}) &=- e^{ip} \, \gen{Q}_{\smallR}, \qquad &
  \mathscr{S} (\gen{S}_{\smallR}) &=- e^{-ip} \, \gen{S}_{\smallR}.
\end{aligned}
\end{equation}
where $\mathscr{S}$ denotes the antipode. Notice that the right-hand side of~\eqref{eq:chargeconj} corresponds to the analog of \eqref{eq:antio} calculated in the string frame (\textit{cf.} equation~\eqref{eq:antipode-rel} below).

One can check that, due to the centrality of the elements $e^{ip}\equiv \gen{U}$ and $e^{-ip}\equiv \gen{U}^{-1}$, the coproduct described above is a Lie algebra homomorphism. As described in the main text, the antiparticle representation is the same as the right-moving representation in \cite{Borsato:2012ud}, were it not for a different length-changing action. However, by the simple state redefinition $\ket{\chi} = \ket{\bar{\phi} \, Z^+}$ \cite{Beisert:2005tm,Janik:2006dc,Torrielli:2007mc} the coproduct on the right movers can be made coincide with \eqref{eq:coprod}, which makes the Hopf algebra completely consistent. The R-matrices one finds reproduce the transmission matrices in \cite{Borsato:2012ud}, decorated with suitable momentum-dependent phases to account for the above state-redefinition.
A similar coproduct has recently appeared in \cite{Hoare:2013pma} (see appendix B in that article) starting from a worldsheet perspective. As noticed in \cite{Hoare:2013pma} and in analogy with the AdS$_5$ case \cite{Plefka:2006ze}, cocommutativity of the central charges is satisfied if one accounts for the dependence on $p$ of the central charges' eigenvalues. 

\item \textit{Most symmetric frame}. This frame exactly matches the one used in \cite{Arutyunov:2009ga}. To this purpose, we choose the parameters $\gamma$ and $\delta$ in \cite{Borsato:2012ud}, formula (4.21) and (4.22),  to satisfy
\begin{equation}
\gamma + \delta = 0, \qquad \gamma - \delta = - \pi. 
\end{equation}
We then perform a twist of the coproduct and a rescaling of the generators:
\begin{equation}
  \begin{gathered}
    T \equiv \sum_{i,j=1}^2 \lambda_{ij} \, E_{ii} \otimes E_{jj}, \qquad 
    \Delta \to T^{-1} \Delta \, T, 
    \\
    \lambda_{11} = e^{- i \frac{q}{4}}, \qquad 
    \lambda_{12} = e^{- i \frac{(p+q)}{4}}, \qquad 
    \lambda_{21} = 1, 
    \qquad \lambda_{22} = e^{- i \frac{p}{4}},
    \\
    \gen{Q}_{\smallL} \to \gen{Q}_{\smallL}, \qquad
    \gen{Q}_{\smallR} \to e^{i \frac{p}{2}} \gen{Q}_{\smallR}, \qquad
    \gen{S}_{\smallL} \to \gen{S}_{\smallL}, \qquad
    \gen{S}_{\smallR} \to e^{- i \frac{p}{2}} \gen{S}_{\smallR}.
  \end{gathered}
\end{equation}
Recalling that $\Delta(e^{i p}) = e^{i p} \otimes e^{i p}$, we obtain
\begin{equation}
  \label{eq:coprodp}
  \begin{aligned}
    \Delta(\gen{Q}_{\smallL}) &= \gen{Q}_{\smallL} \otimes e^{- i \frac{p}{4}} + e^{i \frac{p}{4}} \otimes \gen{Q}_{\smallL}, \qquad &
    \Delta(\gen{S}_{\smallL}) &= \gen{S}_{\smallL} \otimes e^{i \frac{p}{4}} + e^{- i \frac{p}{4}} \otimes \gen{S}_{\smallL}, \\
    \Delta(\gen{Q}_{\smallR}) &= \gen{Q}_{\smallR} \otimes e^{- i \frac{p}{4}} + e^{i \frac{p}{4}} \otimes \gen{Q}_{\smallR}, \qquad &
    \Delta(\gen{S}_{\smallR}) &= \gen{S}_{\smallR} \otimes e^{i \frac{p}{4}} + e^{- i \frac{p}{4}} \otimes \gen{S}_{\smallR}.
  \end{aligned}
\end{equation}
The new coproduct on the bosonic generators can be calculated by anti-commuting the supercharges' coproducts. With the choices made above, our generators and coproducts exactly coincide with the ones reported in \cite{Arutyunov:2009ga}, for the choice $\xi = - \frac{p}{4}$ and $g=h$. We also set $s=1$ for the rest of this section to achieve perfect matching.
The new R-matrix is unaffected by the rescaling (since $\Delta(e^{i p}) = e^{i p} \otimes e^{i p}$ is a symmetry) but picks up the twist in the following fashion~\cite{Drinfeld:1989st}:
\begin{equation}
\label{eq:twisto}
  R^{LL} \to T_{21}^{-1} \, R^{LL} \, T. 
\end{equation}
The other R-matrices ($RL$, $LR$ and $RR$) are then directly derived by imposing invariance under the same form of the coproduct \eqref{eq:coprodp} in all mixed and non-mixed representations. Twists similar to \eqref{eq:twisto} can then be shown to connect the R-matrices found in this way to those in the spin-chain frame.
\end{itemize}
As these are simply different manifestations of a same underlying quantum group structure, we will from the rest of this section focus on the most symmetric frame. Let us also remark that, because of the factorized nature of the symmetry algebra \eqref{eq:S-mat-tensor-prod}, we can focus on one copy of the $\algSU(1|1)^2$ R-matrix, as this is enough to recover the entire $\Torus^4$ scattering problem. The antipode $\mathscr{S}$ is easily found, as it needs to respect
\begin{equation}
  \label{eq:rule}
  \mu \, (\mathscr{S} \otimes \matId) \, \Delta = \eta \, \epsilon
\end{equation}
where $\mu : A \otimes A \to A$ is the multiplication on the Hopf algebra $A$, $\epsilon : A \to \mathbbmss{C}$ is the counit and $\eta : \mathbbmss{C} \to A$ the unit. In our case\begin{equation}
\epsilon (\mathfrak{J}) = 0\quad \forall \, \, \, \mathfrak{J} \in \algSU(1|1)^2, \qquad \epsilon (\matId) = 1,
\end{equation} 
hence one straightforwardly obtains
\begin{equation}
  \label{eq:podo}
  \mathscr{S} (\gen{Q}_{\smallL}) = -\gen{Q}_{\smallL}, \qquad 
  \mathscr{S} (\gen{S}_{\smallL}) = -\gen{S}_{\smallL}, \qquad
  \mathscr{S} (\gen{Q}_{\smallR}) = -\gen{Q}_{\smallR}, \qquad
  \mathscr{S} (\gen{S}_{\smallR}) = -\gen{S}_{\smallR}.
\end{equation}
Let us now impose the antiparticle relation~\cite{Janik:2006dc} on a generator $\gen{J}$
\begin{equation}
  \label{eq:underl}
  \mathscr{S} \big(\gen{J}(p)\big) = \mathscr{C}^{-1} \big(\underline{\gen{J}}(\bar{p})\big)^{\str} \mathscr{C},
\end{equation}
where $\mathscr{C}$ is a charge-conjugation matrix, the apex ${}^{\str}$ denotes supertransposition, the upper bar the particle-to-antiparticle transformation in the representation parameters and $\underline{\gen{J}}$ is generator in the antiparticle representation,
\begin{equation}
  \label{eq:unders}
  \underline{\gen{Q}}_{\smallL} = \gen{Q}_{\smallR}, \qquad 
  \underline{\gen{S}}_{\smallL} = \gen{S}_{\smallR}, \qquad 
  \underline{\gen{Q}}_{\smallR} = \gen{Q}_{\smallL}, \qquad 
  \underline{\gen{S}}_{\smallR} = \gen{S}_{\smallL},
\end{equation}
so that the particle-to-antiparticle transformation is, for instance
\begin{equation}
  \label{eq:antipode-rel}
  \mathscr{S} \big(\gen{Q}_{\smallL}(p) \big) = \mathscr{C}^{-1} \, \big(\gen{Q}_{\smallR}(\bar{p})\big)^{\str} \, \mathscr{C},
\end{equation}
and so on. By solving the resulting equation for the four generators we obtain that\footnote{%
  The charge conjugation matrix transforms the left moving basis into the right moving one. The matrix \eqref{eq:C} is written in the basis $(\phi,\psi) \rightarrow (\bar{\phi},\bar{\psi})$ where we denote, with an abuse of notation, the antiboson $|\chi\rangle = |\bar{\phi} Z^+\rangle$ again by the symbol $|\bar{\phi}\rangle$, and the antifermion by $|\bar{\psi}\rangle$.%
} %
\begin{equation}
  \label{eq:C}
\mathscr{C} = \begin{pmatrix}1&0\\0&i\end{pmatrix},
\end{equation}
and that the particle-to-antiparticle transformation can be obtained by crossing the rapidities of all representation parameters, \ie by sending 
\begin{equation}
  \label{eq:cross}
  x^\pm \to \frac{1}{x^\pm}.
\end{equation}
in term of the Zhukovski variables. The R-matrices for $LL$, $RR$, and mixed $RL$ and $LR$ scattering are then obtained by imposing
\begin{equation}
  \Delta^{\text{op}} (\mathfrak{J}) \, \mathcal{R} = \mathcal{R} \, \Delta (\mathfrak{J}) \qquad \forall \, \, \, \mathfrak{J} \in \algSU(1|1)^2.
\end{equation}
In the above formula, $\mathcal{R}$ is the universal R-matrix of (the Yangian of) $\algSU(1|1)$ (see next section). According to whether we project the coproduct and its opposite in the $I \otimes J$ representation, with $I$ and $J$ each assuming values $L$ or $R$, we obtain four specific $4 \times 4$ R-matrices $R^{IJ}$. 
One finds for $LL$
\begin{equation}
  \begin{aligned}
    R^{LL}_{pq} \ket{\phi} \otimes \ket{\phi} &= {\kappa^{LL}_{pq}} \, \frac{x_q^+ - x_p^-}{x_q^- - x_p^+} e^{i \frac{p - q}{4}} \ket{\phi} \otimes \ket{\phi}, \\
    R^{LL}_{pq} \ket{\phi} \otimes \ket{\psi} &= {\kappa^{LL}_{pq}} \, \frac{x_q^+ - x_p^+}{x_q^- - x_p^+} e^{- i \frac{p + q}{4}} \ket{\phi} \otimes \ket{\psi} + {\kappa^{LL}_{pq}} \, \frac{x_q^+ - x_q^-}{x_q^- - x_p^+} \frac{\eta_p}{\eta_q} \ket{\psi} \otimes \ket{\phi}, \\
    R^{LL}_{pq} \ket{\psi} \otimes \ket{\phi} &= {\kappa^{LL}_{pq}} \, \frac{x_q^- - x_p^-}{x_q^- - x_p^+} e^{i \frac{p + q}{4}} \ket{\psi} \otimes \ket{\phi} + {\kappa^{LL}_{pq}} \, \frac{x_p^+ - x_p^-}{x_q^- - x_p^+} \frac{\eta_q}{\eta_p} \ket{\phi} \otimes \ket{\psi}, \\
    R^{LL}_{pq} \ket{\psi} \otimes \ket{\psi} &= {\kappa^{LL}_{pq}} \, e^{- i \frac{p - q}{4}} \ket{\psi} \otimes \ket{\psi},
\end{aligned}
\end{equation}
while the mixed $RL$ R-matrix reads
\begin{equation}
  \begin{aligned}
    R^{RL}_{pq} \ket{\bar{\phi}} \otimes \ket{\phi} &= A_{pq} \ket{\bar{\phi}} \otimes \ket{\phi} + B_{pq} \ket{\bar{\psi}} \otimes \ket{\psi}, \quad &
    R^{RL}_{pq} \ket{\bar{\phi}} \otimes \ket{\psi} &= C_{pq} \ket{\bar{\phi}} \otimes \ket{\psi}, \\
    R^{RL}_{pq} \ket{\bar{\psi}} \otimes \ket{\psi} &= E_{pq} \ket{\bar{\psi}} \otimes \ket{\psi} + F_{pq} \ket{\bar{\phi}} \otimes \ket{\phi}, \quad &
    R^{RL}_{pq} \ket{\bar{\psi}} \otimes \ket{\phi} &= D_{pq} \ket{\bar{\psi}} \otimes \ket{\phi},
  \end{aligned}
\end{equation}
\begin{equation*}
  \begin{aligned}
    A_{pq} &= {\kappa^{RL}_{pq}} \, \frac{x_q^- x_p^+ - 1}{x_q^+ x_p^+ - 1} e^{i \frac{p+q}{4}}, \; &
    B_{pq} &= {\kappa^{RL}_{pq}} \, \frac{i \eta_p \eta_q}{(x_q^+ x_p^+ - 1)} e^{i\frac{p}{2}}, \; &
    C_{pq} &= {\kappa^{RL}_{pq}} \, \frac{x_q^- x_p^- - 1}{x_q^+ x_p^+ - 1} e^{i \frac{3 p + q}{4}}, \\
    E_{pq} &= {\kappa^{RL}_{pq}} \, \frac{x_q^+ x_p^- - 1}{x_q^+ x_p^+ - 1} e^{i \frac{3 p- q}{4}}, \; &
    F_{pq} &= {\kappa^{RL}_{pq}} \, \frac{i \eta_p \eta_q}{x_q^+ x_p^+ - 1} e^{i\frac{p}{2}}, \; &
    D_{pq} &= {\kappa^{RL}_{pq}} e^{i\frac{p-q}{4}}. 
  \end{aligned}
\end{equation*}
The phase factors ${\kappa^{LL}_{pq}}$ and ${\kappa^{RL}_{pq}}$ are overall scalar functions undetermined by the invariance relations (similarly will be for the factors ${\kappa^{LR}_{pq}}$ and ${\kappa^{RR}_{pq}}$ to be introduced below), while 
\begin{equation}
  \eta_p = \sqrt{i (x^-_p - x^+_p)}.
\end{equation}
The crossing relations for an invertible antipode are given by
\begin{equation}
  (\mathscr{S} \otimes \matId) \mathcal{R} = \mathcal{R}^{-1} = (\matId \otimes \mathscr{S}^{-1}) \mathcal{R},
\end{equation}
where $R$ is the (invertible) R-matrix. If we use the antiparticle representation we have equipped ourselves with, we find for the first of the above equations
\begin{equation}
  \label{eq:matricial}
  (\mathscr{C}^{-1} \otimes \matId) \Big[R^{RL}\Big]^{\str_1}\Big(\frac{1}{x_1^\pm}, x_2^\pm\Big) (\mathscr{C} \otimes \matId) R^{LL} (x_1^\pm, x_2^\pm) 
  = \matId \otimes \matId,
\end{equation}
where the apex ${^{\str_1}}$ denotes supertransposition in the first factor. Denoting by  $p$ and $q$ the momenta of the first and second particle, respectively, one obtains that the matricial crossing equation \eqref{eq:matricial} is satisfied provided one imposes 
\begin{equation}
\label{eq:cro1}
  {\kappa^{LL}_{pq}} \kappa^{RL}_{\bar{p} q} = \frac{x_q^+ - x_p^+}{x_q^+ - x_p^-},
\end{equation}
where $\bar{p}$ denotes the map \eqref{eq:cross} We can obtain another equation by directly starting from a mixed R-matrix, and crossing one of the $L$ representations, namely
\begin{equation}
  (\mathscr{C}^{-1} \otimes \matId) \, \Big[R^{RR}\Big]^{\str_1}\Big(\frac{1}{x_1^\pm}, x_2^\pm\Big) \, (\mathscr{C} \otimes \matId) \, R^{LR} (x_1^\pm, x_2^\pm) 
  = \matId \otimes \matId.
\end{equation}
For the mixed representation $LR$ one calculates
\begin{equation}
  \begin{aligned}
    R^{LR}_{pq} \ket{\phi} \otimes \ket{\bar{\phi}} &= A'_{pq} \ket{\phi} \otimes \ket{\bar{\phi}} +  B'_{pq} \ket{\psi} \otimes \ket{\bar{\psi}}, \quad &
    R^{LR}_{pq} \ket{\phi} \otimes \ket{\bar{\psi}} &= C'_{pq} \ket{\phi} \otimes \ket{\bar{\psi}}, \\
    R^{LR}_{pq} \ket{\psi} \otimes \ket{\bar{\psi}} &= E'_{pq} \ket{\psi} \otimes \ket{\bar{\psi}} + F'_{pq} \ket{\phi} \otimes \ket{\bar{\phi}}, \quad &
    R^{LR}_{pq} \ket{\psi} \otimes \ket{\bar{\phi}} &= D'_{pq} \ket{\psi} \otimes \ket{\bar{\phi}},
  \end{aligned}
\end{equation}
with
\begin{equation*}
  \begin{aligned}
    A'_{pq} &= {\kappa^{LR}_{pq}} \frac{x_q^- x_p^+ - 1}{x_q^- x_p^- - 1} e^{- i \frac{p+q}{4}}, \; &
    B'_{pq} &= {\kappa^{LR}_{pq}} \frac{i \eta_p \eta_q}{x_q^- x_p^- - 1} e^{-i \frac{q}{2}}, \; &
    C'_{pq} &= {\kappa^{LR}_{pq}} e^{i \frac{p-q}{4}}, \\
    E'_{pq} &= {\kappa^{LR}_{pq}} \frac{x_q^+ x_p^- - 1}{x_q^- x_p^- - 1} e^{i \frac{p-3q}{4}}, \; &
    F'_{pq} &= {\kappa^{LR}_{pq}} \frac{i \eta_p \eta_q e^{- i \frac{q}{2}}}{x_q^- x_p^- - 1}, \; &
    D'_{pq} &= {\kappa^{LR}_{pq}} \frac{x_p^+ x_q^+ - 1}{x_p^- x_q^- - 1} e^{- i \frac{p+3q}{4}}.
  \end{aligned}
\end{equation*} 
Crossing symmetry on the first factor of the tensor product reads 
\begin{equation}
  (\mathscr{C}^{-1} \otimes \matId) \, \Big[R^{RR}\Big]^{\str_1}\Big(\frac{1}{x_p^\pm}, x_q^\pm\Big) \, (\mathscr{C} \otimes \matId) \, R^{LR} (x_p^\pm, x_q^\pm)  = \matId \otimes \matId,
\end{equation}
The R-matrix $R^{RR}_{pq}$ reads
\begin{equation}
  \label{eq:RRR}
  \begin{aligned}
    R^{RR}_{pq} \ket{\bar{\phi}} \otimes \ket{\bar{\phi}} &= {\kappa^{RR}_{pq}} \frac{x_q^+ - x_p^-}{x_q^- - x_p^+} e^{3 i \frac{p-q}{4}} \ket{\bar{\phi}} \otimes \ket{\bar{\phi}}, \\
    R^{RR}_{pq} \ket{\bar{\phi}} \otimes \ket{\bar{\psi}} &= {\kappa^{RR}_{pq}} \frac{x_q^+ - x_p^+}{x_q^- - x_p^+} e^{i \frac{p-3q}{4}}\ket{\bar{\phi}} \otimes \ket{\bar{\psi}} + {\kappa^{RR}_{pq}} \frac{i \eta_p \eta_q e^{i \frac{p-q}{2}}}{x_q^- - x_p^+} \ket{\bar{\psi}} \otimes \ket{\bar{\phi}}, \\
    R^{RR}_{pq} \ket{\bar{\psi}} \otimes \ket{\bar{\phi}} &= {\kappa^{RR}_{pq}} \frac{x_q^- - x_p^-}{x_q^- - x_p^+} e^{i \frac{3p-q}{4}} \ket{\bar{\psi}} \otimes \ket{\bar{\phi}} + {\kappa^{RR}_{pq}} \frac{i \eta_p \eta_q e^{i \frac{p-q}{2}}}{x_q^- - x_p^+} \ket{\bar{\phi}} \otimes \ket{\bar{\psi}}, \\
    R^{RR}_{pq} \ket{\bar{\psi}} \otimes \ket{\bar{\psi}} &= {\kappa^{RR}_{pq}} e^{i \frac{p-q}{4}} \ket{\bar{\psi}} \otimes \ket{\bar{\psi}},
  \end{aligned}
\end{equation}
from which we deduce 
\begin{equation}
  \label{eq:cro2}
  {\kappa^{LR}_{pq}} \, \kappa^{RR}_{\bar{p}q}= \frac{x_p^+ - \frac{1}{x_q^-}}{x_p^+ - \frac{1}{x_q^+}}.
\end{equation}
The antipode \eqref{eq:podo}is obviously idempotent on the Lie algebra $\algSU(1|1)^2$, from which we can straightforwardly write the second crossing equation as
\begin{equation}
  (\matId \otimes \mathscr{C}^{-1}) \, \Big[R^{LR}\Big]^{\str_2}\Big(x_p^\pm, \frac{1}{x_q^\pm}\Big) \, (\matId \otimes \mathscr{C} ) \, R^{LL} (x_p^\pm, x_q^\pm)= \matId \otimes \matId.
\end{equation}
\begin{equation}
(\matId \otimes \mathscr{C}^{-1}) \, \Big[R^{RR}\Big]^{\str_2}\Big(x_p^\pm, \frac{1}{x_q^\pm}\Big) \, (\matId \otimes \mathscr{C} ) \, R^{RL} (x_p^\pm, x_q^\pm)= \matId \otimes \matId.
\end{equation}
These equations give
\begin{equation}
\label{eq:cro34}  
{\kappa^{LL}_{pq}} \, \kappa^{LR}_{p \bar{q}} = \frac{x_q^- - x_p^-}{x_q^+ - x_p^-}, \qquad 
  \text{and} \qquad 
  {\kappa^{RL}_{pq}}\,  \kappa^{RR}_{p \bar{q}} = \frac{x_q^- - \frac{1}{x_p^+}}{x_q^- - \frac{1}{x_p^-}},
\end{equation}
where $\bar{q}$ means $x_q^\pm \rightarrow \frac{1}{x_q^\pm}$. Braiding unitarity 
\begin{equation}
  \check{R}^{IJ} \, R^{JI} = \matId \otimes \matId,
\end{equation}
with $\check{R}_{pq} = \left( \Pi^g \, \mathcal{R}\right)^{IJ}_{qp}$,
implies
\begin{equation}
\label{eq:unitiar}
  {\kappa^{LL}_{qp}} \, {\kappa^{LL}_{pq}} = 1, \quad 
  \kappa^{RR}_{qp} \, {\kappa^{RR}_{pq}} = 1, \quad 
  \kappa^{RL}_{qp} \, {\kappa^{LR}_{pq}} = 1, \quad 
  \kappa^{LR}_{qp} \, {\kappa^{RL}_{pq}} = 1.
\end{equation}

Let us now make contact with the formulation in the main text, where the LR-symmetry is implemented. To this purpose, we define
\begin{equation}
  \kappa^{LL}_{pq} = \kappa^{RR}_{pq} = S^{LL}_{pq}, \qquad 
  \kappa^{RL}_{pq} = \frac{1}{\zeta_{pq}} e^{-i \frac{p-q}{4}} S^{LR}_{pq}, \qquad 
  \kappa^{LR}_{pq} = \zeta_{pq} e^{-i \frac{p-q}{4}} S^{LR}_{pq},
\end{equation}
where
\begin{equation}
  \zeta_{pq} = \sqrt{\frac{1-\frac{1}{x_p^- \, x_q^-}}{1-\frac{1}{x_p^+ \, x_q^+}}}.
\end{equation}
Unitarity \eqref{eq:unitiar} is then solved by imposing
\begin{equation}
  S^{LL}_{qp} \, S^{LL}_{pq} = 1, \quad S^{LR}_{qp} \, S^{LR}_{pq} = 1,
\end{equation}
while the four crossing relations \eqref{eq:cro1}, \eqref{eq:cro2} and \eqref{eq:cro34} reduce to
\begin{equation}
  \begin{aligned}
    \label{eq:see}
    S^{LR}_{\bar{p} q} \, S^{LL}_{pq} = S^{LR}_{p \bar{q}} \, S^{LL}_{pq} &= \frac{x_q^- - x_p^-}{x_q^+ - x_p^-} \sqrt{\frac{x_q^+ - x_p^+}{x_q^- - x_p^-}} \, e^{- i \frac{p-q}{4}}=f(x_p,x_q), \\
    S^{LL}_{p \bar{q}} \, S^{LR}_{pq} = S^{LL}_{\bar{p} q} \, S^{LR}_{pq} &= \frac{x_q^- \, x_p^+ - 1}{x_q^+ \, x_p^+ - 1} \sqrt{\frac{x_q^+ \, x_p^+ - 1}{x_q^- \, x_p^- - 1}} \, e^{- i \frac{p-q}{4}}=g(x_p,x_q).
  \end{aligned}
\end{equation}
It is worth noticing that these equations are compatible with antisymmetry of $S^{LL},S^{LR}$, provided that crossing amounts to shifts of the torus rapidity in opposite direction in either argument of the S-matrix. For instance, denoting by $\omega$ half of the imaginary period of the $z$-torus, we have that from
\begin{equation}
S^{LR}_{p_1\,\bar{p}_2} \, S^{LL}_{p_1\,p_2}=S^{LR}(z_1,z_2\pm\omega) \, S^{LL}(z_1,z_2) =f(x_1,x_2)\,,
\end{equation}
we use antisymmetry to find
\begin{equation}
S^{LR}(z_1\pm\omega,z_2) \, S^{LL}(z_1,z_2) =\frac{1}{f(x_2,x_1)}\,,
\end{equation}
and analytic continuation in $z_1$ gives
\begin{equation}
S^{LL}(z_1\mp\omega,z_2)\,S^{LR}(z_1,z_2) =  S^{LL}_{\bar{p}_1\,p_2}\,S^{LR}_{p_1\,p_2} =\frac{1}{f(x_2,1/x_1)}=g(x_1,x_2)\,,
\end{equation}
where the last equality is manifestly a property of $f,g$. Similar relations can be found starting from any of the four equations in~\eqref{eq:see}.

Finally, by means of the redefinition
\begin{equation}
  S^{LL}_{pq} = S_{pq} \,  \frac{x_q^- - x_p^+}{x_q^+ - x_p^-}, \qquad S^{LR}_{pq}  = \tilde{S}_{pq}\, e^{i \frac{p-q}{4}},
\end{equation}
we see that the relations \eqref{eq:see} take the same form as \eqref{eq:crossingeq-chain} in the main text.

%%%%%%%%%%%%%%%

\section{The S-matrix from the universal R-matrix of the \texorpdfstring{$\algGL(1|1)$}{gl(1|1)} Yangian}
\label{sec:Yangian}

In this section, we show how to formulate the AdS$_3$ scattering problem in terms of Yangians, following \cite{Khoroshkin:1994uk,Cai:1997,Beisert:2005wm,Arutyunov:2009ce}. We will work in the spin-chain frame for simplicity, as it has a $\algGL(1|1)$ subsector (the left-moving) which is un-braided, hence it does not introduce additional complications when comparing with the Yangian. 

The super-Yangian double $DY(\algGL(1|1))$ is generated by $\Ygen{e}_n$, $\Ygen{f}_n$, $\Ygen{h}_n$, $\Ygen{k}_n$, with $n\in\Integers$, satisfying the following relations typical of Drinfeld's second realization \cite{Drinfeld:1987sy},
\begin{equation}
  \label{eq:Lie}
  \begin{gathered}
    \comm{\Ygen{h}_0}{\Ygen{e}_n} = -2\Ygen{e}_n , \qquad
    \comm{\Ygen{h}_0}{\Ygen{f}_n} = +2\Ygen{f}_n, \qquad
    \acomm{\Ygen{e}_m}{\Ygen{f}_n} = -\Ygen{k}_{m+n} , \\
    \comm{\Ygen{h}_m}{\Ygen{h}_n} = 
    \comm{\Ygen{h}_m }{\Ygen{k}_n} =
    \comm{\Ygen{k}_m }{\Ygen{k}_n} = 
    \comm{\Ygen{k}_m }{\Ygen{e}_n} =
    \comm{\Ygen{k}_m }{\Ygen{f}_n} =
    \acomm{\Ygen{e}_m}{\Ygen{e}_n} = 
    \acomm{\Ygen{f}_m}{\Ygen{f}_n} = 0 , \\
    \comm{\Ygen{h}_{m+1}}{\Ygen{e}_n} - \comm{\Ygen{h}_m}{\Ygen{e}_{n+1}} + \acomm{\Ygen{h}_m}{\Ygen{e}_n} = 0, \qquad
    \comm{\Ygen{h}_{m+1}}{\Ygen{f}_n} - \comm{\Ygen{h}_m}{\Ygen{f}_{n+1}} - \acomm{\Ygen{h}_m}{\Ygen{f}_n} = 0.
  \end{gathered}
\end{equation}
Drinfeld's currents are given by
\begin{equation}
  \begin{aligned}
    E^{\pm}(t) &= \pm \sum_{\substack{n \ge 0 \\ n<0}} \Ygen{e}_n t^{-n-1} , \qquad &
    K^{\pm}(t) &= 1\pm \sum_{\substack{n \ge 0 \\ n<0}} \Ygen{k}_n t^{-n-1} , \\
    F^{\pm}(t) &= \pm \sum_{\substack{n \ge 0 \\ n<0}} \Ygen{f}_n t^{-n-1} , \qquad &
    H^{\pm}(t) &= 1\pm \sum_{\substack{n \ge 0 \\ n<0}} \Ygen{h}_n t^{-n-1} .
  \end{aligned}
\end{equation}
The universal $R$-matrix reads
\begin{equation}
  \label{eq:rmatrix}
  \mathcal{R}=\mathcal{R}_+\mathcal{R}_1\mathcal{R}_2\mathcal{R}_-, 
\end{equation}
where one defines
\begin{equation}
  \begin{gathered}
    \mathcal{R}_+=\prod_{n\ge 0}^{\rightarrow}\exp(- \Ygen{e}_n\otimes \Ygen{f}_{-n-1}),\qquad  
    \mathcal{R}_-=\prod_{n\ge 0}^{\leftarrow}\exp(\Ygen{f}_n\otimes \Ygen{e}_{-n-1}), \\ 
    \begin{aligned}
      \mathcal{R}_1 &= \prod_{n\ge 0} \exp \left\{ \res_{t=z}\left[(-1) \frac{d}{dt}(\log H^+(t)) \otimes \log K^-(z+2n+1)\right]\right\}, \\
      \mathcal{R}_{2} &= \prod_{n\ge 0} \exp \left\{ \res_{t=z}\left[(-1) \frac{d}{dt}(\log K^+(t))\otimes \log H^-(z+2n+1)\right]\right\},
    \end{aligned}
  \end{gathered}
\end{equation}
and the residue is given by the following formula,
\begin{equation}
  \res_{t=z}\left[A(t)\otimes B(z)\right] = \sum_k a_k\otimes b_{-k-1},
\end{equation}
after achieving for the currents the expansions $A(t)=\sum_k a_k t^{-k-1}$, $B(z)=\sum_k b_k z^{-k-1}$.

The following representation satisfies the whole of\eqref{eq:Lie} for the left-moving representation, in terms of an evaluation parameter $\lambda$:
\begin{equation}
  \Ygen{e}_n = \lambda^n \, \gen{Q}_{\smallL}, \qquad  
  \Ygen{f}_n = \lambda^n \, \gen{S}_{\smallL}, \qquad
  \Ygen{k}_n = -\lambda^n \, \{\gen{Q}_{\smallL}, \gen{S}_{\smallL}\}, \qquad 
  \Ygen{h}_n = \lambda^n (-)^F,
\end{equation}
with $F$ the fermionic number. Because of the fermionic nature of the generators, one readily obtains
\begin{equation}
  \begin{aligned}
    \mathcal{R}_- &= 1+\sum_{n\geq0}\Ygen{f}_n\otimes \Ygen{e}_{-n-1}=1-\frac{\gen{S}_{\smallL}\otimes \gen{Q}_{\smallL}}{\lambda_p - \lambda_q} , \\
    \mathcal{R}_+ &= 1-\sum_{n\geq0}\Ygen{e}_n\otimes \Ygen{f}_{-n-1}=1+\frac{\gen{Q}_{\smallL}\otimes \gen{S}_{\smallL}}{\lambda_p - \lambda_q} .
\end{aligned}
\end{equation}
Concerning the Cartan generators, in suitable convergency domains we have \cite{Arutyunov:2009ce}
\begin{equation}
  -\frac{d}{dt}\log H^+ = \sum_{m=1}^{\infty} \left\{{\lambda^m} -{(\lambda-\Ygen{h}_0)^m}\right\} t^{-m-1}
\end{equation}
and
\begin{multline}
  \log K^-(z+2n+1) = \log K^-(2n+1) + \\
  +\sum_{m=1}^{\infty} \left\{\frac{1}{(\lambda-1-2n)^m} -\frac{1}{(\lambda-1-2n-\Ygen{k}_0)^m}\right\} \frac{z^{m}}{m}.
\end{multline}
and similarly for $\mathcal{R}_2$.
By applying the procedure of \cite{Arutyunov:2009ce} one then immediately obtains, in the basis $\{\phi \otimes \phi,  \psi \otimes \phi, \phi \otimes \psi, \psi \otimes \psi\}$,
\begin{equation}
  \label{eq:uni}
  \mathcal{R} = 
  \begin{pmatrix}
    1 & 0 & 0 & 0 \\
    0 & 1-\frac{a_2 b_2 }{\delta \lambda+a_1 b_1 } & \frac{a_1 b_2 }{\delta \lambda+a_1
      b_1 } & 0 \\
    0 & \frac{a_2 b_1 }{\delta \lambda+a_1 b_1 } & \frac{\delta \lambda}{\delta \lambda+a_1 b_1 } &
    0 \\
    0 & 0 & 0 & \frac{\delta \lambda-a_2 b_2 }{\delta \lambda+a_1 b_1 }
  \end{pmatrix},
\end{equation}
where we have disregarded an overall scalar factor.
As already noticed in \cite{Beisert:2005wm,Arutyunov:2009ce}, one has to choose $\delta \lambda = \frac{i h}{2}(x_p^+ - x_q^+)$ to reproduce the LL R-matrix (up to an overall normalization factor). 

Similarly, the mixed RL R-matrix can be shown to originate from the same universal R-matrix we just described, in the mixed left-right representation, provided one requests
for the left mover 
\begin{equation}
  \lambda_2 = \frac{i h}{2} x_2^+ + c
\end{equation}
as expected from the above $LL$ analysis (with $c$ a constant independent on the representation), and 
\begin{equation}
  \label{eq:found}
  \lambda_1 = \frac{i h}{2} \frac{1}{x_1^-} + c
\end{equation}
for the right mover (for the same constant $c$).
This curious observation was a puzzle in~\cite{Arutyunov:2009ce}, however we will provide an explanation of this fact in the following section.

\subsection{Level 1 crossing}
\label{sec:Yangian-crossing}
The Yangian algebra is uniquely determined when one knows the level $0$ and $1$ generators. In the previous appendix we have focussed our attention on the level $0$. Because of the peculiar level $0$ coproduct (nontrivially braided on one of the two $\algSU(1|1)$'s of $\algSU(1|1)^2$) the level $0$ symmetry fixes the matricial form of the scattering matrix in these representations. However, one can ask what the level $1$ Yangian has to say in this respect. It turns out that one can fix the crossing transformation of the Yangian evaluation parameter $\lambda$ by studying the level $1$ coproduct, in a way that makes the Yangian symmetry completely consistent with what found purely by exploring the level $0$ algebra.

One can show that the following level $1$ Yangian coproducts satisfies the algebra defining relations and is also a symmetry of the S-matrix:
\begin{equation}
  \begin{aligned}
    \Delta(\Ygen{e}_1) &= \Ygen{e}_1 \otimes \matId + \matId \otimes \Ygen{e}_1 + \Ygen{k}_0 \otimes \Ygen{e}_0, \\
    \Delta(\Ygen{f}_1) &= \Ygen{f}_1 \otimes \matId + \matId \otimes \Ygen{f}_1 + \Ygen{f}_0 \otimes \Ygen{k}_0, \\
    \Delta(\Ygen{k}_1) &= \Ygen{k}_1 \otimes \matId + \matId \otimes \Ygen{k}_1 + \Ygen{k}_0 \otimes \Ygen{k}_0.
  \end{aligned}
\end{equation}
From \eqref{eq:rule} we obtain 
\begin{equation}
  \mathscr{S} (\Ygen{e}_1) = - \Ygen{e}_1 + \Ygen{e}_0 \, \Ygen{k}_0, \qquad 
  \mathscr{S} (\Ygen{f}_1) = - \Ygen{f}_1 + \Ygen{f}_0 \, \Ygen{k}_0, \qquad 
  \mathscr{S} (\Ygen{k}_1) = - \Ygen{k}_1 + \Ygen{k}_0^2,
\end{equation}
which we can feed into~\eqref{eq:underl}, where $\gen{J}$ is any of $\Ygen{e}_1$, $\Ygen{f}_1$ and $\Ygen{k}_1$. What we still need to specify is what value of the evaluation parameter $ \bar{\underline{\lambda}}$ one has to choose for the right moving representation. Evaluation parameters appear in the above formula as $\Ygen{e}_1 = \lambda \, \Ygen{e}_0$, \textit{etc}.\@ and, respectively, $\bar{\underline{\Ygen{e}}}_1 = \bar{\underline{\lambda}} \, \bar{\underline{\Ygen{e}}}_0$, \textit{etc}. We find that it must be
\begin{equation}
  \label{eq:splend}
  \lambda = i \frac{h}{2} x^+ + c, \qquad 
   \bar{\underline{\lambda}} = i \frac{h}{2} x^- + c,
\end{equation}
with $c$ a constant independent of the representation, for \eqref{eq:underl} to be satisfied, consistently with the result \eqref{eq:found} found from matching the universal R-matrix.

\bibliographystyle{nb}
\bibliography{refs,upcoming}

\makeatletter \@ifundefined{Sphere}{\newcommand{\Sphere}{S}}{}
  \@ifundefined{AdS}{\newcommand{\AdS}{\mathrm{AdS}}}{}
  \@ifundefined{CFT}{\newcommand{\CFT}{\mathrm{CFT}}}{}
  \@ifundefined{CP}{\newcommand{\CP}{\mathrm{CP}}}{}
  \@ifundefined{Torus}{\newcommand{\Torus}{\mathrm{T}}}{}
  \@ifundefined{superN}{\newcommand{\superN}{\mathcal{N}}}{}
  \@ifundefined{grOSp}{\newcommand{\grOSp}{\mathrm{OSp}}}{}
  \@ifundefined{grPSU}{\newcommand{\grPSU}{\mathrm{PSU}}}{}
  \@ifundefined{grSU}{\newcommand{\grSU}{\mathrm{SU}}}{}
  \@ifundefined{grU}{\newcommand{\grU}{\mathrm{U}}}{}
  \@ifundefined{grSL}{\newcommand{\grSL}{\mathrm{SL}}}{}
  \@ifundefined{grSp}{\newcommand{\grSp}{\mathrm{Sp}}}{}
  \@ifundefined{grSO}{\newcommand{\grSO}{\mathrm{SO}}}{}
  \@ifundefined{grO}{\newcommand{\grO}{\mathrm{O}}}{}
  \@ifundefined{algOSp}{\newcommand{\algOSp}{\mathrm{osp}}}{}
  \@ifundefined{algPSU}{\newcommand{\algPSU}{\mathrm{psu}}}{}
  \@ifundefined{algSU}{\newcommand{\algSU}{\mathrm{su}}}{}
  \@ifundefined{algSp}{\newcommand{\algSp}{\mathrm{sp}}}{}
  \@ifundefined{algSL}{\newcommand{\algSL}{\mathrm{sl}}}{}
  \@ifundefined{algGL}{\newcommand{\algGL}{\mathrm{gl}}}{}
  \@ifundefined{algU}{\newcommand{\algU}{\mathrm{u}}}{}
  \@ifundefined{algSO}{\newcommand{\algSO}{\mathrm{so}}}{}
  \@ifundefined{algO}{\newcommand{\algO}{\mathrm{o}}}{}
  \@ifundefined{Integers}{\newcommand{\Integers}{\mathrm{Z}}}{}
  \@ifundefined{Reals}{\newcommand{\Reals}{\mathrm{R}}}{} \makeatother
%bibliography generated by nb.bst v1.01 (C) 2003-2010 Niklas Beisert
\begin{thebibliography}{10}
\ifx\href\asklfhas\newcommand{\href}[2]{#2}\fi
\ifx\arxivref\asklfhas\newcommand{\arxivref}[2]{\href{http://arxiv.org/abs/#1}{#2}}\fi
\ifx\doiref\asklfhas\newcommand{\doiref}[2]{\href{http://dx.doi.org/#1}{#2}}\fi
\raggedright
\small
\parskip 0pt

\bibitem{Maldacena:1997re}
J.~M.~Maldacena,
\textit{``The large {N} limit of superconformal field theories and
  supergravity''},
\textsf{Adv.~Theor.~Math.~Phys.~2,~231~(1998)},
\texttt{\arxivref{hep-th/9711200}{hep-th/9711200}}.
%%CITATION = HEP-TH/9711200;%%

\bibitem{Witten:1998qj}
E.~Witten,
\textit{``Anti-de {S}itter space and holography''},
\textsf{Adv.~Theor.~Math.~Phys.~2,~253~(1998)},
\texttt{\arxivref{hep-th/9802150}{hep-th/9802150}}.
%%CITATION = HEP-TH/9802150;%%

\bibitem{Gubser:1998bc}
S.~S.~Gubser, I.~R.~Klebanov and A.~M.~Polyakov,
\textit{``Gauge theory correlators from non-critical string theory''},
\textsf{\doiref{10.1016/S0370-2693(98)00377-3}{Phys.~Lett.~B428,~105~(1998)}},
\texttt{\arxivref{hep-th/9802109}{hep-th/9802109}}.
%%CITATION = HEP-TH/9802109;%%

\bibitem{Beisert:2010jr}
N.~Beisert et~al.,
\textit{``Review of {AdS/CFT} Integrability: An Overview''},
\textsf{\doiref{10.1007/s11005-011-0529-2}{Lett.Math.Phys.~99,~3~(2010)}},
\texttt{\arxivref{1012.3982}{arxiv:1012.3982}}.
%%CITATION = 1012.3982;%%

\bibitem{Brown:1986nw}
J.~D.~Brown and M.~Henneaux,
\textit{``Central Charges in the Canonical Realization of Asymptotic
  Symmetries: An Example from Three-Dimensional Gravity''},
\textsf{\doiref{10.1007/BF01211590}{Commun.~Math.~Phys.~104,~207~(1986)}}.
%%CITATION = CMPHA,104,207;%%

\bibitem{Maldacena:2000hw}
J.~M.~Maldacena and H.~Ooguri,
\textit{``Strings in {$\AdS_3$} and {$\grSL(2,R)$} {WZW} model. {I}''},
\textsf{\doiref{10.1063/1.1377273}{J.~Math.~Phys.~42,~2929~(2001)}},
\texttt{\arxivref{hep-th/0001053}{hep-th/0001053}}.
%%CITATION = HEP-TH/0001053;%%

\bibitem{Maldacena:2000kv}
J.~M.~Maldacena, H.~Ooguri and J.~Son,
\textit{``Strings in {$\AdS_3$} and the {$\grSL(2,R)$} {WZW} model. {II}:
  {E}uclidean black hole''},
\textsf{\doiref{10.1063/1.1377039}{J.~Math.~Phys.~42,~2961~(2001)}},
\texttt{\arxivref{hep-th/0005183}{hep-th/0005183}}.
%%CITATION = HEP-TH/0005183;%%

\bibitem{Maldacena:2001km}
J.~M.~Maldacena and H.~Ooguri,
\textit{``Strings in {$\AdS_3$} and the {$\grSL(2,R)$} {WZW} model.~{III}:
  Correlation functions''},
\textsf{\doiref{10.1103/PhysRevD.65.106006}{Phys.~Rev.~D65,~106006~(2002)}},
\texttt{\arxivref{hep-th/0111180}{hep-th/0111180}}.
%%CITATION = HEP-TH/0111180;%%

\bibitem{Maldacena:1998bw}
J.~M.~Maldacena and A.~Strominger,
\textit{``{$\AdS_3$} black holes and a stringy exclusion principle''},
\textsf{\doiref{10.1088/1126-6708/1998/12/005}{JHEP~9812,~005~(1998)}},
\texttt{\arxivref{hep-th/9804085}{hep-th/9804085}}.
%%CITATION = HEP-TH/9804085;%%

\bibitem{Seiberg:1999xz}
N.~Seiberg and E.~Witten,
\textit{``The {D1/D5} system and singular {CFT}''},
\textsf{\doiref{10.1088/1126-6708/1999/04/017}{JHEP~9904,~017~(1999)}},
\texttt{\arxivref{hep-th/9903224}{hep-th/9903224}}.
%%CITATION = HEP-TH/9903224;%%

\bibitem{Larsen:1999uk}
F.~Larsen and E.~J.~Martinec,
\textit{``{$\grU(1)$} charges and moduli in the {D1}-{D5} system''},
\textsf{\doiref{10.1088/1126-6708/1999/06/019}{JHEP~9906,~019~(1999)}},
\texttt{\arxivref{hep-th/9905064}{hep-th/9905064}}.
%%CITATION = HEP-TH/9905064;%%

\bibitem{Gauntlett:1998kc}
J.~P.~Gauntlett, R.~C.~Myers and P.~K.~Townsend,
\textit{``Supersymmetry of rotating branes''},
\textsf{\doiref{10.1103/PhysRevD.59.025001}{Phys.~Rev.~D59,~025001~(1999)}},
\texttt{\arxivref{hep-th/9809065}{hep-th/9809065}}.
%%CITATION = HEP-TH/9809065;%%

\bibitem{Elitzur:1998mm}
S.~Elitzur, O.~Feinerman, A.~Giveon and D.~Tsabar,
\textit{``String theory on {$\AdS_3 \times \Sphere^3 \times \Sphere^3 \times
  \Sphere^1$}''},
\textsf{\doiref{10.1016/S0370-2693(99)00101-X}{Phys.~Lett.~B449,~180~(1999)}},
\texttt{\arxivref{hep-th/9811245}{hep-th/9811245}}.
%%CITATION = HEP-TH/9811245;%%

\bibitem{deBoer:1999rh}
J.~de~Boer, A.~Pasquinucci and K.~Skenderis,
\textit{``{AdS/CFT} dualities involving large 2d {$\superN = 4$} superconformal
  symmetry''},
\textsf{Adv.~Theor.~Math.~Phys.~3,~577~(1999)},
\texttt{\arxivref{hep-th/9904073}{hep-th/9904073}}.
%%CITATION = HEP-TH/9904073;%%

\bibitem{Gukov:2004ym}
S.~Gukov, E.~Martinec, G.~W.~Moore and A.~Strominger,
\textit{``The search for a holographic dual to {$\AdS_3 \times \Sphere^3 \times
  \Sphere^3 \times \Sphere^1$}''},
\textsf{Adv.~Theor.~Math.~Phys.~9,~435~(2005)},
\texttt{\arxivref{hep-th/0403090}{hep-th/0403090}}.
%%CITATION = HEP-TH/0403090;%%

\bibitem{Pakman:2009mi}
A.~Pakman, L.~Rastelli and S.~S.~Razamat,
\textit{``A Spin Chain for the Symmetric Product {$\CFT_2$}''},
\textsf{\doiref{10.1007/JHEP05(2010)099}{JHEP~1005,~099~(2009)}},
\texttt{\arxivref{0912.0959}{arxiv:0912.0959}}.
%%CITATION = 0912.0959;%%

\bibitem{Babichenko:2009dk}
A.~Babichenko, B.~Stefa{\'n}ski,~jr. and K.~Zarembo,
\textit{``Integrability and the {$\AdS_3/\CFT_2$} correspondence''},
\textsf{\doiref{10.1007/JHEP03(2010)058}{JHEP~1003,~058~(2010)}},
\texttt{\arxivref{0912.1723}{arxiv:0912.1723}}.
%%CITATION = 0912.1723;%%

\bibitem{Sundin:2012gc}
P.~Sundin and L.~Wulff,
\textit{``Classical integrability and quantum aspects of the {$\AdS_3 \times
  \Sphere^3 \times \Sphere^3 \times \Sphere^1$} superstring''},
\textsf{\doiref{10.1007/JHEP10(2012)109}{JHEP~1210,~109~(2012)}},
\texttt{\arxivref{1207.5531}{arxiv:1207.5531}}.
%%CITATION = ARXIV:1207.5531;%%

\bibitem{Cagnazzo:2012se}
A.~Cagnazzo and K.~Zarembo,
\textit{``{B}-field in {$\AdS_3/\CFT_2$} Correspondence and Integrability''},
\texttt{\arxivref{1209.4049}{arxiv:1209.4049}}.
%%CITATION = ARXIV:1209.4049;%%

\bibitem{OhlssonSax:2011ms}
O.~Ohlsson~Sax and B.~Stefa{\'n}ski,~jr.,
\textit{``Integrability, spin-chains and the {$\AdS_3/\CFT_2$}
  correspondence''},
\textsf{\doiref{10.1007/JHEP08(2011)029}{JHEP~1108,~029~(2011)}},
\texttt{\arxivref{1106.2558}{arxiv:1106.2558}}.
%%CITATION = 1106.2558;%%

\bibitem{Sax:2012jv}
O.~Ohlsson~Sax, B.~Stefa{\'n}ski,~jr. and A.~Torrielli,
\textit{``On the massless modes of the {$\AdS_3/\CFT_2$} integrable systems''},
\textsf{\doiref{10.1007/JHEP03(2013)109}{JHEP~1303,~109~(2013)}},
\texttt{\arxivref{1211.1952}{arxiv:1211.1952}}.
%%CITATION = ARXIV:1211.1952;%%

\bibitem{David:2008yk}
J.~R.~David and B.~Sahoo,
\textit{``Giant magnons in the {D1}-{D5} system''},
\textsf{\doiref{10.1088/1126-6708/2008/07/033}{JHEP~0807,~033~(2008)}},
\texttt{\arxivref{0804.3267}{arxiv:0804.3267}}.
%%CITATION = 0804.3267;%%

\bibitem{Beisert:2005tm}
N.~Beisert,
\textit{``The {$\algSU(2|2)$} dynamic {$S$}-matrix''},
\textsf{Adv.~Theor.~Math.~Phys.~12,~945~(2008)},
\texttt{\arxivref{hep-th/0511082}{hep-th/0511082}}.
%%CITATION = HEP-TH/0511082;%%

\bibitem{Borsato:2012ud}
R.~Borsato, O.~Ohlsson~Sax and A.~Sfondrini,
\textit{``A dynamic {$\algSU(1|1)^2$} {S}-matrix for {$\AdS_3/\CFT_2$}''},
\textsf{\doiref{10.1007/JHEP04(2013)113}{JHEP~1304,~113~(2013)}},
\texttt{\arxivref{1211.5119}{arxiv:1211.5119}}.
%%CITATION = ARXIV:1211.5119;%%

\bibitem{Borsato:2012ss}
R.~Borsato, O.~Ohlsson~Sax and A.~Sfondrini,
\textit{``All-loop {B}ethe ansatz equations for {$\AdS_3/\CFT_2$}''},
\textsf{\doiref{10.1007/JHEP04(2013)116}{JHEP~1304,~116~(2013)}},
\texttt{\arxivref{1212.0505}{arxiv:1212.0505}}.
%%CITATION = ARXIV:1212.0505;%%

\bibitem{Beisert:2007sk}
N.~Beisert and B.~I.~Zwiebel,
\textit{``On Symmetry Enhancement in the {$\algPSU(1,1|2)$} Sector of {$\superN
  = 4$} {SYM}''},
\textsf{\doiref{10.1088/1126-6708/2007/10/031}{JHEP~0710,~031~(2007)}},
\texttt{\arxivref{0707.1031}{arxiv:0707.1031}}.
%%CITATION = 0707.1031;%%

\bibitem{Hoare:2011fj}
B.~Hoare and A.~Tseytlin,
\textit{``Towards the quantum {S}-matrix of the {P}ohlmeyer reduced version of
  {$\AdS_5 \times \Sphere^5$} superstring theory''},
\textsf{\doiref{10.1016/j.nuclphysb.2011.05.016}{Nucl.Phys.~B851,~161~(2011)}},
\texttt{\arxivref{1104.2423}{arxiv:1104.2423}}.
%%CITATION = ARXIV:1104.2423;%%

\bibitem{Hoare:2013pma}
B.~Hoare and A.~A.~Tseytlin,
\textit{``On string theory on {$\AdS_3 \times \Sphere^3 \times \Torus^4$} with
  mixed 3-form flux: tree-level {S}-matrix''},
\texttt{\arxivref{1303.1037}{arxiv:1303.1037}}.
%%CITATION = ARXIV:1303.1037;%%

\bibitem{Rughoonauth:2012qd}
N.~Rughoonauth, P.~Sundin and L.~Wulff,
\textit{``Near {BMN} dynamics of the {$\AdS_3 \times \Sphere^3 \times \Sphere^3
  \times \Sphere^1$} superstring''},
\textsf{\doiref{10.1007/JHEP07(2012)159}{JHEP~1207,~159~(2012)}},
\texttt{\arxivref{1204.4742}{arxiv:1204.4742}}.
%%CITATION = ARXIV:1204.4742;%%

\bibitem{Sundin:2013ypa}
P.~Sundin and L.~Wulff,
\textit{``Worldsheet scattering in {$\AdS_3/\CFT_2$}''},
\texttt{\arxivref{1302.5349}{arxiv:1302.5349}}.
%%CITATION = ARXIV:1302.5349;%%

\bibitem{David:2010yg}
J.~R.~David and B.~Sahoo,
\textit{``{S}-matrix for magnons in the {D1}-{D5} system''},
\textsf{\doiref{10.1007/JHEP10(2010)112}{JHEP~1010,~112~(2010)}},
\texttt{\arxivref{1005.0501}{arxiv:1005.0501}}.
%%CITATION = 1005.0501;%%

\bibitem{Ahn:2012hw}
C.~Ahn and D.~Bombardelli,
\textit{``Exact {S}-matrices for {$\AdS_3/\CFT_2$}''},
\texttt{\arxivref{1211.4512}{arxiv:1211.4512}}.
%%CITATION = ARXIV:1211.4512;%%

\bibitem{Arutyunov:2006yd}
G.~Arutyunov, S.~Frolov and M.~Zamaklar,
\textit{``The {Z}amolodchikov-{F}addeev algebra for {$\AdS_5 \times \Sphere^5$}
  superstring''},
\textsf{\doiref{10.1088/1126-6708/2007/04/002}{JHEP~0704,~002~(2007)}},
\texttt{\arxivref{hep-th/0612229}{hep-th/0612229}}.
%%CITATION = HEP-TH/0612229;%%

\bibitem{Arutyunov:2009ga}
G.~Arutyunov and S.~Frolov,
\textit{``Foundations of the {$\AdS_5 \times \Sphere^5$} Superstring. Part
  {I}''},
\textsf{\doiref{10.1088/1751-8113/42/25/254003}{J.Phys.A~A42,~254003~(2009)}},
\texttt{\arxivref{0901.4937}{arxiv:0901.4937}}.
%%CITATION = ARXIV:0901.4937;%%

\bibitem{Vieira:2010kb}
P.~Vieira and D.~Volin,
\textit{``Review of {AdS/CFT} Integrability, {C}hapter {III.3}: The dressing
  factor''},
\textsf{\doiref{10.1007/s11005-011-0482-0}{Lett.Math.Phys.~99,~231~(2010)}},
\texttt{\arxivref{1012.3992}{arxiv:1012.3992}}.
%%CITATION = 1012.3992;%%

\bibitem{Janik:2006dc}
R.~A.~Janik,
\textit{``The {$\AdS_5 \times \Sphere^5$} superstring worldsheet {S}-matrix and
  crossing symmetry''},
\textsf{\doiref{10.1103/PhysRevD.73.086006}{Phys.~Rev.~D73,~086006~(2006)}},
\texttt{\arxivref{hep-th/0603038}{hep-th/0603038}}.
%%CITATION = HEP-TH/0603038;%%

\bibitem{upcoming}
R.~Borsato, O.~Ohlsson~Sax, A.~Sfondrini, B.~Stefa{\'n}ski,~jr. and
  A.~Torrielli,
to appear.

\bibitem{Arutyunov:2004vx}
G.~Arutyunov, S.~Frolov and M.~Staudacher,
\textit{``{B}ethe ansatz for quantum strings''},
\textsf{\doiref{10.1088/1126-6708/2004/10/016}{JHEP~0410,~016~(2004)}},
\texttt{\arxivref{hep-th/0406256}{hep-th/0406256}}.
%%CITATION = HEP-TH/0406256;%%

\bibitem{Gomez:2006va}
C.~G{\'o}mez and R.~Hern{\'a}ndez,
\textit{``The magnon kinematics of the {A}d{S/CFT} correspondence''},
\textsf{\doiref{10.1088/1126-6708/2006/11/021}{JHEP~0611,~021~(2006)}},
\texttt{\arxivref{hep-th/0608029}{hep-th/0608029}}.
%%CITATION = HEP-TH/0608029;%%

\bibitem{Plefka:2006ze}
J.~Plefka, F.~Spill and A.~Torrielli,
\textit{``On the {H}opf algebra structure of the {A}d{S/CFT} {S}-matrix''},
\textsf{\doiref{10.1103/PhysRevD.74.066008}{Phys.~Rev.~D74,~066008~(2006)}},
\texttt{\arxivref{hep-th/0608038}{hep-th/0608038}}.
%%CITATION = HEP-TH/0608038;%%

\bibitem{Beisert:2005wm}
N.~Beisert,
\textit{``An {$\algSU(1|1)$}-invariant {S}-matrix with dynamic
  representations''},
\textsf{Bulg.~J.~Phys.~33S1,~371~(2006)},
\texttt{\arxivref{hep-th/0511013}{hep-th/0511013}}.
%%CITATION = HEP-TH/0511013;%%

\bibitem{Beisert:2005fw}
N.~Beisert and M.~Staudacher,
\textit{``Long-range {$\grPSU(2,2|4)$} {B}ethe ansaetze for gauge theory and
  strings''},
\textsf{\doiref{10.1016/j.nuclphysb.2005.06.038}{Nucl.~Phys.~B727,~1~(2005)}},
\texttt{\arxivref{hep-th/0504190}{hep-th/0504190}}.
%%CITATION = HEP-TH/0504190;%%

\bibitem{deLeeuw:2007uf}
M.~de~Leeuw,
\textit{``Coordinate {B}ethe Ansatz for the String {S}-Matrix''},
\textsf{\doiref{10.1088/1751-8113/40/48/008}{J.Phys.A~A40,~14413~(2007)}},
\texttt{\arxivref{0705.2369}{arxiv:0705.2369}}.
%%CITATION = ARXIV:0705.2369;%%

\bibitem{Kazakov:2004qf}
V.~A.~Kazakov, A.~Marshakov, J.~A.~Minahan and K.~Zarembo,
\textit{``Classical/quantum integrability in {AdS/CFT}''},
\textsf{\doiref{10.1088/1126-6708/2004//24}{JHEP~5,~24~(2004)}},
\texttt{\arxivref{hep-th/0402207}{hep-th/0402207}}.
%%CITATION = HEP-TH/0402207;%%

\bibitem{Zarembo:2010yz}
K.~Zarembo,
\textit{``Algebraic Curves for Integrable String Backgrounds''},
\texttt{\arxivref{1005.1342}{arxiv:1005.1342}}.
%%CITATION = 1005.1342;%%

\bibitem{Abbott:2012dd}
M.~C.~Abbott,
\textit{``Comment on Strings in {$\AdS_3 \times \Sphere^3 \times \Sphere^3
  \times \Sphere^1$} at One Loop''},
\texttt{\arxivref{1211.5587}{arxiv:1211.5587}}.
%%CITATION = ARXIV:1211.5587;%%

\bibitem{Beccaria:2012kb}
M.~Beccaria, F.~Levkovich-Maslyuk, G.~Macorini and A.~A.~Tseytlin,
\textit{``Quantum corrections to spinning superstrings in {$\AdS_3 \times
  \Sphere^3 \times M^4$}: determining the dressing phase''},
\texttt{\arxivref{1211.6090}{arxiv:1211.6090}}.
%%CITATION = ARXIV:1211.6090;%%

\bibitem{Maldacena:2006rv}
J.~Maldacena and I.~Swanson,
\textit{``Connecting giant magnons to the pp-wave: An interpolating limit of
  {$\AdS_5 \times \Sphere^5$}''},
\textsf{\doiref{10.1103/PhysRevD.76.026002}{Phys.~Rev.~D76,~26002~(2007)}},
\texttt{\arxivref{hep-th/0612079}{hep-th/0612079}}.
%%CITATION = HEP-TH/0612079;%%

\bibitem{Hernandez:2006tk}
R.~Hern{\'a}ndez and E.~L{\'o}pez,
\textit{``Quantum corrections to the string {B}ethe ansatz''},
\textsf{\doiref{10.1088/1126-6708/2006/07/004}{JHEP~0607,~004~(2006)}},
\texttt{\arxivref{hep-th/0603204}{hep-th/0603204}}.
%%CITATION = HEP-TH/0603204;%%

\bibitem{Ahn:2010ka}
C.~Ahn and R.~I.~Nepomechie,
\textit{``Review of {AdS/CFT} Integrability, {C}hapter {III.2}: Exact
  world-sheet {S}-matrix''},
\textsf{\doiref{10.1007/s11005-011-0478-9}{Lett.Math.Phys.~99,~209~(2010)}},
\texttt{\arxivref{1012.3991}{arxiv:1012.3991}}.
%%CITATION = 1012.3991;%%

\bibitem{Dorey:1996gd}
P.~Dorey,
\textit{``Exact {$S$} matrices''},
\texttt{\arxivref{hep-th/9810026}{hep-th/9810026}}.
%%CITATION = HEP-TH/9810026;%%

\bibitem{Torrielli:2010kq}
A.~Torrielli,
\textit{``Review of {AdS/CFT} Integrability, {C}hapter {VI.2}: {Y}angian
  Algebra''},
\textsf{\doiref{10.1007/s11005-011-0491-z}{Lett.Math.Phys.~99,~547~(2010)}},
\texttt{\arxivref{1012.4005}{arxiv:1012.4005}}.
%%CITATION = 1012.4005;%%

\bibitem{Torrielli:2011gg}
A.~Torrielli,
\textit{``{Y}angians, {S}-matrices and {$\AdS/\CFT$}''},
\textsf{\doiref{10.1088/1751-8113/44/26/263001}{J.Phys.~A44,~263001~(2011)}},
\texttt{\arxivref{1104.2474}{arxiv:1104.2474}}.
%%CITATION = ARXIV:1104.2474;%%

\bibitem{Torrielli:2007mc}
A.~Torrielli,
\textit{``Classical r-matrix of the {$\algSU(2|2)$} {SYM} spin-chain''},
\textsf{\doiref{10.1103/PhysRevD.75.105020}{Phys.~Rev.~D75,~105020~(2007)}},
\texttt{\arxivref{hep-th/0701281}{hep-th/0701281}}.
%%CITATION = HEP-TH/0701281;%%

\bibitem{Drinfeld:1989st}
V.~Drinfeld,
\textit{``Quasi {H}opf algebras''},
\textsf{Alg.Anal.~1N6,~114~(1989)}.
%%CITATION = 00040,1N6,114;%%

\bibitem{Khoroshkin:1994uk}
S.~Khoroshkin and V.~Tolstoi,
\textit{``{Y}angian double and rational {R} matrix''},
\textsf{Lett.Math.Phys~36,~373~(1994)},
\texttt{\arxivref{hep-th/9406194}{hep-th/9406194}}.
%%CITATION = HEP-TH/9406194;%%

\bibitem{Cai:1997}
J.~Cai, S.~Wang, K.~Wu and C.~Xiong,
\textit{``Universal {R}-matrix Of The Super Yangian Double
  {$DY(\algGL(1|1))$}''},
\textsf{Comm.~Theor.~Phys~29,~173~(1998)},
\texttt{\arxivref{q-alg/9709038}{q-alg/9709038}}.

\bibitem{Arutyunov:2009ce}
G.~Arutyunov, M.~de~Leeuw and A.~Torrielli,
\textit{``Universal blocks of the {A}d{S/CFT} Scattering Matrix''},
\textsf{\doiref{10.1088/1126-6708/2009/05/086}{JHEP~0905,~086~(2009)}},
\texttt{\arxivref{0903.1833}{arxiv:0903.1833}}.
%%CITATION = ARXIV:0903.1833;%%

\bibitem{Drinfeld:1987sy}
V.~Drinfeld,
\textit{``A New realization of {Y}angians and quantized affine algebras''},
\textsf{Sov.Math.Dokl.~36,~212~(1988)}.
%%CITATION = SVMDA,36,212;%%

\end{thebibliography}

\end{document}